\documentclass[12pt,a4paper,twoside]{article}
\usepackage{amsfonts}
\usepackage{amssymb}
\usepackage{graphicx}
\usepackage{amsthm}
\usepackage{amsmath}
\usepackage{cite}

\def \a {a}
\def \dd {{\mathrm d}}
\newcommand{\scri}{{\mathcal{I}}}
\def \rovno {&\!\!\!=\!\!\!&}

\newcommand{\tg}{\tan}
\newcommand{\cotg}{\cot}
\newcommand{\tgh}{\tanh}
\newcommand{\cotgh}{\coth}
\newcommand{\sign}{\mathop{\rm sign}\nolimits}
\newcommand{\konst}{\mathop{\rm const.}\nolimits}
\newcommand{\y}{\phi}
\newcommand{\yy}{\zeta}
\newcommand{\q}{\psi}
\newcommand{\p}{\vartheta}
\newcommand{\ez}{\epsilon_{0}}
\newcommand{\et}{\epsilon_{2}}
\newcommand{\pul}{{\textstyle\frac{1}{2}}}

\begin{document}

\section*{Yet another family of diagonal metrics \\
for de~Sitter and anti-de~Sitter spacetimes}

\vspace{3mm}

\textbf{Ji\v{r}\'\i\  Podolsk\'y, Ond\v{r}ej Hru\v{s}ka}

\vspace{3mm}

\small{Institute of Theoretical Physics, Faculty of Mathematics and Physics,

Charles University, Prague, V~Hole\v{s}ovi\v{c}k\'ach~2, 180~00 Praha 8, Czech Republic

\texttt{podolsky@mbox.troja.mff.cuni.cz} and \texttt{HruskaOndrej@seznam.cz}
}

\today


\section*{Abstract}

In this work we present and analyze a new class of coordinate representations of de~Sitter and anti-de~Sitter spacetimes for which the metrics are diagonal and (typically) static and axially symmetric. Contrary to the well-known forms of these fundamental geometries, that usually correspond to a ${1+3}$ foliation with the 3-space of a constant spatial curvature, the new metrics are adapted to a ${2+2}$ foliation, and are warped products of two 2-spaces of constant curvature. This new class of (anti-)de~Sitter metrics depends on the value of cosmological constant $\Lambda$ and two discrete parameters ${+1,0,-1}$ related to the curvature of the 2-spaces. The class admits 3 distinct subcases for ${\Lambda>0}$ and 8 subcases for ${\Lambda<0}$. We systematically study all these possibilities. In particular, we explicitly present the corresponding parametrizations of the (anti-)de Sitter hyperboloid, visualize the coordinate lines and surfaces within the global conformal cylinder, investigate their mutual relations, present some closely related forms of the metrics, and give transformations to standard de~Sitter and anti-de~Sitter metrics.
Using these results, we also provide a physical interpretation of $B$-metrics as exact gravitational fields of a tachyon.

\section{Introduction}

Almost exactly 100 years ago in 1917 Willem de Sitter published two seminal papers \cite{deSitter1917a, deSitter1917b} in which he presented his, now famous, exact solution of Einstein's gravitational field equations without matter but with a positive cosmological constant~$\Lambda$. Together with the pioneering work from the same year by Einstein himself \cite{Einstein1917}, they mark the very origin of relativistic cosmology.

Since then, the vacuum de Sitter universe with ${\Lambda>0}$ and its formal antipode with ${\Lambda<0}$, later nicknamed the anti-de Sitter universe, have become truly fundamental spacetimes. They admit a cornucopia of applications, ranging from purely mathematical studies, theoretical investigations in classical general relativity, quantum field theory and string theory in higher dimensions, to contemporary ``inflationary'' and ``dark matter'' cosmologies, modeling naturally the observed accelerated expansion of the universe, both in its distant past and future.

 This remarkable number of applications in various research branches clearly stems from the fact that the (anti-)de~Sitter spacetime is highly symmetric and simple, yet it is non-trivial by being curved everywhere. In fact, it is \emph{maximally symmetric} and has a \emph{constant curvature}, just as the flat Minkowski space. Of course, this greatly simplifies all analyses and studies.

One would be inclined to expect that, after a century of thorough investigation of various aspects of these important spacetimes, there is not much left to add or discover. Quite surprisingly, still there are interesting new properties which deserve some attention. To celebrate the centenary of de Sitter space, we take the liberty of presenting here yet another family of coordinate representations of de Sitter and anti-de Sitter 4-dimensional solutions, in which the metric is \emph{diagonal} and typically \emph{static and axially symmetric}. It naturally arises in the context of $B$-metrics \cite{EhlersKundt1962, Stephani:2003, GriPod2009} with a cosmological constant~$\Lambda$, which belong to a wider class of Pleba\'nski--Demia\'nski non-expanding spacetimes \cite{PleDem1976, GriPod2006b}. As far as we know, this family has not yet been considered and systematically described. We hope that it could find useful applications in general relativity and high energy physics theories.

In particular, in our work we will present and analyze the family of metrics for de Sitter and anti--de Sitter spacetimes which can be written in a unified form
\begin{equation}\label{eq:PDdads}
 \dd s^{2}=p^{2}\Big(\!-Q\,\dd t^{2}+\frac{\dd q^{2}}{Q}\Big)
   +\frac{\dd p^{2}}{P}+P\,a^2\dd \y^{2}\,,
\end{equation}
with
\begin{eqnarray}
Q(q) \rovno \ez-\epsilon_{2}\,q^{2}\,, \label{eq:Q}\\
P(p) \rovno \et-\frac{\Lambda}{3}\, p^{2}\,, \label{eq:P}
\end{eqnarray}
where ${\,\a^2=3/|\Lambda|\,}$ and ${\,\ez, \et = 1,0,-1\,}$ are two independent discrete parameters.

The metric has the geometry of a \emph{warped product} of two \emph{2-spaces of constant curvature}, namely ${{dS}_2, {M}_2, {AdS}_2}$ (according to the sign of $\et$) spanned by ${t,q}$\,, and ${{S}^2, {E}^2, {H}^{2}}$ (according to the sign of $\Lambda$) spanned by ${p,\y}$. The warp factor is $p^2$. Also, there are two obvious symmetries corresponding to the Killing vectors $\partial_t$ and $\partial_\y$. The metric is clearly static when ${Q>0}$. It is also axially symmetric when the axis given by ${P=0}$ is admitted (for ${\Lambda>0, \epsilon_{2}=1}$ and for ${\Lambda<0, \epsilon_{2}=-1}$), in which case this axis is regular for ${\y\in[0,2\pi) }$.

The metric (\ref{eq:PDdads})--(\ref{eq:P})  is a special subcase of the Pleba\'{n}ski--Demia\'{n}ski non-expanding (and thus Kundt) type~D solutions, see metric (16.27) of \cite{GriPod2009}. It is obtained by setting ${\gamma=0}$ and ${e=0=g}$ (so that ${m=0=k}$) with the identification
\begin{equation}\label{eq:PDsubcase}
 Q \equiv \tilde{\mathcal{Q}}=\ez-\epsilon_{2}\,q^{2}\,, \qquad
 P \equiv \frac{\mathcal{P}}{p^2}=\frac{2n}{p}+\et-\frac{\Lambda}{3}\, p^{2}\,.
\end{equation}
This gives vacuum $B$-metrics, as classified by Ehlers and Kundt \cite{EhlersKundt1962}, generalized here to any value of the cosmological constant $\Lambda$. When ${n=0}$, the curvature singularity at ${p=0}$ disappears and the metric reduces to the conformally flat (anti-)de~Sitter space with (\ref{eq:P}). Therefore, the family of de Sitter and anti--de Sitter metrics (\ref{eq:PDdads})--(\ref{eq:P}) can be viewed as a \emph{natural background for the B-metrics with $\Lambda$}, and will thus play a key role in understanding their geometrical and physical properties. This is our main motivation.

The structure of the present work is simple. In the next Section~\ref{sec2} we recall basic properties of (anti-)de Sitter spacetimes, introducing the notation. In Section~\ref{sec3} we identify all distinct subclasses of the general metric (\ref{eq:PDdads})--(\ref{eq:P}). These are studied in detail in subsequent Section~\ref{sc:dS} for the de Sitter case ${\Lambda>0}$, and for the anti-de~Sitter case ${\Lambda<0}$ in Section~\ref{sc:adS}. An interesting physical application is given in Section~\ref{sec6}. There are also four appendices in which we present a unified form of the new parametrizations, their mutual relations, further metric forms, and transformations to standard (anti-)de~Sitter metrics.

\section{The de~Sitter and anti-de~Sitter spacetimes}
\label{sec2}

Minkowski, de~Sitter and anti-de~Sitter spacetimes are the most fundamental and simplest exact solutions of Einstein's field equations. They are maximally symmetric, have constant curvature, and are the only vacuum solutions with a vanishing Weyl tensor --- they are conformally flat and satisfy the field equations ${R_{\alpha\beta}=\Lambda g_{\alpha\beta}}$, where~$\Lambda$ is a cosmological constant, so that in four spacetime dimensions the curvature is ${R=4\Lambda}$. Minkowski, de~Sitter and anti-de~Sitter spacetimes are thus Einstein spaces with vanishing, positive, and negative cosmological constant~$\Lambda$, respectively.

Specific properties of these classic spacetimes have been investigated and employed in various contexts for a century. They were described in many reviews,  such as by Schr\"odinger \cite{Schroedinger1956}, Penrose \cite{Penrose1968a}, Weinberg \cite{Weinberg1972}, M{\o}ller \cite{Moler1972}, Hawking and Ellis \cite{HawkingEllis1973}, Birell and Davies \cite{BirellDavies1982}, Bi\v{c}\'ak \cite{Bicak2000a}, Bi\v{c}\'ak and Krtou\v{s} \cite{BicakKrtous2005}, Gr\o{n} and Hervik \cite{GronHervik2007}, or Griffiths and Podolsk\'y \cite{GriPod2009}. Let us recall here only those expressions and properties that will be important for our analysis.

\subsection{The de~Sitter spacetime}
\label{sec2a}

\begin{figure}[htp] 
\begin{center}
   \includegraphics[scale=1, trim=-20 10 0 10]{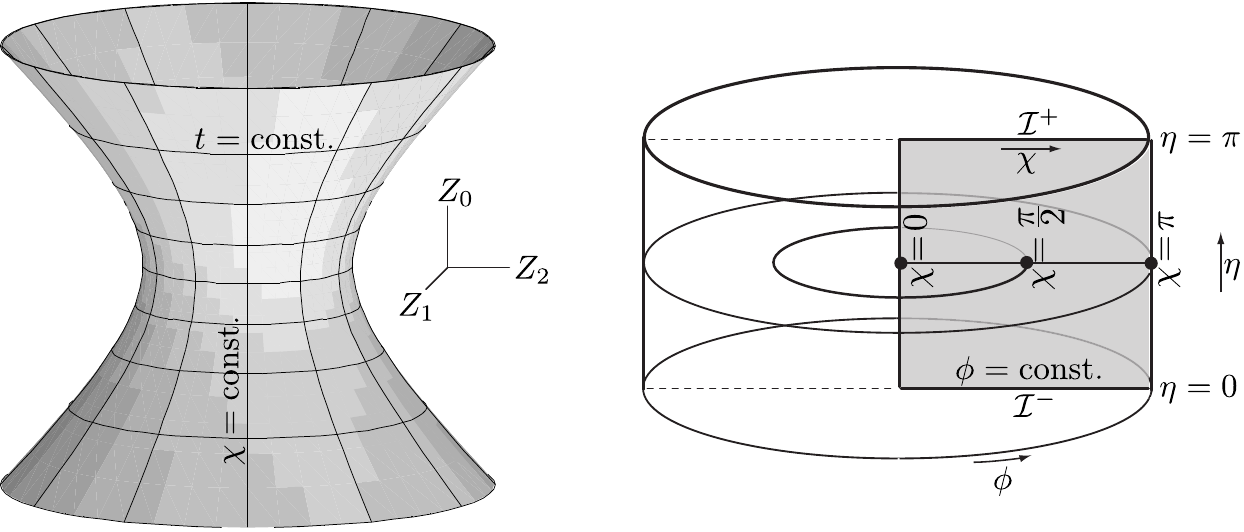}
\end{center}
\caption{\small
    Left: The de Sitter spacetime as a hyperboloid embedded in a flat 5D spacetime with the global coordinates ${(t,\chi,\theta,\phi)}$, drawn for ${Z_3=0=Z_4}$ (i.e., ${\theta=0, \pi}$). With the full range of $\theta$, $\phi$ reintroduced, each point on this hyperboloid represents a 2D hemisphere. Right: The global structure of de~Sitter spacetime (for ${\theta=\frac{\pi}{2}}$). It is conformal to the region ${\eta\in(0,\pi)}$ of the Einstein static universe, represented as an embedded solid cylinder whose radius and length are both equal to~$\pi$. The centre ${\chi=0}$ is the North pole of the 3-sphere $S^3$, whereas the outer boundary ${\chi=\pi}$ is its South pole. The conformal infinities $\scri^-$ and $\scri^+$ are spacelike. The shaded region is the Penrose diagram of the de~Sitter space for a fixed value of ${\phi\in[0,2\pi)}$. }
\label{fig:dshyp1}
\end{figure}

Since the original works by de~Sitter \cite{deSitter1917a, deSitter1917b} and Lanczos \cite{Lanczos1922} it is known that the de~Sitter manifold can be conveniently visualized as the hyperboloid
\begin{equation}
-Z_0^2+Z_1^2+Z_2^2+Z_3^2+Z_4^2=\a^2 \,,\quad \hbox{ where }\quad \a=\sqrt{3/\Lambda}\,,
\label{C1}
\end{equation}
embedded in a flat five-dimensional Minkowski space
\begin{equation}
\dd s^2=-\dd Z_0^2+\dd Z_1^2+\dd Z_2^2+\dd Z_3^2+\dd Z_4^2 \,.
\label{C2}
\end{equation}
This geometrical representation of de~Sitter spacetime is related to its symmetry structure characterised by the ten-parameter group SO(1,4). The entire hyperboloid (\ref{C1}) is covered by coordinates ${t\in(-\infty,+\infty)}$, ${\chi\in[0,\pi]}$, ${\theta\in[0,\pi]}$, ${\phi\in[0,2\pi)}$ such that
\begin{eqnarray}
&&Z_0=\a\sinh\frac{t}{\a} \,, \nonumber\\
&&Z_1=\a\cosh\frac{t}{\a}\,\cos \chi \,, \nonumber\\
&&Z_2=\a\cosh\frac{t}{\a}\,\sin \chi \cos \theta \,, \label{C3}\\
&&Z_3=\a\cosh\frac{t}{\a}\,\sin \chi \sin \theta \cos \phi \,, \nonumber\\
&&Z_4=\a\cosh\frac{t}{\a}\,\sin \chi \sin \theta \sin \phi \,, \nonumber
\end{eqnarray}
in which the de Sitter metric takes the FLRW form
\begin{equation}
\dd s^2=-\dd t^2+\a^2\cosh^2 \frac{t}{\a}\,\Big( \dd\chi^2+\sin^2\chi
 \big(\dd\theta^2+\sin^2\theta\> \dd\phi^2\big)\Big) \,.
\label{C4}
 \end{equation}
The spatial sections at a fixed synchronous time~$t$ are 3-spheres $S^3$ of constant positive curvature which have radius ${\>\a\cosh \frac{t}{\a}\>}$. These contract to a minimum size $\a$ at ${t=0}$, and then re-expand. The de~Sitter spacetime thus has a natural topology $R^1\times S^3$.
This most natural parametrisation of the hyperboloid is illustrated in the left part of Figure~\ref{fig:dshyp1}.

\emph{Global causal structure} of the de~Sitter spacetime can be analyzed by introducing
a conformal time $\eta$ and the conformal factor $\Omega$ by
 \begin{equation}
 \Omega=\sin\eta={\rm sech}\,\frac{t}{a} \,.
 \label{C5}
 \end{equation}
The metric (\ref{C4}) then becomes
\begin{equation}
\dd s^2=\frac{\a^2}{\Omega^2}\Big(-\dd\eta^2+ \dd\chi^2+\sin^2\chi
 \big(\dd\theta^2+\sin^2\theta\> \dd\phi^2\big)\Big) \,,
\label{C6}
\end{equation}
corresponding to the parametrization
\begin{eqnarray}
&&Z_0= \a\,\cot\eta \,, \nonumber\\
&&Z_1=\frac{\a}{\sin\eta}\,\cos \chi \,, \nonumber\\
&&Z_2=\frac{\a}{\sin\eta}\,\sin \chi \cos \theta \,, \label{confDS}\\
&&Z_3=\frac{\a}{\sin\eta}\,\sin \chi \sin \theta \cos \phi \,, \nonumber\\
&&Z_4=\frac{\a}{\sin\eta}\,\sin \chi \sin \theta \sin \phi \,. \nonumber
\end{eqnarray}
The de Sitter spacetime is thus conformal to a part of the Einstein static universe.
Its boundary given by ${\Omega=0}$ is located at ${\eta=0}$ and ${\eta=\pi}$, which correspond to past and future conformal infinities $\scri^-$ and $\scri^+$, respectively, as illustrated in the right part of Figure~\ref{fig:dshyp1}. These infinities have a \emph{spacelike character}. For more details see, e.g., \cite{GriPod2009, GronHervik2007, HawkingEllis1973}.

\subsection{The anti-de~Sitter spacetime}
\label{sec2b}

Analogously, the anti-de~Sitter manifold can be viewed as the hyperboloid
 \begin{equation}
 -Z_0^2+Z_1^2+Z_2^2+Z_3^2-Z_4^2=-\a^2 ,\quad \hbox{ where }\quad \a=\sqrt{-3/\Lambda}\> ,
 \label{D1}
 \end{equation}
 embedded in a flat five-dimensional space
 \begin{equation}
 \dd s^2=-\dd Z_0^2+\dd Z_1^2+\dd Z_2^2+\dd Z_3^2-\dd Z_4^2 \,,
 \label{D2}
 \end{equation}
 which has two temporal dimensions $Z_0$ and $Z_4$. This representation of anti-de~Sitter space reflects its symmetry structure characterised by the ten-parameter group of isometries SO(2,3).

\begin{figure}[htp]
\begin{center}
   \includegraphics[scale=1, trim=0 5 0 5]{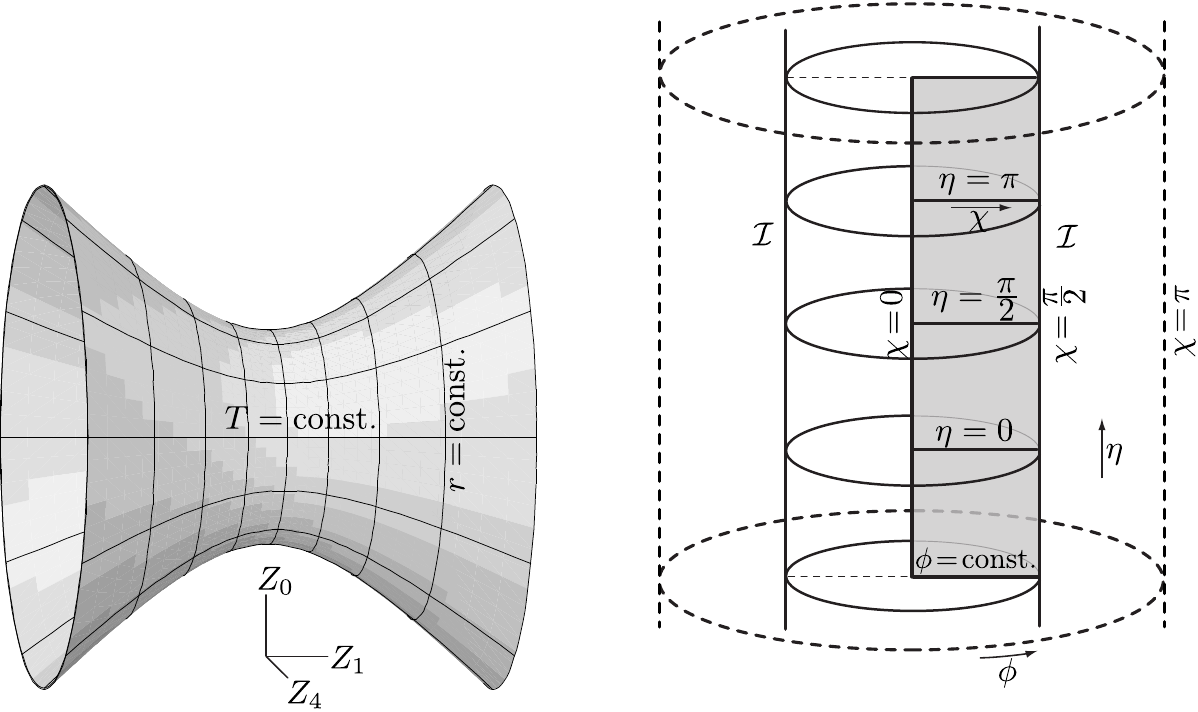}
\end{center}
\caption{\small
    Left: The anti-de Sitter spacetime as a hyperboloid embedded in a flat 5D spacetime in the parametrisation by global static coordinates ${(T,r,\theta,\phi)}$. The surface drawn is that for ${Z_2=0=Z_3}$ (that is ${\theta=0, \pi}$). With $\theta$ and $\phi$ reintroduced, each point represents a 2D hemisphere.
    Right: The global conformal structure of anti-de~Sitter spacetime (with ${\theta=\frac{\pi}{2}}$). It is conformal to the region ${\chi \in [0,\frac{\pi}{2})}$ of the Einstein static universe represented here (by dashed lines) as a solid cylinder of radius~$\pi$ and infinite length. The boundary ${\chi=\frac{\pi}{2}}$ is the anti-de~Sitter conformal infinity~${\cal I}$, which has a timelike character and topology ${R^1\times S^2}$. The Penrose diagram, corresponding to 2D shaded section, is obtained by fixing the coordinate~$\phi$.}
\label{fig:adshyp1}
\end{figure}

The most natural static coordinates ${(T,r,\theta,\phi)}$ covering the entire hyperboloid are\footnote{Notice the swap ${Z_0 \leftrightarrow Z_4}$ with respect to the parametrization (5.3) employed in \cite{GriPod2009}.}
\begin{eqnarray}
&&Z_0 = \a\,\cosh{r}\cos\frac{T}{\a} \,, \nonumber\\
&&Z_1 = \a\,\sinh{r}\cos \theta \,,\nonumber\\
&&Z_2 = \a\,\sinh{r}\sin \theta \cos \phi \,,\label{D3}\\
&&Z_3 = \a\,\sinh{r}\sin \theta \sin \phi \,,\nonumber\\
&&Z_4 = \a\,\cosh{r}\sin\frac{T}{\a} \,,\nonumber
\end{eqnarray}
see the left part of Figure~\ref{fig:adshyp1}, in which the anti-de Sitter metric reads
\begin{equation}
\dd s^2=-\cosh^2 r\> \dd T^2+\a^2 \Big(\dd r^2+\sinh^2 r\>\big(\dd\theta^2+\sin^2\theta\> \dd\phi^2\big)\Big)\,.
\label{D4}
\end{equation}
Any spatial section ${T=}$~const.~is a 3-space of constant negative curvature spanned by ${r\in[0,\infty)}$, ${\theta\in[0,\pi]}$, ${\phi\in[0,2\pi)}$. The singularities at ${r=0}$ and ${\theta=0, \pi}$ are only coordinate singularities.

The two temporal dimensions  $Z_0$ and $Z_4$ in the 5D flat space are parametrised using (\ref{D3}) by a \emph{single} time coordinate $T$ that is \emph{periodic}: values of $T$ which differ by a multiple of  ${2\pi a}$ represent the same points on the hyperboloid. Thus, the anti-de Sitter spacetime defined in this way has the topology $S^1\times R^3$, and contains closed timelike worldlines. This periodicity of $T$ is not evident in the four-dimensional metric (\ref{D4}), and it is possible to take ${T\in(-\infty,+\infty)}$. Such a range of coordinates corresponds to an infinite number of turns around the hyperboloid. It is usual to unwrap the circle $S^1$ and extend it to the whole $R^1$ instead, without reference to the parametrisation (\ref{D3}). One thus obtains a \emph{universal covering space} of the anti-de~Sitter universe with topology $R^4$ without closed timelike curves.

To represent \emph{global causal structure} of the anti-de~Sitter spacetime, a conformal spatial coordinate~$\chi$ and the conformal factor $\Omega$ may be introduced as
\begin{equation}
\tan\chi= \sinh r \,, \qquad
 \Omega=\cos\chi\,.
\label{confactantiDS}
\end{equation}
Writing ${\>\eta\equiv T/\a\>}$, the metric (\ref{D4}) then takes the form
\begin{equation}
\dd s^2=\frac{\a^2}{\Omega^2}\Big(-\dd\eta^2
 + \dd\chi^2+\sin^2\chi \big(\dd\theta^2+\sin^2\theta\> \dd\phi^2\big)\Big) \,,
\label{D6}
\end{equation}
corresponding to the parametrization
\begin{eqnarray}
&&Z_0= \a\,\frac{\cos\eta}{\cos\chi} \,, \nonumber\\
&&Z_1= \a\,\tan\chi\,\cos \theta \,, \nonumber\\
&&Z_2= \a\,\tan\chi\,\sin \theta \cos \phi \,, \label{confADS}\\
&&Z_3= \a\,\tan\chi\,\sin \theta \sin \phi \,, \nonumber\\
&&Z_4= \a\,\frac{\sin\eta}{\cos\chi} \,. \nonumber
\end{eqnarray}
The whole (universal) anti-de Sitter spacetime is thus conformal to the region ${\chi \in [0,\frac{\pi}{2})}$ of the Einstein static universe, see the right part of Figure~\ref{fig:adshyp1}. The anti-de~Sitter conformal infinity $\scri$, given by ${\Omega=0}$, is located at the boundary ${\chi=\frac{\pi}{2}}$  (corresponding to ${r=\infty}$). In contrast to the de~Sitter space, the conformal infinity~$\scri$ in the anti-de~Sitter spacetime forms a \emph{timelike surface} ${\chi=\frac{\pi}{2}}$.

\section{Subcases of the new (anti-)de Sitter metric for distinct values of $\Lambda$, $\epsilon_{2}$, $\epsilon_{0}$}
\label{sec3}

First, we are going to summarize different possible forms of the metric (\ref{eq:PDdads}) with (\ref{eq:Q}), (\ref{eq:P}) for the (anti-)de Sitter space, depending on the cosmological constant ${\Lambda>0}$ or ${\Lambda<0}$, and the two discrete parameters ${\epsilon_{2}= +1,0,-1\,}$ and ${\epsilon_{0}= +1,0,-1\,}$.

To keep the correct signature of the metric, the metric function $P(p)$ \emph{must be positive}, ${P>0}$ (otherwise there would be two additional temporal coordinates $p$ and $\y$). This puts a constraint on the parameter $\epsilon_{2}$ and the range of $p$. On the other hand, the function $Q(q)$ \emph{can be both positive and negative}, depending on $\epsilon_{0}$, $\epsilon_{2}$, and the range of $q$. For ${Q>0}$, the coordinate $t$ is temporal and $q$ is spatial, while for ${Q<0}$, $q$ is temporal and $t$ is spatial. The boundary ${Q(q)=0}$ localizes the \emph{Killing horizon} related to the Killing vector field $\partial_{t}$. In the case of de~Sitter universe, this coincides with the \emph{cosmological horizon} separating the static and dynamic regions.

The coordinate singularity at ${P(p)=0}$, where the norm of the Killing vector field $\partial_{\phi}$ vanishes, localizes the \emph{axis of symmetry}. This occurs either when ${\Lambda>0, \epsilon_{2}=1}$ or ${\Lambda<0, \epsilon_{2}=-1}$. In both these cases such an axis is given by ${p=\pm a}$, and it is regular for the range ${\phi\in[0,2\pi)}$.

The list of all possible subcases of the general metric (\ref{eq:PDdads})--(\ref{eq:P}), with the admitted ranges of $p$ and $q$, are summarized in Table~\ref{tbl:dS} for ${\Lambda>0}$, and in Table~\ref{tbl:adS} for ${\Lambda<0}$.

\begin{table}[h]
\begin{center}
\begin{tabular}{| c | c | c || c | c|| c | c |}
\hline
$\epsilon_{2}$ & $\epsilon_{0}$ & $\exists$ & $P$ & range of $p$ & $Q$ & range of $q$ \\
\hline
\hline
1 & 1 & \checkmark & $1-p^{2}/a^{2}$ & $(-a,a)$ & $1-q^{2}$ & $\mathbb{R}\setminus\{\pm 1\}$ \\
\hline
1 & 0 & \checkmark & $1-p^{2}/a^{2}$ & $[0,a)$  & $-q^{2}$ & $\mathbb{R}\setminus\{0\}$ \\
\hline
1 & $-1$ & \checkmark & $1-p^{2}/a^{2}$ & $[0,a)$ & $-1-q^{2}$ & $\mathbb{R}$ \\
\hline
0 & 0,$\pm1$ & $\times$ &  &  &  &  \\
\hline
$-1$ & 0,$\pm1$ & $\times$ &  &  &  &  \\
\hline
\end{tabular}
\caption{\small
Summary of all possible subcases for $\Lambda>0$ and different values of $\et$, $\ez$. The symbol \checkmark\, indicates admitted solutions (their geometry is  ${\,p^2{dS}_2 \times {S}^2}$), while those with $\times$ are forbidden.
}
\label{tbl:dS}
\end{center}
\end{table}

\vspace{-6mm}

\begin{table}[h]
\begin{center}
\begin{tabular}{| c | c | c || c | c || c | c |}
\hline
$\epsilon_{2}$ & $\epsilon_{0}$ & $\exists$ & $P$ & range of $p$ & $Q$ & range of $q$ \\
\hline
\hline
1 & 1 & \checkmark & $1+p^{2}/a^{2}$ & $\mathbb{R}$ & $1-q^{2}$ & $\mathbb{R}\setminus\{\pm 1\}$ \\
\hline
1 & 0 & \checkmark & $1+p^{2}/a^{2}$ & $[0,\infty)$ & $-q^{2}$ & $\mathbb{R}\setminus\{0\}$ \\
\hline
1 & $-1$ & \checkmark & $1+p^{2}/a^{2}$ & $[0,\infty)$ & $-1-q^{2}$ & $\mathbb{R}$ \\
\hline
0 & $ 1$ & \checkmark & $p^{2}/a^{2}$ & $\mathbb{R}$ & $ 1$ & $\mathbb{R}$ \\
\hline
0 & 0 & $\times$ &  &  &  &  \\
\hline
0 & $-1$ & \checkmark & $p^{2}/a^{2}$ & $\mathbb{R}$ & $-1$ & $\mathbb{R}$ \\
\hline
$-1$ & 1 & \checkmark & $-1+p^{2}/a^{2}$ & $[a,\infty)$ & $1+q^{2}$ & $\mathbb{R}$ \\
\hline
$-1$ & 0 & \checkmark & $-1+p^{2}/a^{2}$ & $[a,\infty)$ & $q^{2}$ & $\mathbb{R}\setminus\{0\}$ \\
\hline
$-1$ & $-1$ & \checkmark & $-1+p^{2}/a^{2}$ & $\mathbb{R}\setminus(-a,a)$ & $-1+q^{2}$ & $\mathbb{R}\setminus\{\pm 1\}$ \\
\hline
\end{tabular}
\caption{\small
Summary of all possible subcases for $\Lambda<0$ and different values of $\et$, $\ez$. Their geometry is
 ${\,p^2{dS}_2 \times {H}^2}$ for ${\epsilon_{2}=1}$,
 ${\,p^2{M}_2 \times {H}^2}$ for ${\epsilon_{2}=0}$, and
 ${\,p^2{AdS}_2 \times {H}^2}$ for ${\epsilon_{2}=-1}$.}
\label{tbl:adS}
\end{center}
\end{table}

We conclude that for ${\Lambda>0}$ there are 3 distinct subcases (namely ${\et=1}$ with any $\ez$) while for ${\Lambda<0}$ we must discuss 8 subcases (namely all combinations of ${\et, \ez}$, except ${\et=0=\ez}$). These will now be analyzed in the following sections \ref{sc:dS} and \ref{sc:adS}, respectively.

\section{New parametrizations of the de~Sitter spacetime}\label{sc:dS}

\subsection{Subcase ${\Lambda>0}$, ${\epsilon_{2}=1}$, ${\epsilon_{0}=1}$}
\label{ch:11a}

In this case, the metric (\ref{eq:PDdads})--(\ref{eq:P}) of de~Sitter universe has the form
\begin{equation}\label{eq:dS11ds}
\dd s^{2}=p^{2}\Big(\!-(1-q^{2})\,\dd t^{2}+\frac{\dd q^{2}}{1-q^{2}}\Big)
    +\frac{a^{2}\,\dd p^{2}}{a^{2}-p^{2}}+(a^{2}-p^{2})\,\dd \y^{2}\,,
\end{equation}
where ${|p|<a\equiv\sqrt{3/\Lambda}}$, ${q\in\mathbb{R}\setminus\{\pm 1\}}$, ${t\in\mathbb{R}}$, ${\y\in[0,2\pi)}$, with ${|p|= a}$ representing the axis of symmetry. Here ${Q=0}$ is the horizon: for ${Q>0\Leftrightarrow|q|<1}$ the coordinate $q$ is spatial and $t$ is temporal, and vice versa for ${Q<0\Leftrightarrow|q|>1}$. These two cases must be discussed separately:

\noindent
$\bullet$ \textbf{For} ${|q|<1}$, the coordinates of the metric (\ref{eq:dS11ds}) parametrize the de~Sitter hyperboloid (\ref{C1}) as
 \begin{equation}
\left. \begin{array}{l}
Z_0 = {\displaystyle p\,\sqrt{1-q^{2}}\,\sinh t}\,, \\[8pt]
Z_1 = {\displaystyle p\,\sqrt{1-q^{2}}\,\cosh t}\,, \\[8pt]
Z_2 = {\displaystyle |p|\,q}\,, \\[8pt]
Z_3 = {\displaystyle \sqrt{a^2-p^{2}}\,\cos\y}\,, \\[8pt]
Z_4 = {\displaystyle \sqrt{a^2-p^{2}}\,\sin\y}\,,
 \end{array} \!\right\} \ \Leftrightarrow \ \left\{ \!
 \begin{array}{l}
\tgh t  = {\displaystyle \frac{Z_0}{Z_1}}\,, \\[8pt]
\tan \y = {\displaystyle \frac{Z_4}{Z_3}}\,, \\[6pt]
p =  {\displaystyle \sign(Z_1)\,\sqrt{-Z_0^2+Z_1^2+Z_2^2}} \\[2pt]
\quad =  {\displaystyle \sign(Z_1)\,\sqrt{a^2-Z_3^2-Z_4^2}}\,, \\[6pt]
q =  {\displaystyle \frac{Z_2}{\sqrt{-Z_0^2+Z_1^2+Z_2^2}}}\,.
 \end{array} \right.
 \label{eq:dS11para}
 \end{equation}
Actually, the parametrization (\ref{eq:dS11para}) represents \emph{two maps} covering the de~Sitter manifold: the first one for ${p>0}$ covers the part ${Z_1>0}$, while the other for ${p<0}$ covers ${Z_1<0}$. Moreover, ${q>0}$ corresponds to ${Z_2>0}$, while ${q<0}$ corresponds to ${Z_2<0}$.

\noindent
$\bullet$ \textbf{For} ${|q|>1}$, the parametrization is the same as (\ref{eq:dS11para}), except that now
 \begin{equation}
\left. \begin{array}{l}
Z_0 = {\displaystyle p\,\sqrt{q^{2}-1}\,\cosh t}\,, \\[10pt]
Z_1 = {\displaystyle p\,\sqrt{q^{2}-1}\,\sinh t}\,,
 \end{array} \!\right\} \ \Leftrightarrow \ \left\{ \!
 \begin{array}{l}
\tgh t  = {\displaystyle \frac{Z_1}{Z_0}}\,, \\[6pt]
p =  {\displaystyle \sign(Z_0)\,\sqrt{-Z_0^2+Z_1^2+Z_2^2}} \,.
 \end{array} \right.
 \label{eq:dS11bpar}
 \end{equation}
Again, these are \emph{two maps}:  ${p>0}$ covers the part ${Z_0>0}$, while ${p<0}$ covers ${Z_0<0}$.

\begin{figure}[h!]
\begin{center}
\includegraphics[scale=1]{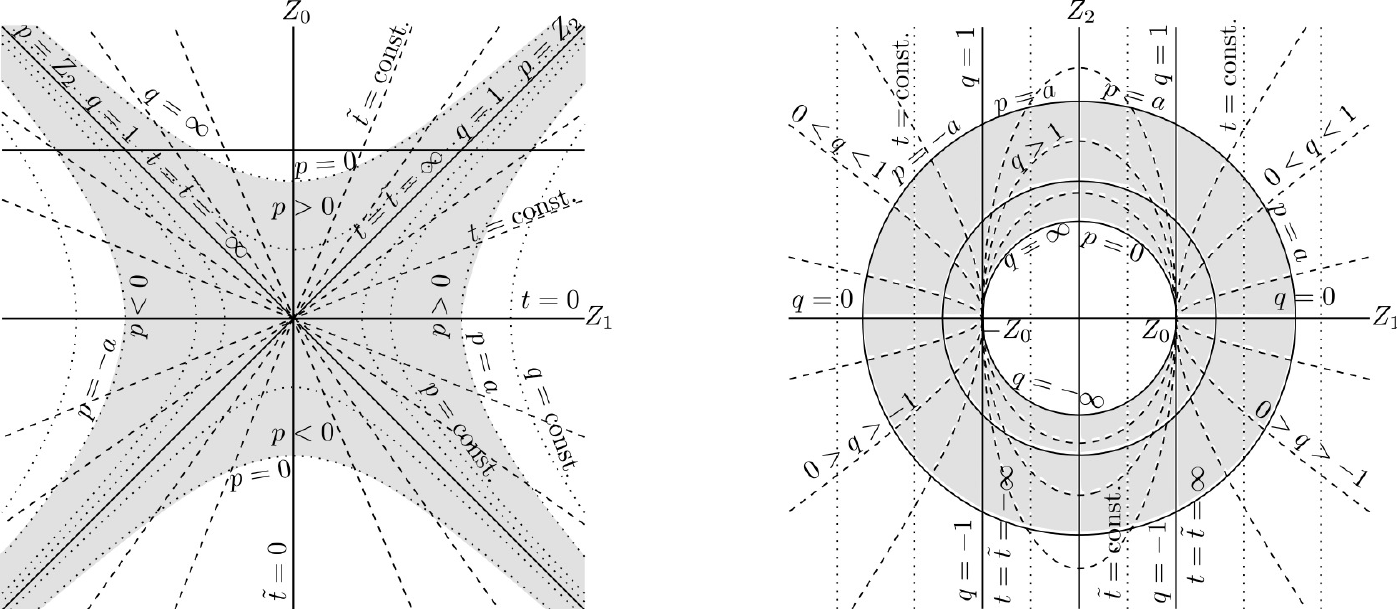}
\caption{\small
Left:
Section ${Z_2=\konst>0}$ for ${\y=\konst}$ (namely ${\y=\frac{\pi}{2}}$ corresponding to  ${Z_3=0}$). The region covered by all the coordinates is shaded. Sections through the surfaces ${|p|=\konst}$ and ${q=\konst}$ are hyperbolae, whereas ${t=\konst}$ are radial straight lines since ${\tgh t=Z_{0}/Z_{1}}$ and ${\cotgh \tilde{t}=Z_{0}/Z_{1}}$ (where $\tilde{t}$ is the relabeled coordinate $t$ in the chart ${|q|>1}$).
Right:
Section ${Z_{0}=\konst>0}$ (indicated by the horizontal line on the left part of this Figure). The curves ${|p|=\konst}$ are circles, which are also sections through the de~Sitter hyperboloid for changing~$Z_4$.
The curves ${q=\konst}$ (dashed lines) are hyperbolas for ${|q|<1}$, straight lines for ${|q|=1}$, and  ellipses for ${|q|>1}$, respectively.
}\label{img:dS11aZ0Z1}
\end{center}
\end{figure}

Specific character of these coordinates covering the de~Sitter hyperboloid (\ref{C1}) is illustrated for two different sections in Figure \ref{img:dS11aZ0Z1}.

\textbf{Global conformal representation:} To understand the \emph{global} character of the coordinates ${(t, q, p, \y)}$ of (\ref{eq:dS11ds}), it is best to plot them in the \emph{conformal representation of de~Sitter spacetime}, see the cylinder shown in the right part of Figure~\ref{fig:dshyp1}. This is achieved by comparing the 5D-parametrization (\ref{eq:dS11para}), (\ref{eq:dS11bpar}) of the de Sitter hyperboloid with the standard conformal parametrization (\ref{confDS})  corresponding to the metric (\ref{C6}). The explicit relations are
\begin{eqnarray}\label{eq:dS11cfinv}
\begin{array}{rll}
a\,\cotg\eta&\!\!\!=p\,\sqrt{1-q^{2}}\,\sinh t&\ \textrm{for} \ |q|<1\,,\qquad
a\,\cotg\eta=p\,\sqrt{q^{2}-1}\,\cosh t \hspace{7mm} \textrm{for} \  |q|>1\,,\\
\cotg\chi&\!\!\!=\frac{\displaystyle p\,\sqrt{1-q^{2}}\,\cosh t}{\displaystyle \sqrt{a^{2}-p^{2}(1-q^{2})}}&\ \textrm{for} \  |q|<1\,,\hspace{11mm}
\cotg\chi=\frac{\displaystyle p\,\sqrt{q^{2}-1}\,\sinh t}{\displaystyle \sqrt{a^{2}-p^{2}(1-q^{2})}} \quad \textrm{for} \  |q|>1\,,\\
\cotg\theta&\!\!\!=\frac{\displaystyle |p|\,q}{\displaystyle \sqrt{a^{2}-p^{2}}}\,.&
\end{array}
\end{eqnarray}
or, inversely,
\begin{eqnarray}\label{eq:dS11conf}
\tgh t \rovno \frac{\cos\eta}{\cos\chi}\quad \textrm{for} \quad |q|<1\,,\,\qquad
\cotgh t = \frac{\cos\eta}{\cos\chi}\quad \textrm{for} \quad |q|>1\,,\nonumber\\
q \rovno \frac{\sin\chi\cos\theta}{\sqrt{\sin^{2}\eta-\sin^{2}\chi\sin^{2}\theta}}\,,\qquad
\frac{p^2}{a^2} = 1-\frac{\sin^{2}\chi\sin^{2}\theta}{\sin^{2}\eta}\,.
\end{eqnarray}

\begin{figure}[h!]
\begin{center}
\includegraphics[scale=0.97]{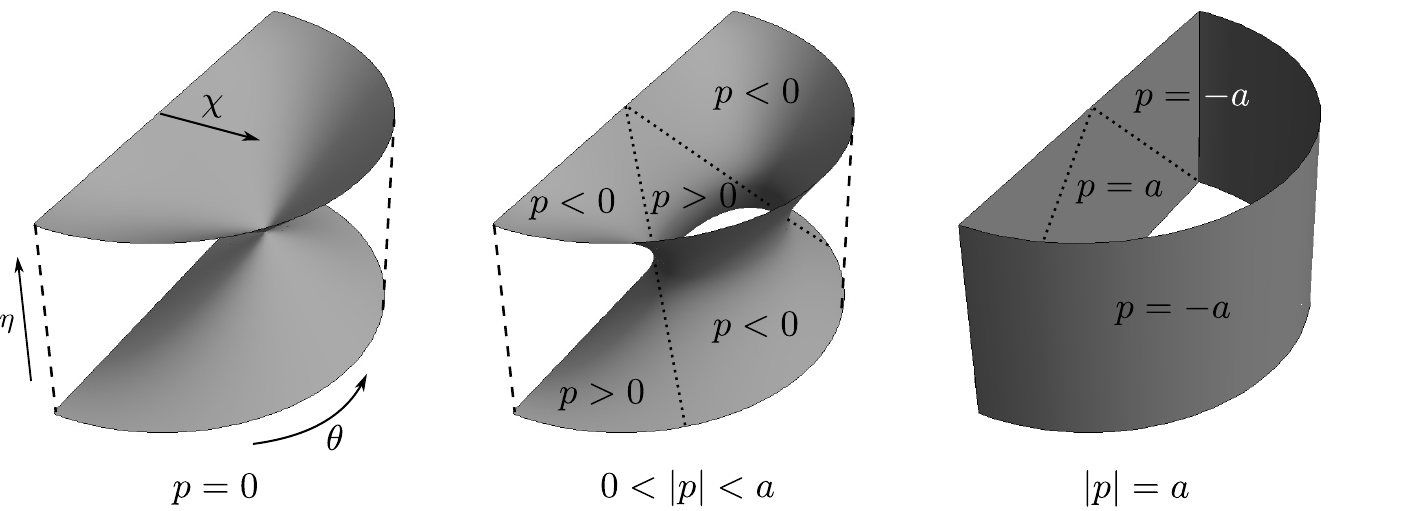}
\caption{\small
The surfaces ${p=\konst}$ for ${p=0}$ (left), a generic $p$ (middle), and ${|p|=a}$ (right) in the global conformal representation of de~Sitter spacetime (fixing ${\y}$). For ${p=0}$, there are two surfaces touching each other at a \emph{single point} ${\chi=\frac{\pi}{2},\eta=\frac{\pi}{2},\theta=\frac{\pi}{2}}$. As $|p|$ increases, the throat of the surface ${|p|=\konst}$ widens.
The outer boundary surface ${|p|=a}$ (which is the axis of symmetry) corresponds to a coordinate singularity ${\sin\theta=0}$ in the metric (\ref{C6}), i.e., the North and South poles ${\theta=0,\pi}$ of the de~Sitter space, respectively. The dotted lines separate the regions ${p>0}$ and ${p<0}$.}\label{img:ConfdS11p}
\end{center}
\end{figure}

\begin{figure}[h!]
\begin{center}
\includegraphics[scale=0.9]{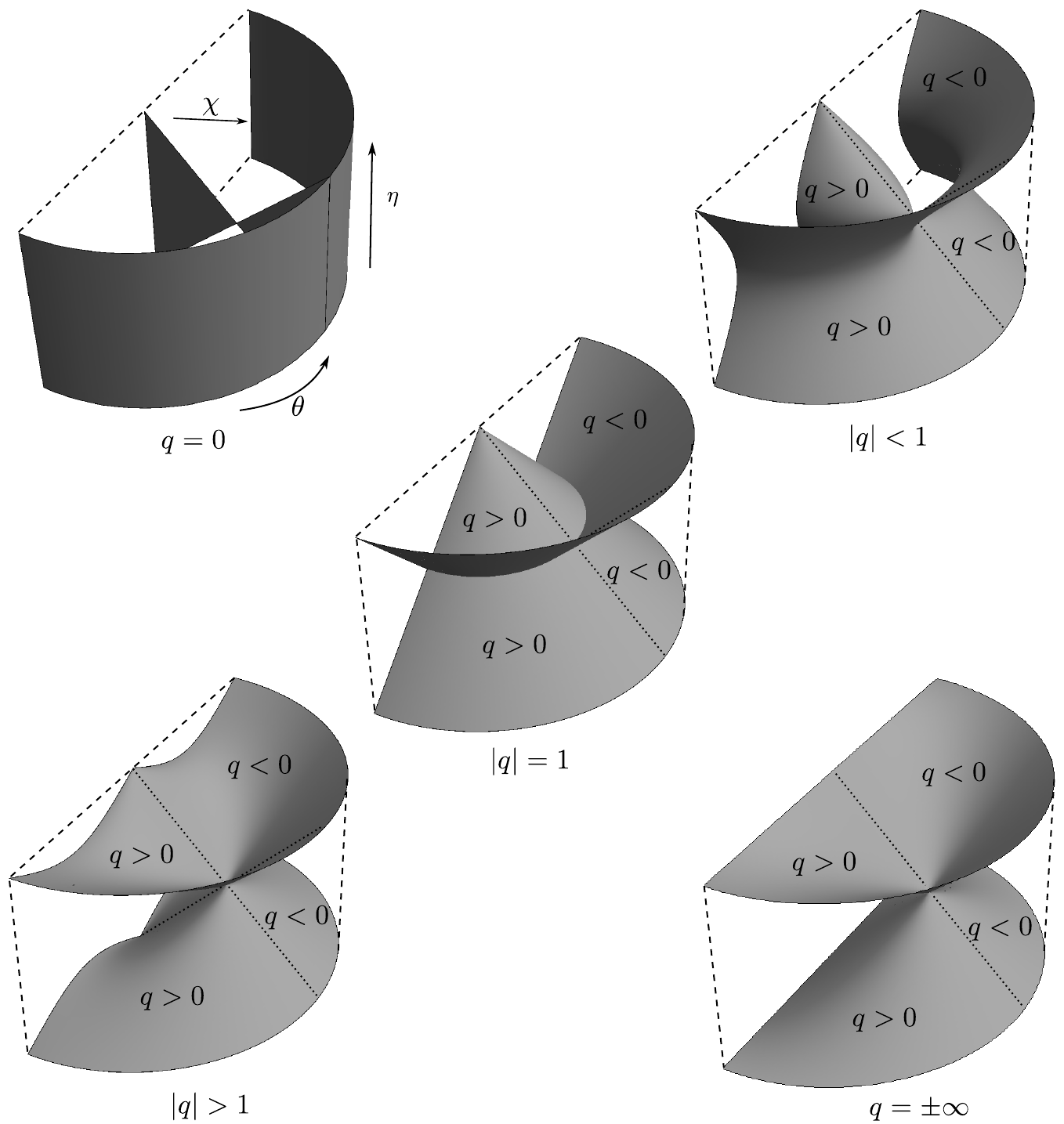}
\caption{\small
The surfaces ${q=\konst}$ for various values of $q$ in the global conformal representation of the de~Sitter spacetime (fixing ${\y}$). For ${|q|<1}$, the surface ${q=\konst}$ is split into an inner and an outer part which are connected at the special single point ${\chi=\frac{\pi}{2},\eta=\frac{\pi}{2},\theta=\frac{\pi}{2}}$. As $q$ increases, the gap between the inner and outer surface narrows. For ${|q|=1}$, there are two intersecting rotational cones that represent cosmological horizons. For ${|q|>1}$, the surface ${q=\konst}$ splits into the upper and lower part, which are again connected at the same point ${\chi=\frac{\pi}{2},\eta=\frac{\pi}{2},\theta=\frac{\pi}{2}}$ (as clearly seen also in Figure~\ref{img:ConfdS11qb}). For ${q=\pm\infty}$, the surface coincides with the surface ${p=0}$, cf. Figure \ref{img:ConfdS11p}.}\label{img:ConfdS11q}
\end{center}
\end{figure}

\begin{figure}[h]
\vspace{-6mm}
\begin{center}
\includegraphics[scale=1.05]{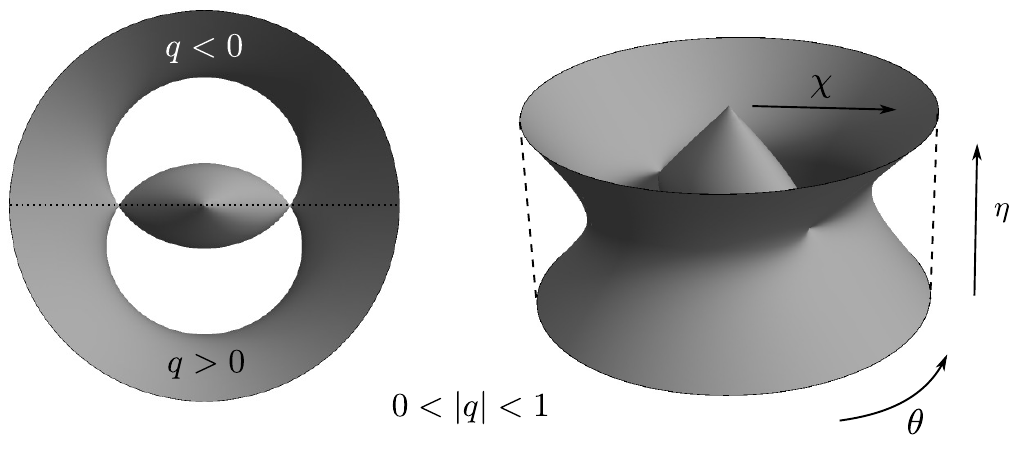}
\caption{\small
A view from top (left) and a general view (right) on the typical surface ${q=\konst}$ for ${0<|q|<1}$ in the global conformal coordinates of the de~Sitter universe.
The ``full'' cylindrical surface drawn here was obtained by ``gluing'' two half-cylinders together, namely one for the pole ${\y=0}$ (the left half) and the other for the pole ${\y=\pi}$ (the right half).
The outer dashed lines indicate the boundaries of the global de~Sitter cylinder (Figure~\ref{fig:dshyp1}). The dotted line separates the region ${q>0\Leftrightarrow \cos\theta>0}$ from ${q<0\Leftrightarrow \cos\theta<0}$.}\label{img:ConfdS11qa}
\end{center}
\end{figure}

\begin{figure}[h]
\begin{center}
\includegraphics[scale=1.05]{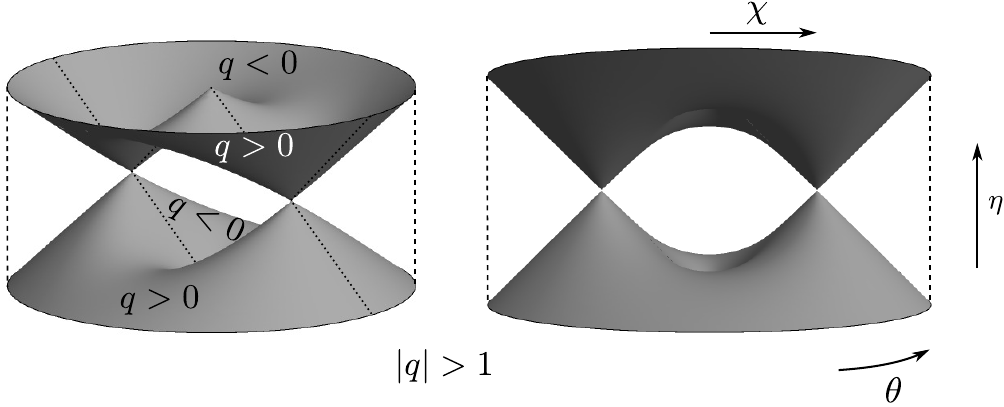}
\caption{\small
A general view (left) and a side view (right) on the surface ${q=\konst}$ for a generic ${|q|>1}$. Its upper and lower parts are connected at two points --- the vertices of the  cones.}\label{img:ConfdS11qb}
\end{center}
\end{figure}

Using (\ref{eq:dS11cfinv}) we can now visualize the main surfaces ${p=\konst}$ and  ${q=\konst}$ in the conformal de~Sitter cylinder, see Figure~\ref{img:ConfdS11p}--Figure~\ref{img:ConfdS11qb}. It is usual that the coordinate $\theta$ is suppressed and the full cylinder is drawn with the angular coordinate $\y\in[0,2\pi)$, see Figure~\ref{fig:dshyp1}. However, in (\ref{eq:dS11conf}) the new coordinates $p,q$ depend on $\theta$, while they are independent of $\y$. Therefore, in the subsequent figures we will suppress $\y$ instead of $\theta$, so we will plot half-cylinders with the angular coordinate ${\theta\in[0,\pi]}$. Indeed, the global conformal metric (\ref{C6}) is the same both for ${\theta=\frac{\pi}{2}}$ and for ${\y=\konst}$ if we relabel $\y\leftrightarrow\theta$, only the domain of these coordinates differ.


\newpage
\
\newpage
\subsection{Subcase ${\Lambda>0}$, ${\epsilon_{2}=1}$, ${\epsilon_{0}=0}$}
\label{ch:10}

\begin{figure}[h!]
\begin{center}
\includegraphics[scale=1]{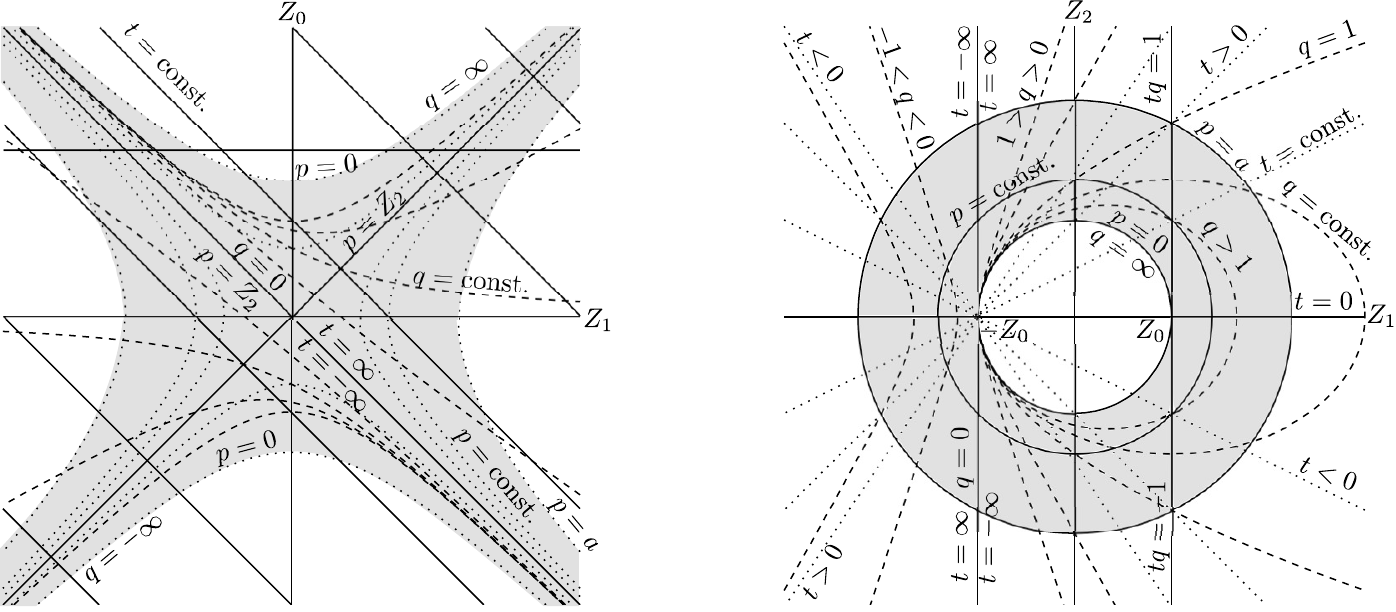}
\caption{\small
Visualization of the coordinates in section ${Z_{2}=\konst>0}$ with ${Z_3=0}$ (left), and section ${Z_{0}=\konst>0}$ (right), marked by the horizontal line on the left part of this figure. The lines ${|tq|=1}$ represent the horizon.
}\label{img:dS10Z0Z1}
\end{center}
\end{figure}

In this case, the de~Sitter metric (\ref{eq:PDdads})--(\ref{eq:P}) has the form
\begin{equation}\label{eq:dS10ds}
\dd s^{2}=p^{2}\Big(-\frac{\dd q^{2}}{q^{2}}+q^{2}\,\dd t^{2}\,\Big)
    +\frac{a^{2}\,\dd p^{2}}{a^{2}-p^{2}}+(a^{2}-p^{2})\,\dd \y^{2}\,,
\end{equation}
where ${p\in[0,a)}$, ${q\in\mathbb{R}\setminus\{0\}}$, ${t\in\mathbb{R}}$, ${\y\in[0,2\pi)}$, with ${p= a}$ representing the axis of symmetry. Clearly, $t$ is now a \emph{spatial} coordinate, whereas $q$ is \emph{temporal}. These coordinates cover the de~Sitter hyperboloid (\ref{C1}) by
\begin{equation}
\left. \begin{array}{l}
Z_0 = {\displaystyle \pul \,p\,q\big(1+t^{2}-q^{-2}\big)}\,, \\[4pt]
Z_1 = {\displaystyle \pul \,p\,q\big(1-t^{2}+q^{-2}\big)}\,, \\[4pt]
Z_2 = {\displaystyle p\,q\,t}\,, \\[4pt]
Z_3 = {\displaystyle \sqrt{a^2-p^{2}}\cos\y}\,, \\[4pt]
Z_4 = {\displaystyle \sqrt{a^2-p^{2}}\,\sin\y}\,,
 \end{array} \!\right\} \ \Leftrightarrow \ \left\{ \!
 \begin{array}{l}
 t  = {\displaystyle \frac{Z_2}{Z_0+Z_1}}\,, \\[8pt]
\tan \y = {\displaystyle \frac{Z_4}{Z_3}}\,, \\[6pt]
p =  {\displaystyle \sqrt{-Z_0^2+Z_1^2+Z_2^2}}\,, \\[6pt]
q =  {\displaystyle \frac{Z_0+Z_1}{\sqrt{-Z_0^2+Z_1^2+Z_2^2}}}\,.
 \end{array} \right.
 \label{eq:dS10par}
\end{equation}
They are illustrated for specific sections in Figure~\ref{img:dS10Z0Z1}.

\begin{figure}[h!]
\begin{center}
\vspace{5mm}
\includegraphics[scale=1]{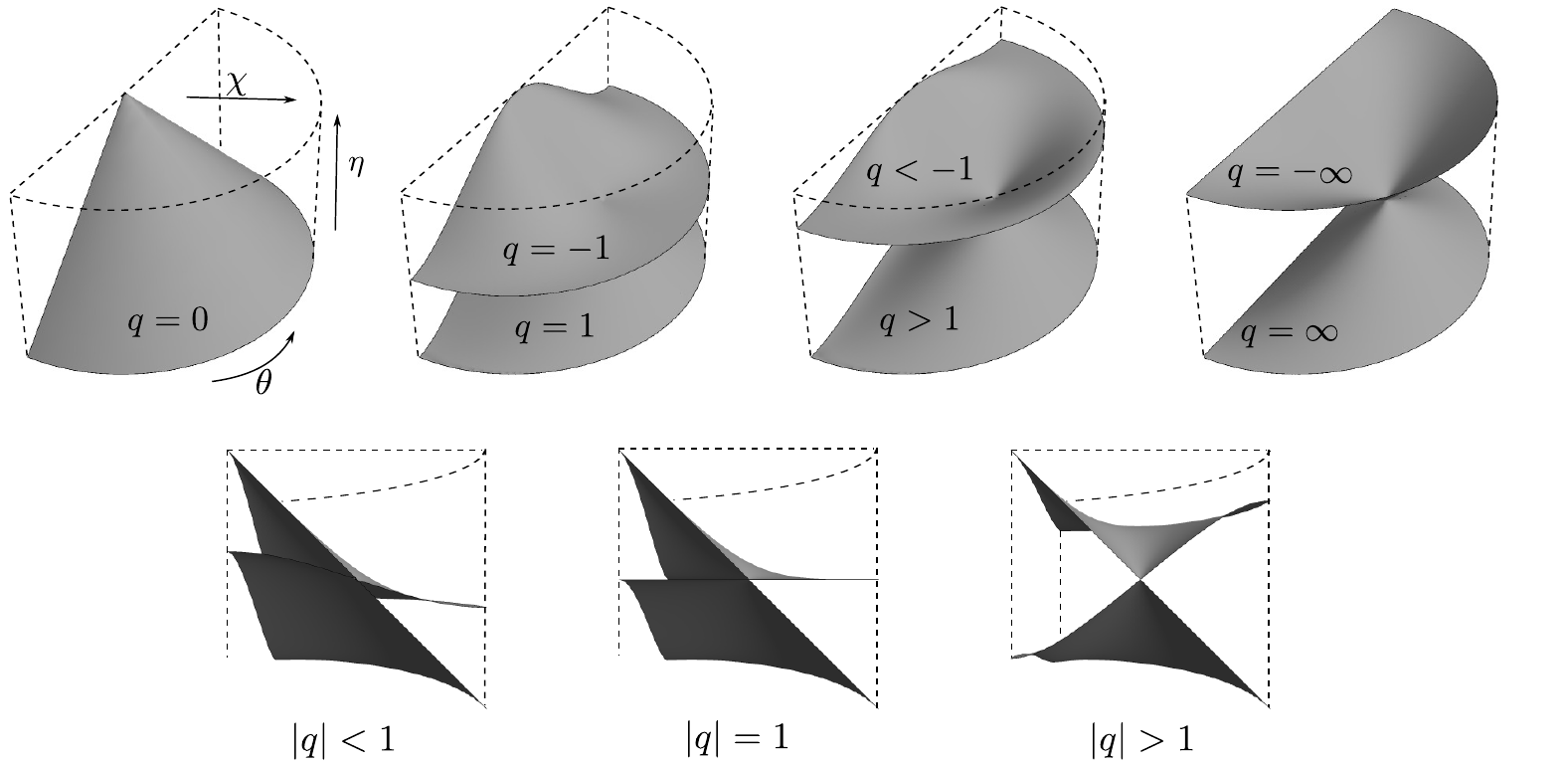}
\caption{\small
Top: The surfaces ${q=\konst}$ drawn in the global conformal representation of the de~Sitter universe (${\y=\konst}$). For ${q=0}$ there is only one conical surface, whereas for  ${|q|>0}$ we observe two surfaces (the lower one for ${q>0}$ and the upper one for ${q<0}$) which are always connected at the single point ${\chi=\frac{\pi}{2},\eta=\frac{\pi}{2},\theta=\frac{\pi}{2}}$ ``in the middle'' of the conformal (half)cylinder. In the limit $q=\pm\infty$ the surfaces are the same as in the previous case shown in Figure~\ref{img:ConfdS11q}. Moreover, they coincide with the surface $p=0$.
Bottom: A side view on these surfaces ${q=\konst}$  for various values of $q$. These pictures are ``sections'' obtained by plotting ${\theta\in[\frac{\pi}{2},\pi]}$ only.
}\label{img:ConfdS10q}
\end{center}
\end{figure}

\clearpage

\textbf{Global conformal representation} is obtained using the transformation
  \begin{equation}
\left. \begin{array}{l}
\cotg\eta = {\displaystyle  \frac{1}{2a}\,p\,q\big(1+t^{2}-q^{-2}\big)}\,, \\[7pt]
\cotg\chi = {\displaystyle  \frac{1}{2}\frac{\displaystyle p\,q\left(1-t^{2}+q^{-2}\right)}{\sqrt{a^2+p^2(q^2\,t^2-1)}}}\,, \\[9pt]
\cotg\theta = {\displaystyle \frac{p\,q\,t}{\sqrt{a^2-p^2}}}\,,
 \end{array} \!\right\} \ \Leftrightarrow \ \left\{ \!
 \begin{array}{l}
t  = {\displaystyle \frac{\sin\chi\cos\theta}{\cos\chi+\cos\eta}}\,, \\[10pt]
{\displaystyle \frac{p^2}{a^2} =  1-\frac{\sin^{2}\chi\sin^{2}\theta}{\sin^{2}\eta}}\,, \\[10pt]
q =  {\displaystyle \frac{\cos\chi+\cos\eta}{\sqrt{\sin^{2}\eta-\sin^{2}\chi\sin^{2}\theta}}}\,,
 \end{array} \right.
 \label{eq:dS10conf}
 \end{equation}
relating (\ref{eq:dS10ds}) to the global de~Sitter metric (\ref{C6}) --- compare (\ref{eq:dS10par}) to (\ref{confDS}). This enables us to understand the global character of the coordinates of the metric (\ref{eq:dS10ds}). Because in both cases ${\ez=1}$ and ${\ez=0}$ the surfaces  ${p=\konst>0}$ are given by the same relations (compare (\ref{eq:dS11conf}) and (\ref{eq:dS10conf})), their form is the same as in Figure~\ref{img:ConfdS11p}. However, the surfaces ${q=\konst}$ are now different --- they are shown in Figure~\ref{img:ConfdS10q}.

\subsection{Subcase ${\Lambda>0}$, ${\epsilon_{2}=1}$, ${\epsilon_{0}=-1}$}
\label{ch:1-1}
\begin{figure}[h]
\begin{center}
\includegraphics[scale=1]{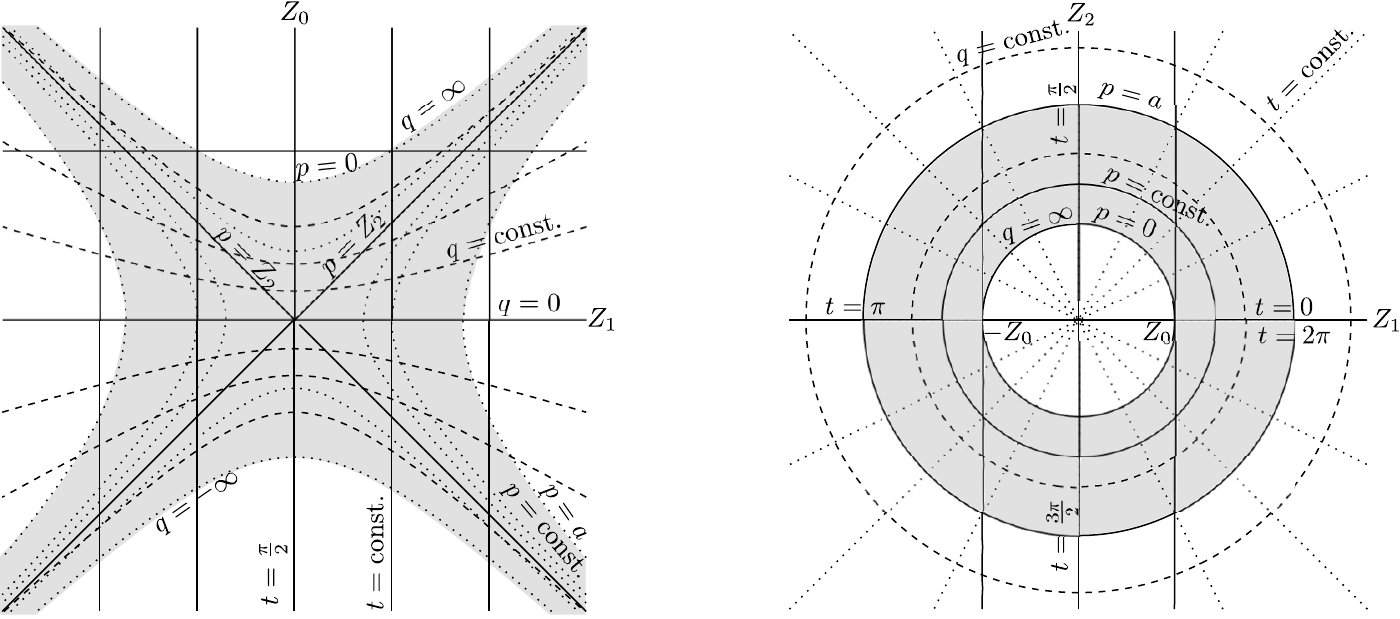}
\caption{\small
The coordinates in section ${Z_{2}=\konst>0}$ with ${Z_3=0}$ (left), where all the lines ${p=\konst}$ and ${q=\konst}$ are hyperbolas, and section ${Z_{0}=\konst>0}$ (right) marked by the horizontal line on the left part of this Figure, where the lines ${p=\konst}$ and ${q=\konst}$ are concentric circles. Straight lines ${Z_1=\pm Z_0}$ indicate the cosmological horizon.}\label{img:dS1-1Z0Z1a}
\end{center}
\end{figure}

The metric (\ref{eq:PDdads})--(\ref{eq:P}) for this third (and last) de~Sitter subcase reads
\begin{equation}\label{eq:dS1-1ds}
\dd s^{2}=p^{2}\Big(\!-\frac{\dd q^{2}}{1+q^{2}}+(1+q^{2})\,\dd t^{2}\Big)
    +\frac{a^{2}\,\dd p^{2}}{a^{2}-p^{2}}+(a^{2}-p^{2})\,\dd \y^{2}\,,
\end{equation}
where ${p\in[0,a)}$, ${q\in\mathbb{R}}$, ${t\in[0,2\pi)}$, ${\y\in[0,2\pi)}$.
Since ${-Q=1+q^2\not =0}$, there is no Killing horizon related to the vector field $\partial_t$ in this metric. In fact, $t$ is everywhere a \emph{spatial angular} coordinate, while $q$ is \emph{temporal}. These coordinates cover the (part of) de~Sitter hyperboloid (\ref{C1}) as
\begin{equation}
\left. \begin{array}{ll}
 Z_0 = p\,q \,, \\[4pt]
 Z_1 = p\,\sqrt{1+q^2}\,\cos t \,, \\[4pt]
 Z_2 = p\,\sqrt{1+q^2}\,\sin t\,, \\[4pt]
 Z_3 = \sqrt{a^2-p^2}\,\cos\,\y\,, \\[4pt]
 Z_4 = \sqrt{a^2-p^2}\,\sin\,\y\,,
 \end{array}\right\} \quad\Leftrightarrow\quad
 \left\{ \begin{array}{l}
 \tan t  = {\displaystyle \frac{Z_2}{Z_1}}\,, \\[8pt]
 \tan \y = {\displaystyle \frac{Z_4}{Z_3}}\,, \\[8pt]
 p = \sqrt{-Z_{0}^{2}+Z_{1}^{2}+Z_{2}^{2}}\,, \\[6pt]
 q = {\displaystyle \frac{Z_0}{\sqrt{-Z_0^2+Z_1^2+Z_2^2}} }\,.
 \end{array} \right.
\label{eq:dS1-1par}
\end{equation}
This parametrization is visualized in different sections of the de~Sitter hyperboloid in Figure~\ref{img:dS1-1Z0Z1a}.

\begin{figure}[h]
\begin{center}
\includegraphics[scale=1]{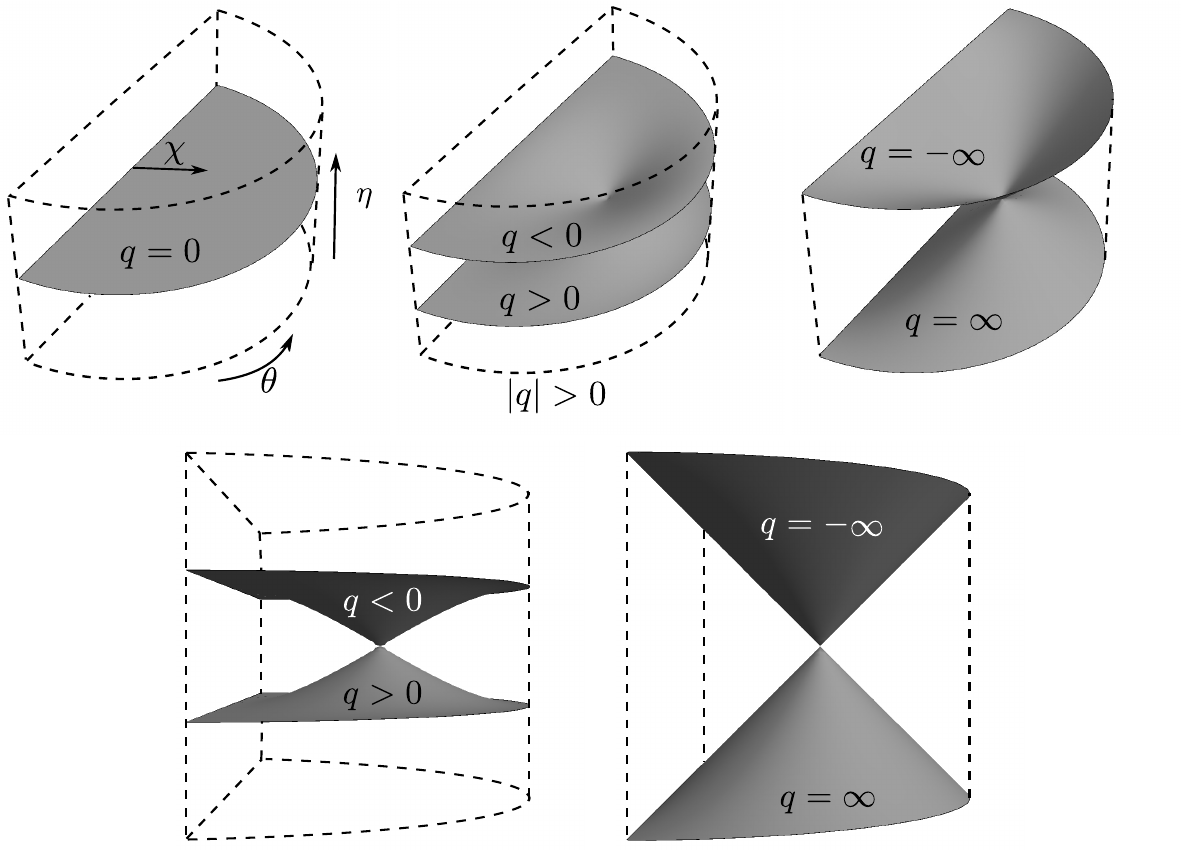}
\caption{\small
Top: The surfaces ${q=\konst}$ plotted in the global conformal representation of the de~Sitter universe (setting ${\y=\konst}$). For ${q=0}$ this is a single flat plane ${\eta=\frac{\pi}{2}}$ in the middle of the (half)cylinder, whereas for  ${|q|>0}$ we observe two curved surfaces (the lower one for ${q>0}$ and the upper one for ${q<0}$). In the limit ${q=\pm\infty}$ the surfaces are the same as in both previous cases shown in Figures~\ref{img:ConfdS11q} and \ref{img:ConfdS10q}.
Bottom: A side view on these surfaces ${q=\konst}$ for a generic value of $|q|$ and for ${|q|=\infty}$, clearly indicating that the two cones remain connected at the single point ${\chi=\frac{\pi}{2},\eta=\frac{\pi}{2},\theta=\frac{\pi}{2}}$, and are mutual mirror images.
}\label{img:ConfdS1-1q}
\end{center}
\end{figure}

\textbf{Global conformal representation} is obtained by combining (\ref{eq:dS1-1par}) with (\ref{confDS}):
  \begin{equation}
\left. \begin{array}{l}
\cotg\eta = {\displaystyle  \frac{p\,q}{a}}\,, \\[7pt]
\cotg\chi = {\displaystyle  \frac{p\sqrt{1+q^2}\cos t}{\sqrt{p^2\,(1+q^2)\,\sin^2 t-p^2+a^2}}}\,, \\[9pt]
\cotg\theta = {\displaystyle \frac{p\sqrt{1+q^2}\sin t}{\sqrt{a^2-p^2}}}\,,
 \end{array} \!\right\} \ \Leftrightarrow \ \left\{ \!
 \begin{array}{l}
\tg t  = {\displaystyle \tan\chi\cos\theta}\,, \\[10pt]
{\displaystyle \frac{p^2}{a^2} =  1-\frac{\sin^{2}\chi\sin^{2}\theta}{\sin^{2}\eta}}\,, \\[10pt]
q =  {\displaystyle \frac{\cos\eta}{\sqrt{\sin^{2}\eta-\sin^{2}\chi\sin^{2}\theta}}}\,.
 \end{array} \right.
 \label{eq:dS1-1conf}
 \end{equation}
Specific character of the coordinates of metric (\ref{eq:dS1-1ds}) can thus be visualized in terms of the global de~Sitter metric (\ref{C6}). Again, the surfaces  ${p=\konst>0}$ are the same as in Figure~\ref{img:ConfdS11p} because the  relations (\ref{eq:dS11conf}) and (\ref{eq:dS1-1conf}) are identical. The different surfaces ${q=\konst}$ are shown in Figure~\ref{img:ConfdS1-1q}.

\clearpage

\section{New parametrizations of the anti-de~Sitter spacetime}\label{sc:adS}

\subsection{Subcase ${\Lambda<0}$, ${\epsilon_{2}=1}$, ${\epsilon_{0}=1}$}
\label{ch:11b}

The metric (\ref{eq:PDdads})--(\ref{eq:P}) of the anti-de~Sitter universe for this choice of
${\epsilon_{2},\epsilon_{0}}$ takes the form
\begin{equation}\label{eq:adS11ds}
\dd s^{2}=p^{2}\Big(\!-(1-q^{2})\,\dd t^{2}+\frac{\dd q^{2}}{1-q^{2}}\Big)
    +\frac{a^{2}\,\dd p^{2}}{a^{2}+p^{2}}+(a^{2}+p^{2})\,\dd \yy^{2}\,,
\end{equation}
where ${\a=\sqrt{-3/\Lambda}\,}$ and ${\,p, t, \yy \in \mathbb{R}}$, ${q\in\mathbb{R}\setminus\{\pm 1\}}$. Since $\y$ employed in (\ref{eq:PDdads}) now does \emph{not} play the role of an angular coordinate, we have relabeled it as $\yy$. The condition ${Q=1-q^2=0}$ identifies the Killing horizon associated with the vector field $\partial_t$: for ${|q|<1}$ the coordinate $q$ is spatial and~$t$ is temporal, while for ${|q|>1}$ the coordinate $q$ is temporal and $t$ is spatial. However, contrary to the analogous de~Sitter case discussed in Section~\ref{ch:11a}, there is no cosmological horizon in the anti-de~Sitter universe since there exists another Killing vector field which is everywhere timelike (see, e.g.,\cite{GriPod2009}). The distinct cases ${|q|<1}$ and ${|q|>1}$ are:

\noindent
$\bullet$ \textbf{For} ${|q|<1}$, the coordinates of (\ref{eq:adS11ds}) parametrize the anti-de~Sitter hyperboloid (\ref{D1}) as
 \begin{equation}
\left. \begin{array}{l}
Z_0 = {\displaystyle p\,\sqrt{1-q^{2}}\,\sinh t}\,, \\[8pt]
Z_1 = {\displaystyle p\,\sqrt{1-q^{2}}\,\cosh t}\,, \\[8pt]
Z_2 = {\displaystyle |p|\,q}\,, \\[8pt]
Z_3 = {\displaystyle \pm \sqrt{a^2+p^{2}}\,\sinh\yy}\,, \\[8pt]
Z_4 = {\displaystyle \pm \sqrt{a^2+p^{2}}\,\cosh\yy}\,,
 \end{array} \!\right\} \ \Leftrightarrow \ \left\{ \!
 \begin{array}{l}
\tgh t  = {\displaystyle \frac{Z_0}{Z_1}}\,, \\[8pt]
\tgh \yy = {\displaystyle \frac{Z_3}{Z_4}}\,, \\[6pt]
p =  {\displaystyle \sign(Z_1)\,\sqrt{-Z_0^2+Z_1^2+Z_2^2}} \\[2pt]
\quad =  {\displaystyle \sign(Z_1)\,\sqrt{Z_4^2-Z_3^2-a^2}}\,, \\[6pt]
q =  {\displaystyle \frac{Z_2}{\sqrt{-Z_0^2+Z_1^2+Z_2^2}}}\,.
 \end{array} \right.
 \label{eq:adS11apar}
 \end{equation}
This parametrization gives \emph{two maps} covering the anti-de~Sitter manifold, namely
the coordinate map ${Z_4\geq a}$ for the ``$+$'' sign, and   ${Z_4\leq a}$ for the ``$-$'' sign (and again two maps ${p>0}$ and ${p<0}$). Moreover, ${q>0}$ corresponds to ${Z_2>0}$, while ${q<0}$ corresponds to ${Z_2<0}$.

\noindent
$\bullet$ \textbf{For} ${|q|>1}$, the parametrization is the same as (\ref{eq:adS11apar}), except that now
 \begin{equation}
\left. \begin{array}{l}
Z_0 = {\displaystyle p\,\sqrt{q^{2}-1}\,\cosh t}\,, \\[10pt]
Z_1 = {\displaystyle p\,\sqrt{q^{2}-1}\,\sinh t}\,,
 \end{array} \!\right\} \ \Leftrightarrow \ \left\{ \!
 \begin{array}{l}
\tgh t  = {\displaystyle \frac{Z_1}{Z_0}}\,, \\[6pt]
p =  {\displaystyle \sign(Z_0)\,\sqrt{-Z_0^2+Z_1^2+Z_2^2}} \,.
 \end{array} \right.
 \label{eq:adS11bpar}
 \end{equation}
Again, these are \emph{two maps}:  ${p>0}$ covers the part ${Z_0>0}$, while ${p<0}$ covers ${Z_0<0}$.

\begin{figure}[h!]
\begin{center}
\includegraphics[scale=1]{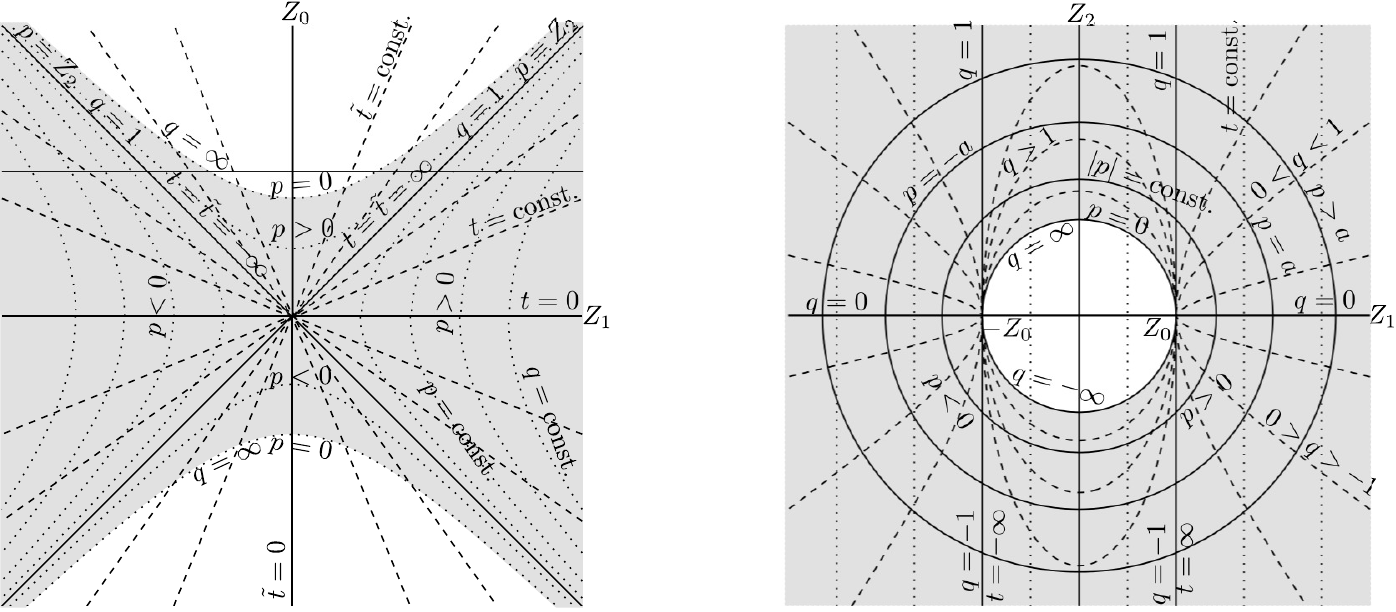}
\caption{\small
Left:
Section ${Z_2=\konst>0}$ for ${\yy=\konst}$ (namely ${\yy=0\Leftrightarrow Z_3=0}$). The region covered by all the coordinates is shaded. Sections through the surfaces ${|p|=\konst}$ and ${q=\konst}$ are hyperbolae, whereas ${t=\konst}$ and ${\tilde t=\konst}$ are radial straight lines.
Right:
Section ${Z_{0}=\konst>0}$. The curves ${|p|=\konst}$ are circles, while ${|q|=\konst<1}$  are hyperbolae and ${|q|=\konst>1}$ are ellipses. These lines are also sections through the anti-de~Sitter hyperboloid for $Z_4$ changing. They resemble those in Figure~\ref{img:dS11aZ0Z1} for the analogous de~Sitter case.
}\label{img:adS11aZ0Z1}
\end{center}
\end{figure}

We immediately observe that the coordinates ${t, p, q}$ depend on the coordinates ${Z_0, Z_1, Z_2}$ in exactly the same way as in the analogous case of the de~Sitter spacetime, cf. (\ref{eq:adS11apar}) with (\ref{eq:dS11para}), and (\ref{eq:adS11bpar}) with (\ref{eq:dS11bpar}). Therefore, sections through the anti-de~Sitter spacetime in the subspace ${Z_0, Z_1, Z_2}$ are the same as the corresponding sections through the de~Sitter spacetime, except that now the hyperbolas ${p=\konst}$ are not bounded by ${p=a}$. Character of these coordinates covering the anti-de~Sitter hyperboloid (\ref{D1}) is illustrated for two such sections in Figure~\ref{img:adS11aZ0Z1}.
\clearpage

\textbf{Global conformal representation:} To visualize the \emph{global} character of these coordinates ${(t, q, p, \yy)}$ of (\ref{eq:adS11ds}), we will plot them in the standard \emph{conformal representation of anti-de~Sitter spacetime}, see the right part of Figure~\ref{fig:adshyp1}. This is achieved by comparing the 5D-parametrization (\ref{eq:adS11apar}), (\ref{eq:adS11bpar}) with the standard conformal parametrization (\ref{confADS}) corresponding to the metric (\ref{D6}). We thus obtain the following explicit relations
\begin{eqnarray}\label{eq:adS11cfinv}
\begin{array}{rl}
\cotg\eta&\!\!\!={\displaystyle \pm \frac{ p\,\sqrt{1-q^{2}}\,\sinh t}{\sqrt{p^{2}+a^{2}}\,\cosh\yy}} \hspace{5mm} \textrm{for} \ |q|<1\,,\\[10pt]
\cotg\eta&\!\!\!={\displaystyle \pm \frac{ p\,\sqrt{q^{2}-1}\,\cosh t}{\sqrt{p^{2}+a^{2}}\,\cosh\yy}} \hspace{5mm} \textrm{for} \  |q|>1\,,\\[10pt]
a\,\tg\chi&\!\!\!=\sqrt{p^2\left(\cosh^2 t-q^2\sinh^2 t\right)+(p^2+a^2)\sinh^2 \yy} \hspace{5mm} \textrm{for}\    |q|<1\,,\\[10pt]
a\,\tg\chi&\!\!\!=\sqrt{p^2\left(q^2\cosh^2 t-\sinh^2 t\right)+(p^2+a^2)\sinh^2 \yy} \hspace{5mm} \textrm{for} \  |q|>1\,,\\[10pt]
\cotg\theta&\!\!\!={\displaystyle \frac{p\,\sqrt{1-q^2}\,\cosh t}{\sqrt{p^2q^2+(p^2+a^2)\sinh^2\yy}}} \hspace{5mm} \textrm{for} \ |q|<1\,,\\[10pt]
\cotg\theta&\!\!\!={\displaystyle \frac{p\,\sqrt{q^2-1}\,\sinh t}{\sqrt{p^2q^2+(p^2+a^2)\sinh^2\yy}}} \hspace{5mm} \textrm{for} \ |q|>1\,,\\[10pt]
\cotg\y &\!\!\!={\displaystyle \pm \frac{|p|\,q}{\sqrt{p^2+a^2}\,\sinh\yy}}\,,
\end{array}
\end{eqnarray}
where the signs ``$\pm$'' again correspond to two coordinate maps ${Z_4>0}$ and ${Z_4<0}$, respectively.

\begin{figure}[h]
\begin{center}
\includegraphics[scale=1]{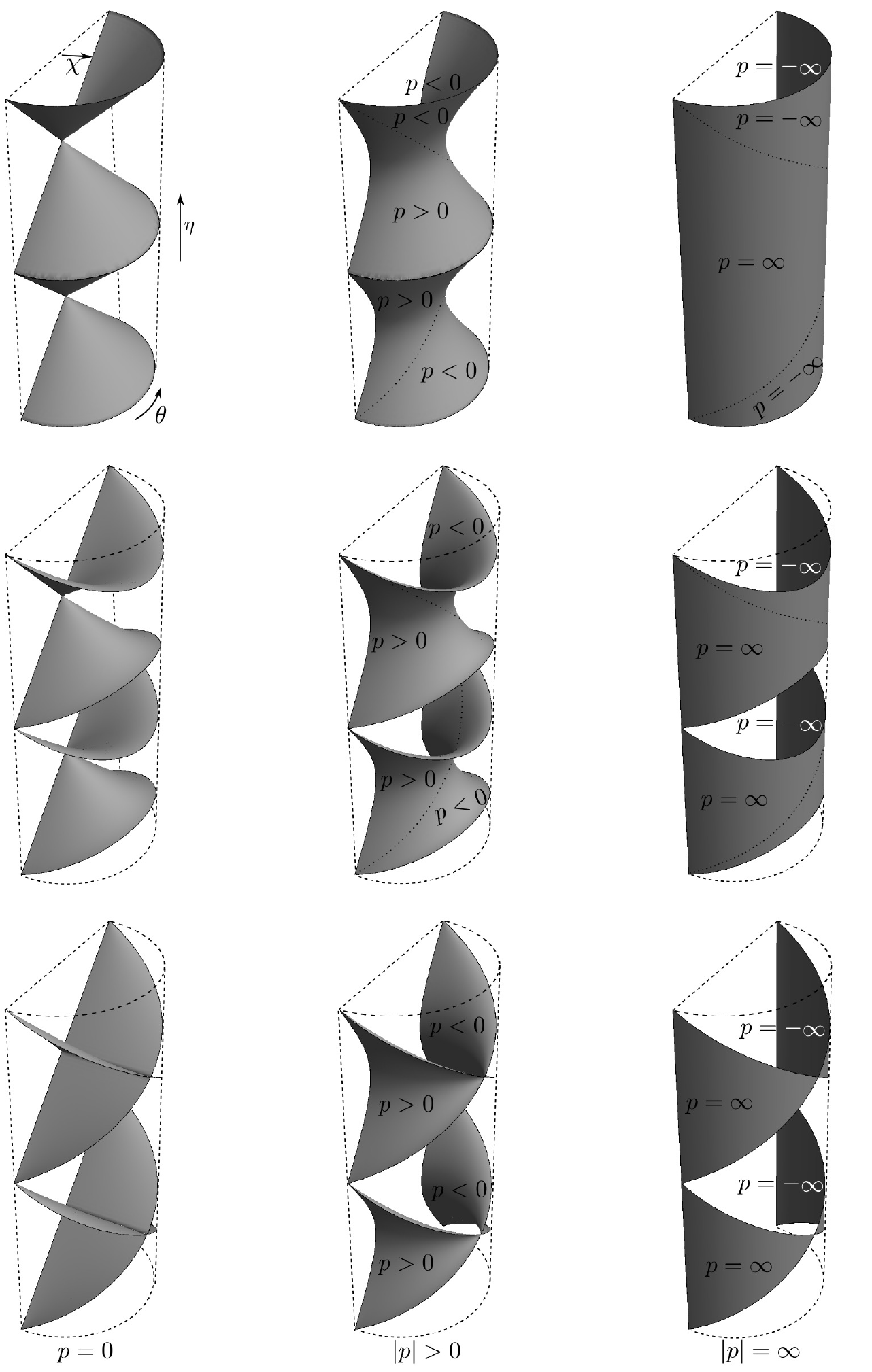}
\caption{\small
The surfaces ${p=\konst}$ for various values of $p$ in the conformal coordinates ${\eta\in[-\pi,\pi]}$, ${\chi\in[0,\frac{\pi}{2})}$, ${\theta\in[0,\pi]}$ for ${\y=0}$ (top row), ${\y=\frac{\pi}{4}}$ (middle row) and ${\y=\frac{\pi}{2}}$ (bottom row), respectively. The surfaces ${|p|=\infty}$ form the outer cylindrical boundary ${\chi=\frac{\pi}{2}}$ which is a null conformal infinity $\mathcal{I}$ of the anti-de~Sitter universe. The dotted lines separate ${p>0}$ from ${p<0}$. The coordinate $p$ does not cover the whole anti-de~Sitter space-time.}\label{img:ConfadS11p}
\end{center}
\end{figure}

\begin{figure}[h]
\begin{center}
\includegraphics[scale=1]{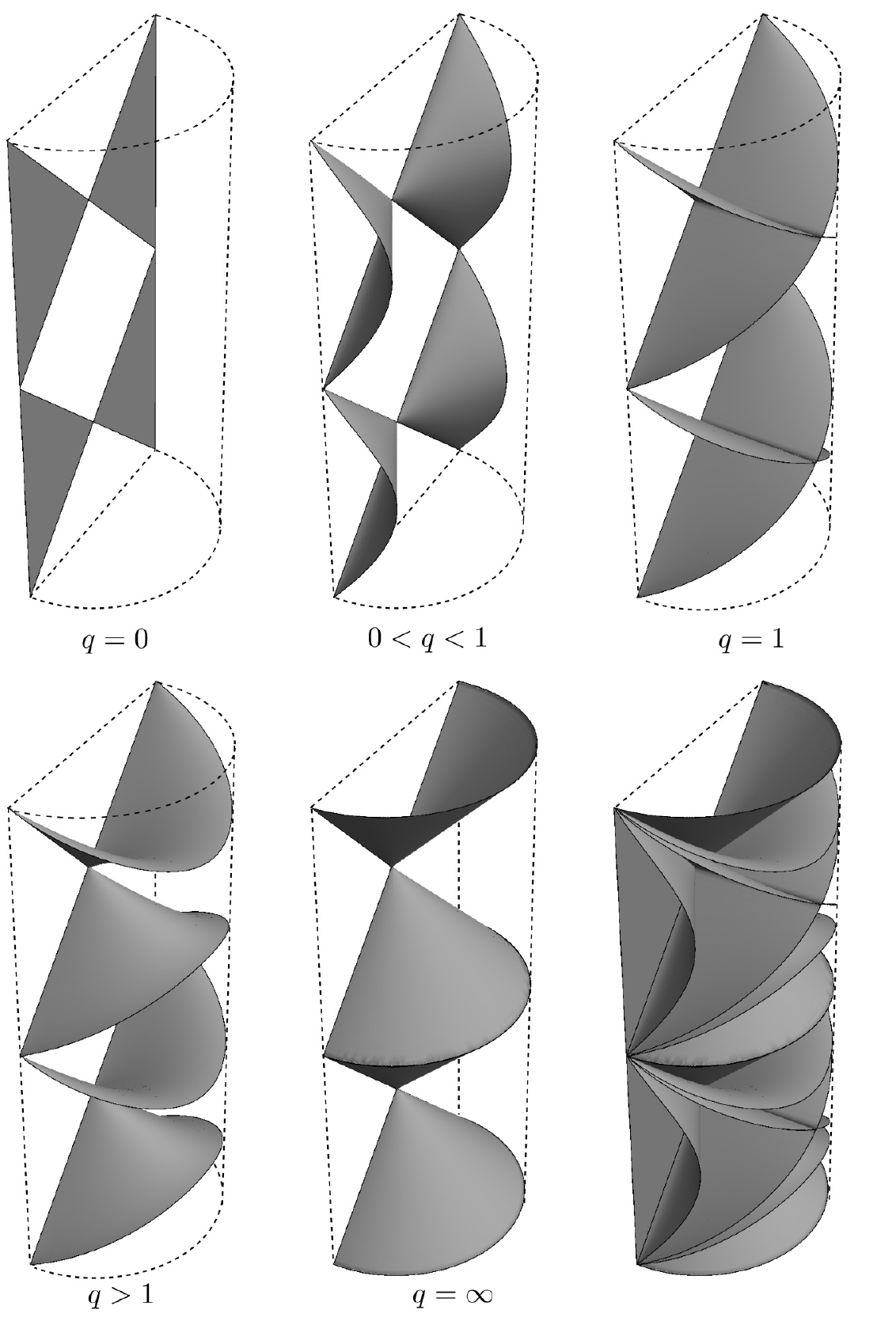}
\caption{\small
The surfaces ${q=\konst}$ for five distinct values of $q$, plotted in the global conformal coordinates ${\eta\in[-\pi,\pi]}$, ${\chi\in[0,\frac{\pi}{2})}$, ${\theta\in[0,\pi]}$ for ${\y=0}$. The surface ${q=0}$ corresponds to ${\theta=0,\pi}$, the surface ${|q|=1}$ (representing the Killing horizon $\partial_t$) resembles  ${p=0}$ for ${\y=\frac{\pi}{2}}$, while the surface ${q=\pm\infty}$ resembles ${p=0}$ for ${\y=0}$ (see Figure~\ref{img:ConfadS11p}). In the bottom right picture, all these five surfaces are plotted together, indicating the foliation of the global anti-de~Sitter cylinder. The coordinate $q$ does not cover the whole anti-de~Sitter spacetime, namely, it does not cover the regions inside the null half-cones ${q=\infty}$.}\label{img:ConfadS11q0}
\end{center}
\end{figure}

\clearpage

The inverse relations  to (\ref{eq:adS11cfinv}) are
\begin{eqnarray}\label{eq:adS11conf}
\tgh t \rovno \frac{\cos\eta}{\sin\chi\cos\theta}\quad \textrm{for} \quad |q|<1\,,\,\qquad
\cotgh t = \frac{\cos\eta}{\sin\chi\cos\theta}\quad \textrm{for} \quad |q|>1\,,\nonumber\\
p \rovno \frac{a\,\sign(\cos\theta)}{\cos\chi}\,\sqrt{\sin^2\chi(1-\sin^2\theta\sin^2\y)-\cos^2\eta}\quad \textrm{for} \quad |q|<1,\nonumber\\
p \rovno \frac{a\,\sign(\cos\eta)}{\cos\chi}\,\sqrt{\sin^2\chi(1-\sin^2\theta\sin^2\y)-\cos^2\eta}\quad \textrm{for} \quad |q|>1,\\
q \rovno \frac{\sin\chi\sin\theta\cos\y}{\sqrt{\sin^2\chi(1-\sin^2\theta\sin^2\y)-\cos^2\eta}}\,,
\qquad \tgh\yy = \frac{\sin\chi\sin\theta\sin\y}{\sin\eta}\,.\nonumber
\end{eqnarray}

\begin{figure}[h]
\begin{center}
\includegraphics[scale=1]{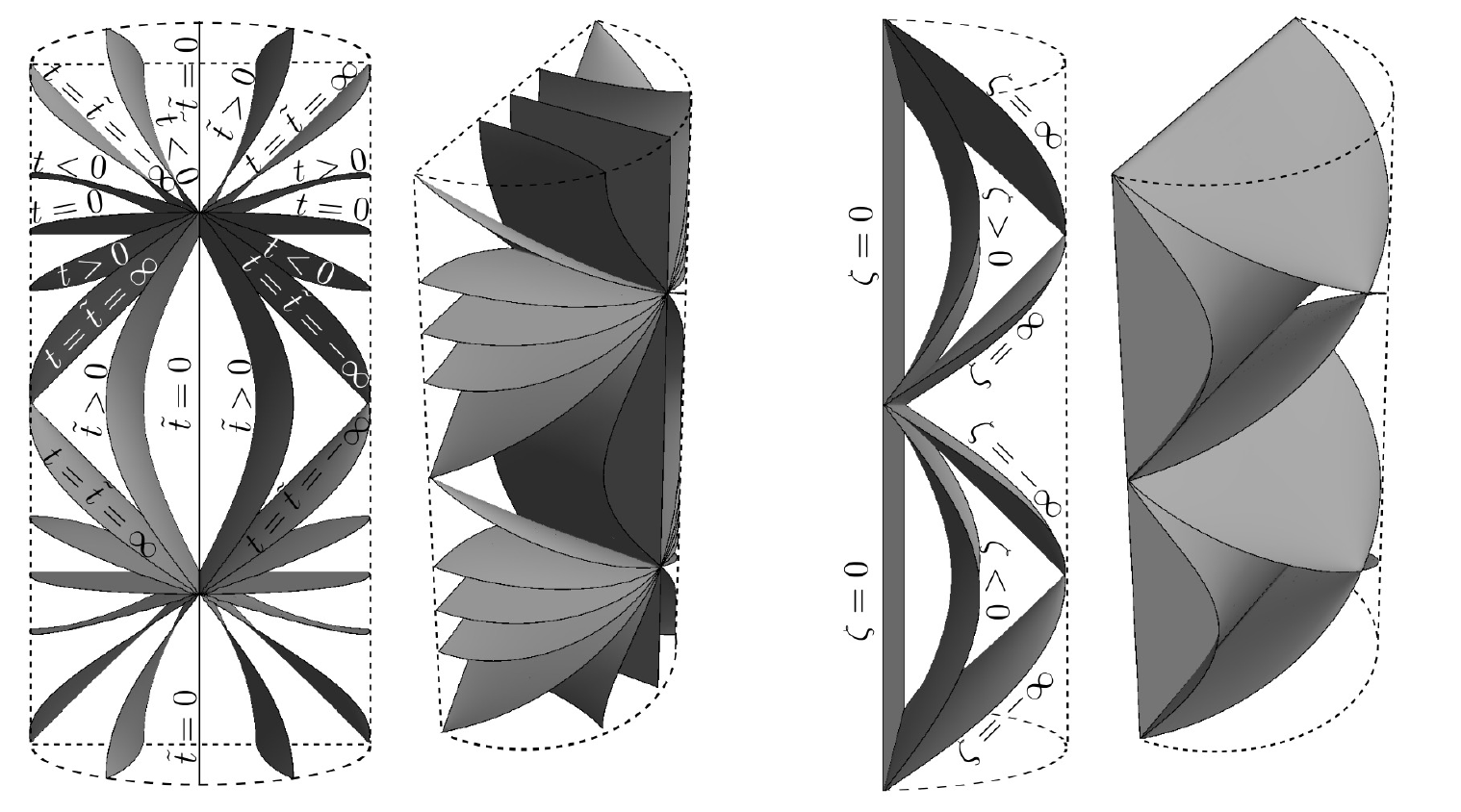}
\caption{\small
Left: A front view  and a general view on the surfaces ${t=\konst}$ for various $t$ in the global conformal coordinates of  anti-de~Sitter universe for ${\eta\in[-\pi,\pi]}$, ${\chi\in[0,\frac{\pi}{2})}$, ${\theta\in[0,\pi]}$, ${\y=\konst}$ For ${|q|>1}$, the coordinate $t$ has been relabeled to $\tilde{t}$. The surfaces ${t=0}$ correspond to ${\eta=\pm\frac{\pi}{2}}$ while the surfaces ${\tilde{t}=0}$ correspond to  ${\theta=\frac{\pi}{2}}$. Right: A side view and a general view on the surfaces ${\yy=\konst}$ for various  $\yy$, drawn in the conformal coordinates for ${\y=\frac{\pi}{2}}$. The surface ${\yy=0}$ corresponds to  ${\theta=0,\pi}$. Moreover, ${\y=0}$ implies ${\yy=0}$. The coordinate $\yy$ also does not cover the whole anti-de~Sitter space-time.}
\label{img:ConfadS11t}
\end{center}
\end{figure}

We use the relations (\ref{eq:adS11cfinv}) to draw the surfaces  ${p=\konst}$ (Figure~\ref{img:ConfadS11p}), ${q=\konst}$ (Figure~\ref{img:ConfadS11q0}), ${t=\konst}$ (left part of Figure~\ref{img:ConfadS11t}), and $\yy=\konst$ (right part of Figure~\ref{img:ConfadS11t}), respectively. In order to draw these global conformal pictures, it was necessary to suppress one coordinate. Because  $p,q$ and $\yy$ now explicitly depend on $\y$, we cannot simply suppress it (unlike in the previous cases of  de~Sitter universe). It can be seen from (\ref{eq:adS11conf}) that for larger $\y$ there is a greater constraint on the ranges of $p$ and $q$. Therefore, the coordinates $p$ and $q$ cover a smaller portion of the section. As an important illustration we draw the section ${\y=0}$ (and also ${\y=\frac{\pi}{4}}$ and ${\frac{\pi}{2}}$ in Figure~\ref{img:ConfadS11p}). For $\y$ shifted by $\pi$ we would obtain the same pictures (possibly with $\pm$).

\subsection{Subcase ${\Lambda<0}$, ${\epsilon_{2}=1}$, ${\epsilon_{0}=0}$}
\label{sbs:adS10}

\begin{figure}[h!]
\begin{center}
\includegraphics[scale=1]{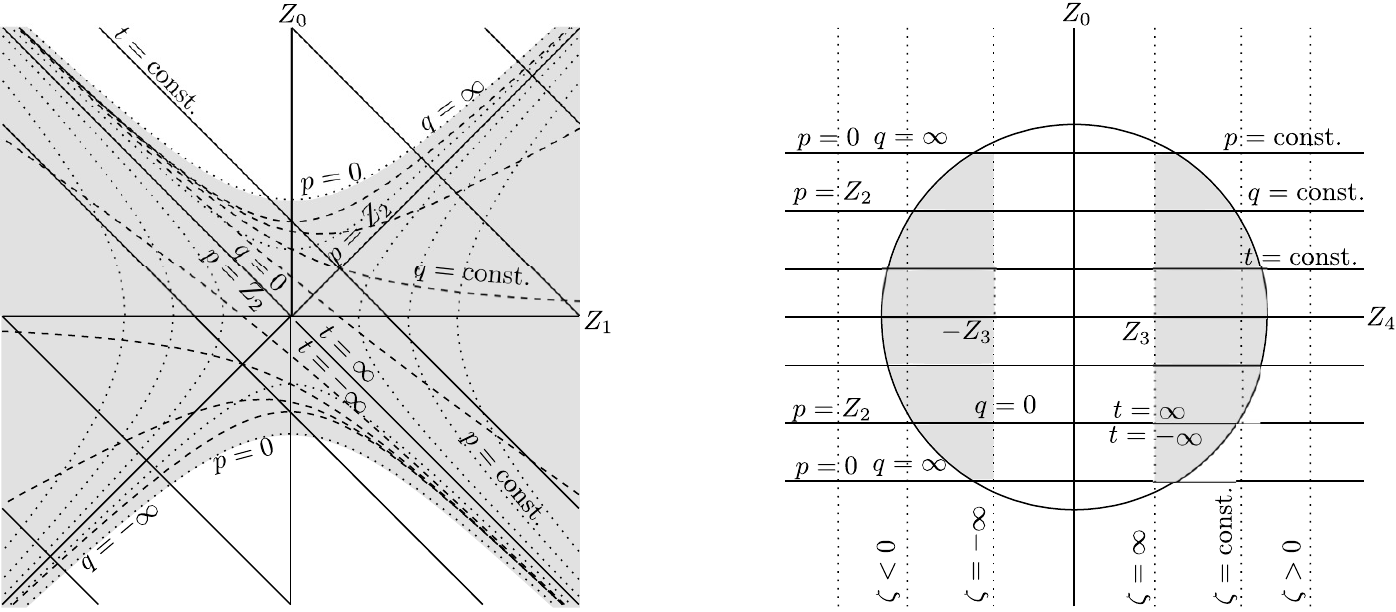}
\caption{\small
Left: Section ${Z_{2}=\konst>0}$ for ${\yy=0}$. The coordinate lines in this section look the same as in the left part of Figure~\ref{img:dS10Z0Z1} for ${\Lambda>0}$ since the parameterizations (\ref{eq:adS10par}) and (\ref{eq:dS10par}) coincide. Therefore, the section ${Z_{0}=\konst>0}$ also looks the same as in the right part of Figure~\ref{img:dS10Z0Z1}. Right: Section through the anti-de~Sitter spacetime spanned by the two temporal coordinates $Z_0$ and $Z_4$ of (\ref{D1}), (\ref{D2}). The circle indicates a section through the anti-de~Sitter hyperboloid (for a growing $Z_1$, the radius of the circle grows, and the lines ${p=0}$ move further away from the $Z_4$ axis). Here the lines ${p=\konst}$ coincide with the lines ${q=\konst}$ and ${t=\konst}$ The lines $\yy=\konst$ only cover the region ${|Z_{4}|>|Z_{3}|}$.
}\label{img:adS10Z0Z1}
\end{center}
\end{figure}

In this case, the anti-de~Sitter metric (\ref{eq:PDdads})--(\ref{eq:P}) with ${\y=\yy}$ reads
\begin{equation}\label{eq:adS10ds}
\dd s^{2}=p^{2}\Big(-\frac{\dd q^{2}}{q^{2}}+q^{2}\,\dd t^{2}\,\Big)
    +\frac{a^{2}\,\dd p^{2}}{a^{2}+p^{2}}+(a^{2}+p^{2})\,\dd \yy^{2}\,,
\end{equation}
where ${p\in[0,\infty)}$, ${q\in\mathbb{R}\setminus\{0\}}$, ${t,\yy\in\mathbb{R}}$. Clearly, $t$ is a \emph{spatial} coordinate, whereas $q$ is \emph{temporal}. Such coordinates cover the anti-de~Sitter hyperboloid (\ref{D1}) as
\begin{equation}
\left. \begin{array}{l}
Z_0 = {\displaystyle \pul \,p\,q\big(1+t^{2}-q^{-2}\big)}\,, \\[4pt]
Z_1 = {\displaystyle \pul \,p\,q\big(1-t^{2}+q^{-2}\big)}\,, \\[4pt]
Z_2 = {\displaystyle p\,q\,t}\,, \\[4pt]
Z_3 = {\displaystyle \pm\sqrt{a^2+p^{2}}\,\sinh\yy}\,, \\[4pt]
Z_4 = {\displaystyle \pm\sqrt{a^2+p^{2}}\,\cosh\yy}\,,
 \end{array} \!\right\} \ \Leftrightarrow \ \left\{ \!
 \begin{array}{l}
 t  = {\displaystyle \frac{Z_2}{Z_0+Z_1}}\,, \\[8pt]
\tanh \yy = {\displaystyle \frac{Z_3}{Z_4}}\,, \\[6pt]
p =  {\displaystyle \sqrt{-Z_0^2+Z_1^2+Z_2^2}}\,, \\[6pt]
q =  {\displaystyle \frac{Z_0+Z_1}{\sqrt{-Z_0^2+Z_1^2+Z_2^2}}}\,,
 \end{array} \right.
 \label{eq:adS10par}
\end{equation}
where the ``$+$'' sign corresponds to the coordinate chart ${Z_4>0}$, while the ``$-$'' sign corresponds to ${Z_4<0}$. These coordinates are visualized in Figure~\ref{img:adS10Z0Z1}.

\clearpage

\textbf{Global conformal representation} of the anti-de~Sitter metric (\ref{eq:adS10ds}) is given by the transformation (obtained by comparing (\ref{eq:adS10par}) with (\ref{confADS}))
\begin{eqnarray}\label{eq:adS10cfinv}
\begin{array}{rl}
\cotg\eta&\!\!\!={\displaystyle \pm \frac{p\,q\,\big(1+t^{2}-q^{-2}\big)}{2\,\sqrt{p^{2}+a^{2}}\cosh\yy}}\,,\\[10pt]
a\,\tg\chi&\!\!\!=\sqrt{p^2q^2\Big(\frac{1}{4}\big(1-t^{2}+q^{-2}\big)^2+t^2\Big)+(p^{2}+a^{2})\sinh^2\yy}\,,\\[10pt]
\tg\theta&\!\!\!={\displaystyle \frac{2\,\sqrt{p^2q^2t^2+(p^2+a^2)\sinh^2\yy}}{p\,q\,\big(1-t^{2}+q^{-2}\big)}} \,,\\[10pt]
\cotg\y &\!\!\!={\displaystyle \pm \frac{p\,q\,t}{\sqrt{p^2+a^2}\sinh\yy}}\,,
\end{array}
\end{eqnarray}
where the signs ``$\pm$'' correspond to the coordinate charts ${Z_4>0}$ and ${Z_4<0}$, respectively.
An inverse transformation then reads
\begin{eqnarray}\label{eq:adS10conf}
t \rovno \frac{\sin\chi\sin\theta\cos\phi}{\sin\chi\cos\theta+\cos\eta}\,,\nonumber\\
p \rovno \frac{a}{\cos\chi}\,\sqrt{\sin^2\chi(1-\sin^2\theta\sin^2\y)-\cos^2\eta}\,,\nonumber\\
q \rovno \frac{\cos\eta+\sin\chi\cos\theta}{\sqrt{\sin^2\chi(1-\sin^2\theta\sin^2\y)-\cos^2\eta}}\,,\\
 \tanh\yy\rovno \frac{\sin\chi\sin\theta\sin\y}{\sin\eta}\,.\nonumber
\end{eqnarray}

\begin{figure}[h!]
\begin{center}
\includegraphics[scale=0.98]{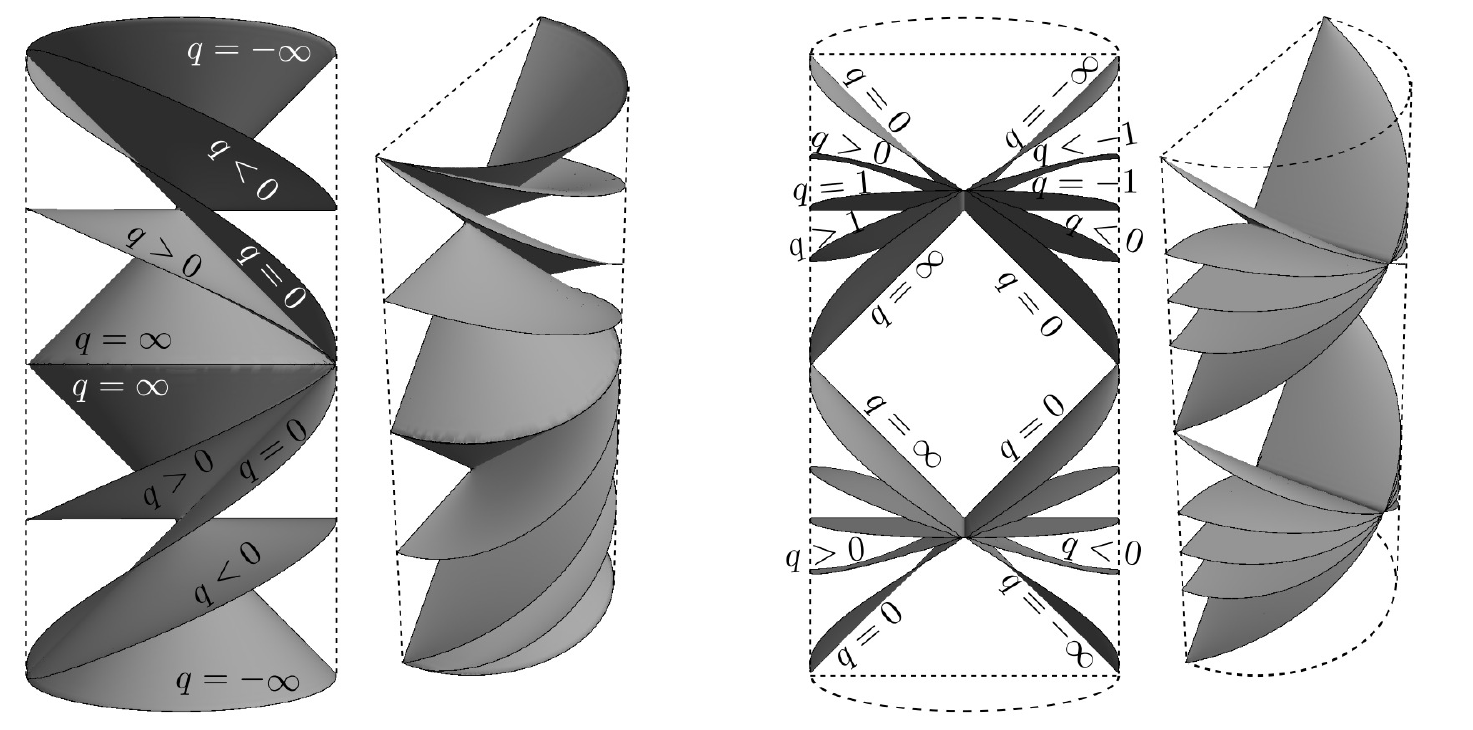}
\caption{\small
A front view and a general view on the surfaces ${q=\konst}$ for various $q$ in the global conformal coordinates for ${\y=0}$ (left two pictures) and for ${\y=\frac{\pi}{2}}$ (right two pictures). The coordinate $q$ does not cover the regions of  anti-de Sitter spacetime inside the half-cones ${q=\pm\infty}$. For ${\y=\frac{\pi}{2}}$ even a smaller part of the spacetime is covered.}\label{img:ConfadS10q0}
\end{center}
\end{figure}

The surfaces  ${p=\konst}$ and ${\yy=\konst}$ are given by the same relations in both cases ${\ez=1}$ and ${\ez=0}$ (compare (\ref{eq:adS11conf}) and (\ref{eq:adS10conf})) so that their form is the same as in Figures~\ref{img:ConfadS11p} and~\ref{img:ConfadS11t}. The surfaces ${q=\konst}$ are now different --- they are shown in Figure~\ref{img:ConfadS10q0} for ${\y=0}$ and ${\y=\frac{\pi}{2}}$.

\subsection{Subcase ${\Lambda<0}$, ${\epsilon_{2}=1}$, ${\epsilon_{0}=-1}$}
\label{sbs:adS1-1}

\begin{figure}[h]
\begin{center}
\includegraphics[scale=1]{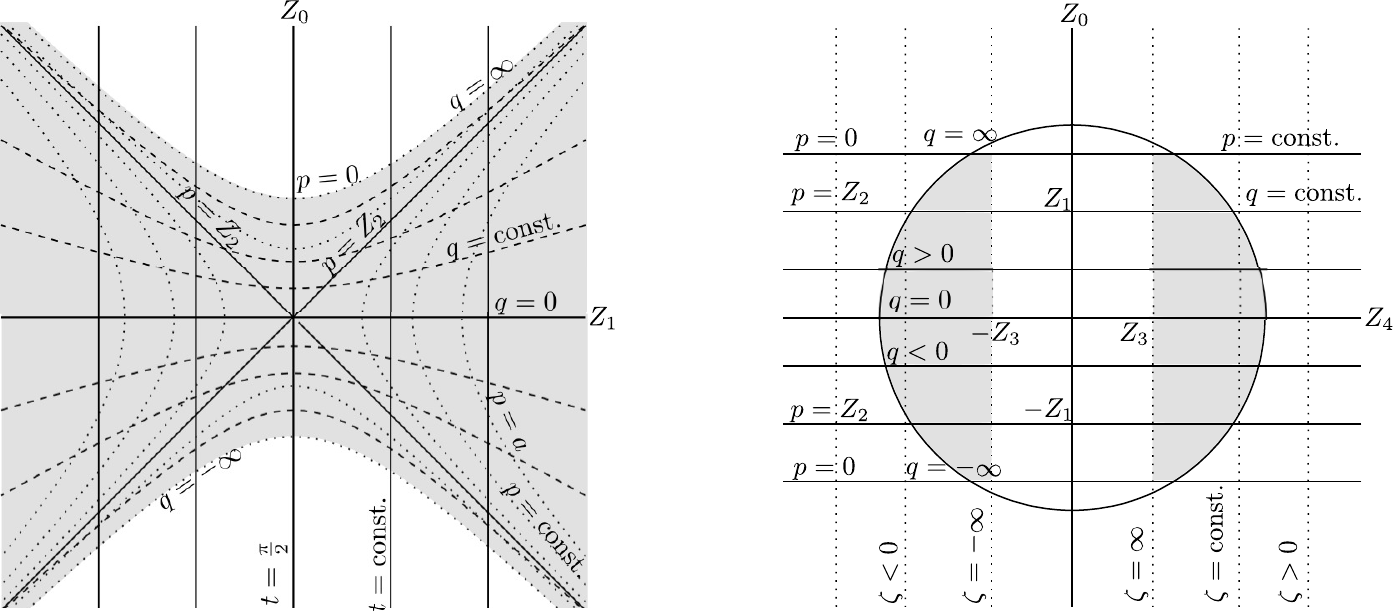}
\caption{\small
Left: Section ${Z_{2}=\konst>0}$ for ${\yy=0}$. The coordinate lines in this section are the same as in the left part of Figure~\ref{img:dS1-1Z0Z1a} for ${\Lambda>0}$ due to the similarity of the parameterizations (\ref{eq:adS1-1par}) and (\ref{eq:dS1-1par}). The section ${Z_{0}=\konst>0}$ also looks the same as in the right part of Figure~\ref{img:dS1-1Z0Z1a}.
Right: Section spanned by $Z_0$ and $Z_4$ through the anti-de~Sitter spacetime resembles the analogous section in the right part of Figure~\ref{img:adS10Z0Z1}.
}\label{img:adS1-1Z0Z1a}
\end{center}
\end{figure}

The anti-de~Sitter metric (\ref{eq:PDdads})--(\ref{eq:P}) with ${\y=\yy}$ now takes the form
\begin{equation}\label{eq:adS1-1ds}
\dd s^{2}=p^{2}\Big(-\frac{\dd q^2}{1+q^2}+(1+q^2)\,\dd t^2\,\Big)
    +\frac{a^2\,\dd p^2}{a^2+p^2}+(a^2+p^2)\,\dd \yy^2\,,
\end{equation}
where ${p\in[0,\infty)}$, ${t\in[0,2\pi)}$, ${q,\yy\in\mathbb{R}}$. Again, $t$ is a \emph{spatial angular coordinate}, whereas $q$ is \emph{temporal}. These coordinates cover the anti-de~Sitter hyperboloid (\ref{D1}) as
\begin{equation}
\left. \begin{array}{ll}
 Z_0 = p\,q \,, \\[4pt]
 Z_1 = p\,\sqrt{1+q^2}\,\cos t \,, \\[4pt]
 Z_2 = p\,\sqrt{1+q^2}\,\sin t\,, \\[4pt]
 Z_3 = \pm\sqrt{a^2+p^2}\,\sinh\,\yy\,, \\[4pt]
 Z_4 = \pm\sqrt{a^2+p^2}\,\cosh\,\yy\,,
 \end{array}\right\} \quad\Leftrightarrow\quad
 \left\{ \begin{array}{l}
 \tan t  = {\displaystyle \frac{Z_2}{Z_1}}\,, \\[8pt]
 \tanh \yy = {\displaystyle \frac{Z_3}{Z_4}}\,, \\[8pt]
 p = \sqrt{-Z_{0}^{2}+Z_{1}^{2}+Z_{2}^{2}}\,, \\[6pt]
 q = {\displaystyle \frac{Z_0}{\sqrt{-Z_0^2+Z_1^2+Z_2^2}} }\,.
 \end{array} \right.
\label{eq:adS1-1par}
\end{equation}
This is visualized in different sections through the anti-de~Sitter hyperboloid in Figure~\ref{img:adS1-1Z0Z1a}.

\clearpage

\textbf{Global conformal representation} of the anti-de~Sitter metric (\ref{eq:adS1-1ds}) is obtained by comparing (\ref{eq:adS1-1par}) with (\ref{confADS}):
\begin{eqnarray}\label{eq:adS1-1cfinv}
\begin{array}{rl}
\cotg\eta&\!\!\!={\displaystyle \pm \frac{p\,q}{\sqrt{p^{2}+a^{2}}\cosh\yy}}\,,\\[10pt]
\tg\chi&\!\!\! ={\displaystyle \sqrt{\frac{p^2}{a^2}\big(\cosh^2\yy+q^2\big)+\sinh^2\yy}}\,,\\[10pt]
\tg\theta&\!\!\!={\displaystyle \frac{\sqrt{p^2(1+q^2)\sin^2 t+(p^2+a^2)\sinh^2 \yy}}{p\sqrt{1+q^2}\cos t}} \,,\\[10pt]
\cotg\y &\!\!\!={\displaystyle \pm \frac{p\sqrt{1+q^2}\sin t}{\sqrt{p^2+a^2}\sinh\yy}}\,,
\end{array}
\end{eqnarray}
where the signs ``$\pm$'' correspond to the coordinate charts ${Z_4>0}$ and ${Z_4<0}$, respectively.
An inverse transformation is
\begin{eqnarray}\label{eq:adS1-1conf}
\tan t \rovno \tan\theta\cos\phi\,,\nonumber\\
p \rovno \frac{a}{\cos\chi}\,\sqrt{\sin^2\chi(1-\sin^2\theta\sin^2\y)-\cos^2\eta}\,,\nonumber\\
q \rovno \frac{\cos\eta}{\sqrt{\sin^2\chi(1-\sin^2\theta\sin^2\y)-\cos^2\eta}}\,,\\
 \tanh\yy\rovno \frac{\sin\chi\sin\theta\sin\y}{\sin\eta}\,.\nonumber
\end{eqnarray}

\begin{figure}[h]
\begin{center}
\includegraphics[scale=1]{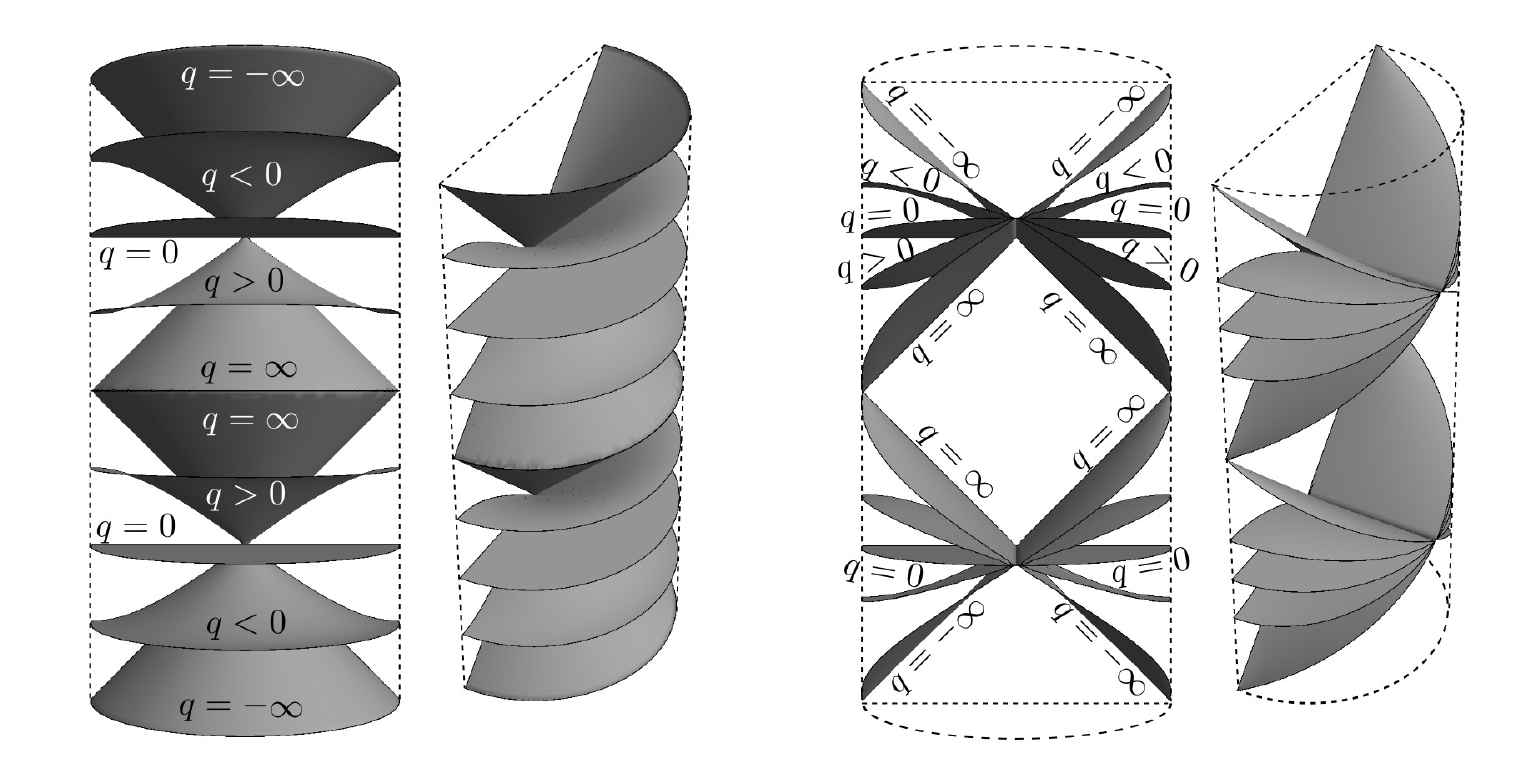}
\caption{\small
A front view and a general view on the surfaces ${q=\konst}$ in the global conformal coordinates for ${\y=0}$ (left two pictures) and for ${\y=\frac{\pi}{2}}$ (right two pictures). As in the previous case of Figure~\ref{img:ConfadS10q0}, the coordinate $q$ does not cover the regions  inside the half-cones ${q=\pm\infty}$.}\label{img:ConfadS1-1q0}
\end{center}
\end{figure}

The surfaces ${p=\konst}$ and ${\yy=\konst}$ are given by the same relations in all the three cases ${\ez=1,0,-1}$, so that their form is the same as in Figure~\ref{img:ConfadS11p} and Figure~\ref{img:ConfadS11t}. However, the surfaces ${q=\konst}$ are now different, as shown in Figure~\ref{img:ConfadS1-1q0} for ${\y=0}$ and ${\y=\frac{\pi}{2}}$.

\subsection{Subcase ${\Lambda<0}$, ${\epsilon_{2}=0}$, ${\epsilon_{0}=1}$}
\label{sbs:adS01}

In this case ${Q=1}$, and the corresponding anti-de~Sitter metric (\ref{eq:PDdads})--(\ref{eq:P}) with ${\y=\yy}$ is
\begin{equation}\label{eq:adS01ds}
\dd s^{2}=p^{2}\,(-\dd t^2+\dd q^2)
    +\frac{a^2}{p^2}\,\dd p^2+p^2\,\dd \yy^2\,,
\end{equation}
in which ${p,q,t,\yy\in\mathbb{R}}$. With a simple transformation
\begin{equation}\label{relads01}
p = \frac{a^{2}}{x}\,,\qquad
t = \frac{\eta_{_0}}{a}\,, \qquad
q = \frac{y}{a}\,, \qquad
\yy = \frac{z}{a}\,,
\end{equation}
where ${\eta_{_0},x,y,z\in\mathbb{R}}$, we immediately obtain the metric
\begin{equation}\label{eq:adScoflat}
\dd s^{2}= \frac{a^2}{x^2}\,(-\dd \eta_{_0}^{2}+\dd x^{2}+\dd y^{2}+\dd z^{2})\,.
\end{equation}
This is exactly the well-known \textit{conformally flat Poincar\'e form}, see e.g. the metric (5.14) in~\cite{GriPod2009}.
These coordinates on the anti-de~Sitter hyperboloid are visualized on Fig. 5.5 therein, as well as the explicit parametrization Eq. (5.13). Combining this with  relations (\ref{relads01}) we obtain
\begin{equation}
\left. \begin{array}{ll}
 Z_0 = {\displaystyle \frac{p}{2}\left(1+\frac{s}{a^2}\right)} \,, \\[6pt]
 Z_1 = {\displaystyle \frac{p}{2}\left(1-\frac{s}{a^2}\right)} \,, \\[3pt]
 Z_2 = p\,q  \,, \\[4pt]
 Z_3 = p\,\yy\,, \\[4pt]
 Z_4 = p\,t  \,,
 \end{array}\right\} \quad\Leftrightarrow\quad
 \left\{ \begin{array}{l}
 t  = {\displaystyle \frac{Z_4}{Z_0+Z_1}}\,, \\[8pt]
 \yy = {\displaystyle \frac{ Z_3}{Z_0+Z_1}}\,, \\[8pt]
 p = Z_0+Z_1 \,, \\[6pt]
 q = {\displaystyle \frac{Z_2}{Z_0+Z_1} }\,,
 \end{array} \right.
\label{eq:adS01par}
\end{equation}
where
\begin{equation}
\frac{s}{a^2}=-t^2+q^2+\yy^2+\frac{a^2}{p^2}\,.
\end{equation}

The metric (\ref{eq:adScoflat}) of anti-de~Sitter spacetime has been thoroughly described and employed in literature, for example in the works on AdS/CFT correspondence. Of course, via the simple relations (\ref{relads01}), all this can be equivalently re-expresed in terms of the  alternative metric form (\ref{eq:adS01ds}) presented here.

\begin{figure}[h!]
\begin{center}
\includegraphics[scale=1]{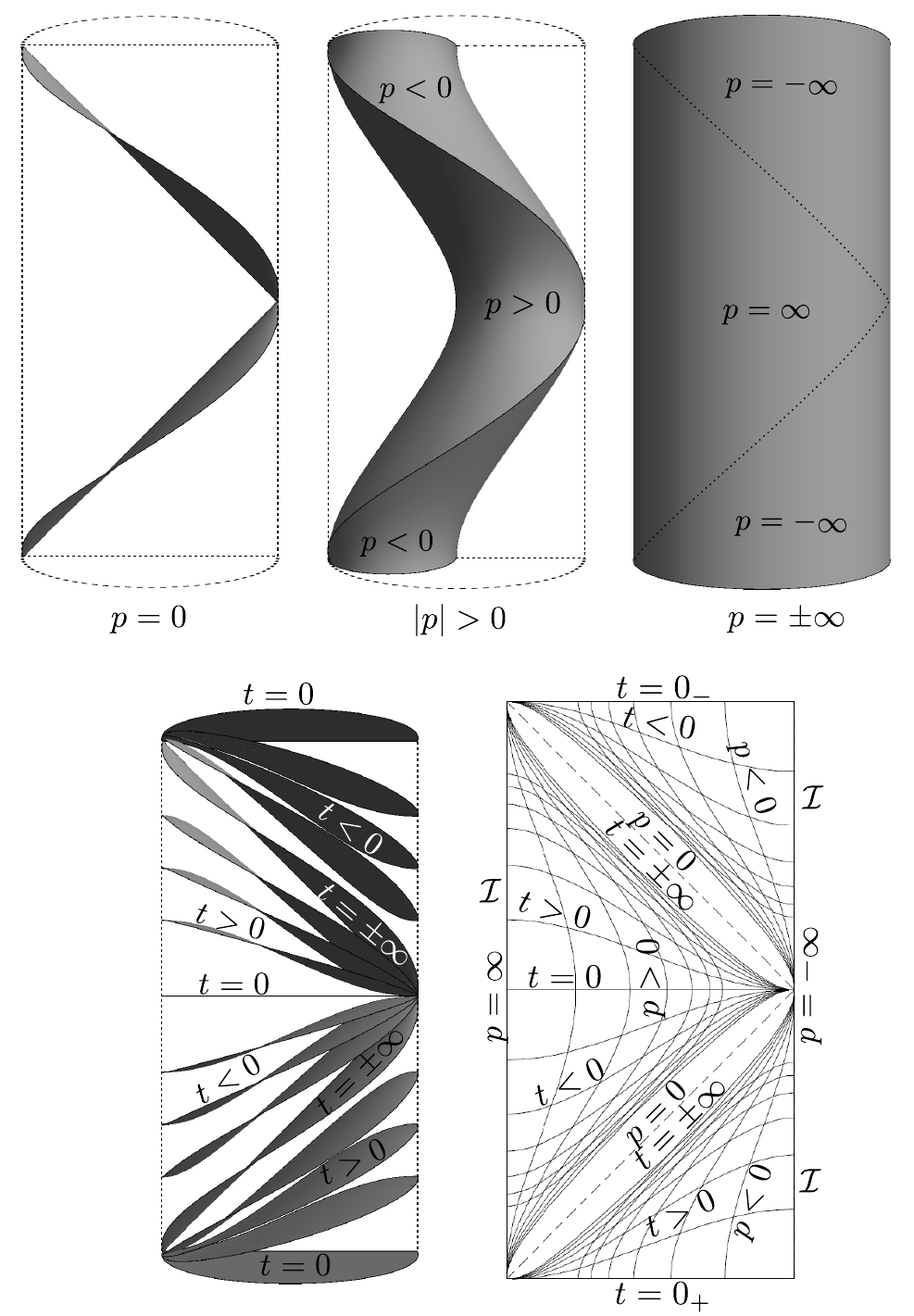}
\caption{\small
The coordinate surfaces ${p=\konst}$ (top row) and ${t=\konst}$ (bottom row on the left) for various values of $p$ and $t$ plotted in the global conformal cylinder of anti-de~Sitter universe. In particular, for section ${q,\yy=0}$ we clearly obtain the standrad two-dimensional Penrose conformal diagram (bottom row on the right).}\label{img:Penrose_adS}
\end{center}
\end{figure}

The coordinate surfaces of constant $p$ and $t$ within the conformal cylinder representing the global structure of anti-de~Sitter spacetime (shown in the right part of Figure~\ref{fig:adshyp1}) are plotted in Figure~\ref{img:Penrose_adS}. Recall that the conformal infinity~${\cal I}$ is the outer boundary ${\chi=\frac{\pi}{2}}$. It is well known that this corresponds to  ${x=0}$ (see e.g. Fig.~5.6 in~\cite{GriPod2009}), which is here equivalent to ${p=\infty}$ in the new coordinates of (\ref{eq:adS01ds}). In fact, it seems more natural to represent the conformal infinity~${\cal I}$ by the \emph{infinite value} of the coordinate $p$ rather than the zero value of the usual coordinate $x$.

The corresponding two-dimensional Penrose conformal diagram of anti-de~Sitter spacetime (the vertical shaded plane in the right part of Figure~\ref{fig:adshyp1}),
with the coordinate lines of $p$ and $t$, is plotted in the bootom right part of Figure~\ref{img:Penrose_adS}.

\subsection{Subcase ${\Lambda<0}$, ${\epsilon_{2}=0}$, ${\epsilon_{0}=-1}$}
In such a case ${Q=-1}$, and the  metric is thus
\begin{equation}\label{eq:adS0-1ds}
\dd s^{2}=p^{2}\,(-\dd q^2+\dd t^2)
    +\frac{a^2}{p^2}\,\dd p^2+p^2\,\dd \yy^2\,.
\end{equation}
This is clearly \emph{the same form as the previous metric} (\ref{eq:adS01ds}) if we swap the coordinates ${t \leftrightarrow q}$.

\subsection{Subcase ${\Lambda<0}$, ${\epsilon_{2}=0}$, ${\epsilon_{0}=0}$}
\label{sbs:adS00}
For this choice ${Q=0}$, so that the  metric (\ref{eq:PDdads})--(\ref{eq:P}) degenerates. This case is thus forbidden.

\subsection{Subcase ${\Lambda<0}$, ${\epsilon_{2}=-1}$, ${\epsilon_{0}=1}$}
\label{sbs:adS-11}

Finally, we will discuss three subcases of the anti-de~Sitter metric (\ref{eq:PDdads})--(\ref{eq:P}) for ${\Lambda<0}$ with ${\et=-1}$ and ${\ez=1,0,-1}$ (which are not possible for ${\Lambda\geq 0}$). For ${\epsilon_{0}=1}$ the metric reads
\begin{equation}\label{eq:adS-11ds}
\dd s^{2}=p^{2}\Big(\!-(1+q^{2})\,\dd t^{2}+\frac{\dd q^{2}}{1+q^{2}}\Big)
    +\frac{a^{2}\,\dd p^{2}}{p^{2}-a^{2}}+(p^{2}-a^{2})\,\dd \y^{2}\,,
\end{equation}
where ${\a=\sqrt{-3/\Lambda}\,}$ and
${p \in [a, \infty)}$, ${q\in\mathbb{R}}$, ${t,\y\in[0,2\pi)}$, with ${p= a}$ representing the axis of symmetry. This arises as the parametrization
\begin{equation}
\left. \begin{array}{ll}
 Z_0 = p\,\sqrt{1+q^2}\,\cos t \,, \\[6pt]
 Z_1 = p\,q \,, \\[3pt]
 Z_2 = \sqrt{p^2-a^2}\,\cos\y  \,, \\[4pt]
 Z_3 = \sqrt{p^2-a^2}\,\sin\y  \,, \\[4pt]
 Z_4 = p\,\sqrt{1+q^2}\,\sin t  \,,
 \end{array}\right\} \quad\Leftrightarrow\quad
 \left\{ \begin{array}{l}
 \tan t  = {\displaystyle \frac{Z_4}{Z_0}}\,, \\[8pt]
 \tan \y = {\displaystyle \frac{Z_3}{Z_2}}\,, \\[8pt]
 p = \sqrt{Z_0^2-Z_1^2+Z_4^{2}}\,, \\[6pt]
 q = {\displaystyle \frac{Z_1}{\sqrt{Z_0^2-Z_1^2+Z_4^2}} }\,,
 \end{array} \right.
\label{eq:adS-11par}
\end{equation}
of the anti-de~Sitter hyperboloid (\ref{D1}) in the flat space (\ref{D2}). Let us recall that $Z_0$ a $Z_4$ are two temporal coordinates expressed here by the natural single temporal coordinate ${t\in[0,2\pi)}$. The covering space is obtained by allowing ${t\in\mathbb{R}}$ in (\ref{eq:adS-11ds}).

\begin{figure}[h]
\begin{center}
\includegraphics[scale=1]{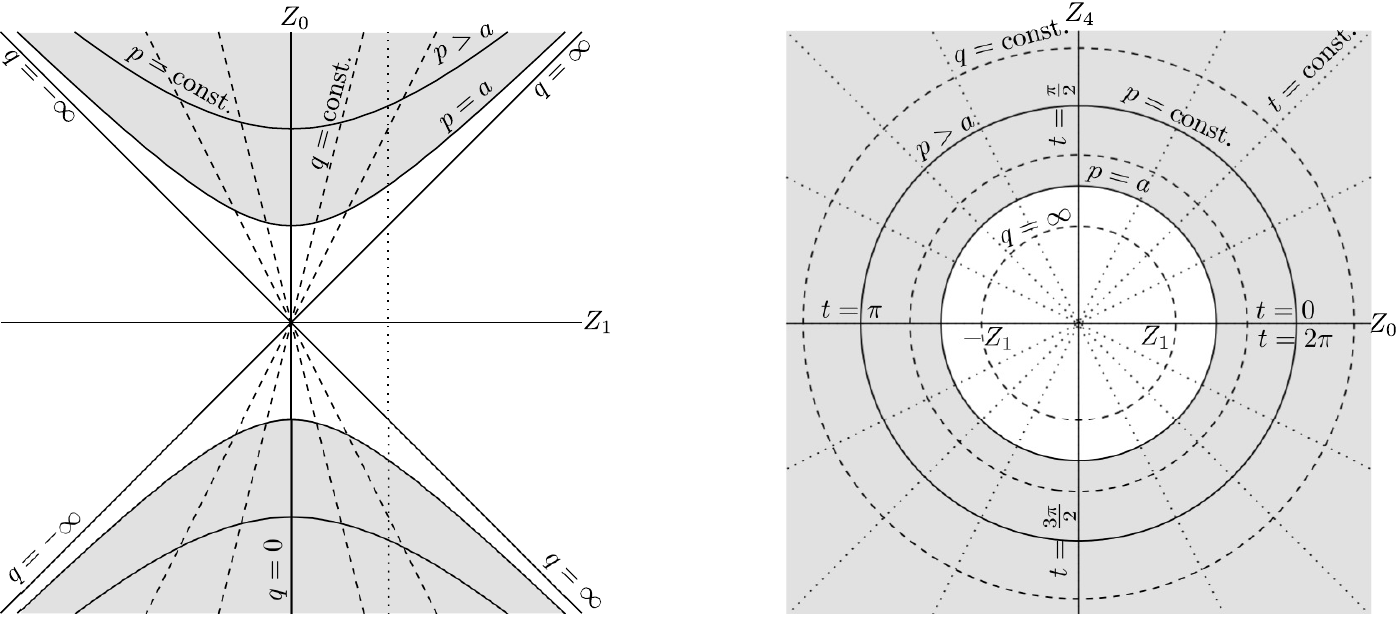}
\caption{\small
Left: Section ${Z_{4}=0}$, in which the coordinate lines ${p=\konst}$ are hyperbolas ${Z_0^2-Z_1^2=p^2}$ while ${q=\konst}$ are radial lines ${Z_0=\sqrt{1+q^{-2}}\,Z_1}$.
Right: Section ${Z_{1}=\konst>0}$ indicated by the vertical dashed line in the left part of this Figure. The lines ${p=\konst}$ are circles through the anti-de~Sitter hyperboloid. Their radius grows as $\sqrt{p^2+Z_1^2}$, and similarly the radius of the dashed circles ${q=\konst}$ grows as $\sqrt{1+q^{-2}}\,Z_1$. An inclination of the dotted straight lines ${t=\konst}$ are the same for all values of $Z_{1}$.}\label{img:adS-11Z0Z1}
\end{center}
\end{figure}

It is interesting to notice that after the formal relabeling $Z_0 \leftrightarrow Z_1$ and ${Z_2 \rightarrow Z_3 \rightarrow Z_4 \rightarrow Z_2}$ we obtain basically the same expressions as (\ref{eq:dS1-1par}) for the de~Sitter subcase ${\Lambda>0}$, ${\epsilon_{2}=1}$, ${\epsilon_{0}=-1}$. Therefore, the sections through the anti-de~Sitter hyperboloid closely resemble those shown in Figure~\ref{img:dS1-1Z0Z1a}, after the relabeling of the axes $Z_a$ and reconsidering different ranges of the coordinates. In particular, in Figure~\ref{img:adS-11Z0Z1} we plot the sections ${Z_4=0}$ and ${Z_1=\konst>0}$, respectively. It can be seen that (unlike in the cases ${\et=1}$) for ${\et=-1}$ the coordinates \emph{cover the whole anti-de~Sitter universe}, which is also true for the related subcases ${\ez=0}$ and ${\ez=-1}$.

\textbf{Global conformal representation} is obtained by combining (\ref{eq:adS-11par}) with (\ref{confADS}):
  \begin{equation}
\left. \begin{array}{l}
\eta = t \,, \\[7pt]
\cos\chi = {\displaystyle  \frac{a}{p\,\sqrt{1+q^2}}}\,, \\[11pt]
\cos\theta = {\displaystyle \frac{p\,q}{\sqrt{p^2\left(1+q^2\right)-a^2}}}\,,
 \end{array} \!\right\} \ \Leftrightarrow \ \left\{ \!
 \begin{array}{l}
t = \eta\,, \\[10pt]
p = a\,\sqrt{1+\tg^2\chi\sin^2\theta}\,, \\[10pt]
q = {\displaystyle \frac{\sin\chi\cos\theta}{\sqrt{1-\sin^2\chi\cos^2\theta}}}\,.
 \end{array} \right.
 \label{eq:adS-11conf}
 \end{equation}

\begin{figure}[h!]
\begin{center}
\includegraphics[scale=1]{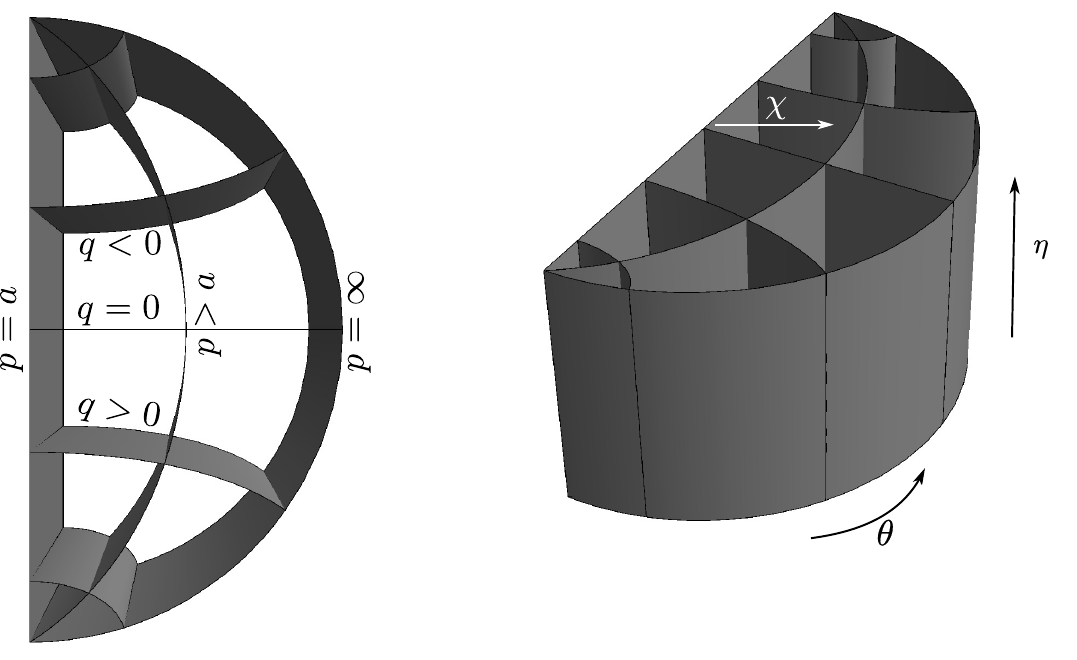}
\caption{\small
A view from top (left) and a general view (right) on the surfaces ${p=\konst}$ and ${q=\konst}$ in terms of the standard global coordinates ${\eta\in[0,\frac{\pi}{2}]}$, ${\chi\in[0,\frac{\pi}{2})}$, ${\theta\in[0,\pi]}$ covering the complete anti-de~Sitter spacetime  for ${\y=\konst}$
The initial surface ${p=a}$ corresponds to coordinate singularities ${\theta=0, \pi}$, while the surface ${p=\infty}$ corresponds to ${\chi=\frac{\pi}{2}}$ which is the conformal infinity $\mathcal{I}$. In fact, the surfaces ${p=\konst}$ are the same for all subcases ${\et=-1}$, that is for ${\ez=1,0, -1}$. The surface ${q=0}$ corresponds to ${\theta=\frac{\pi}{2}}$, while ${q=\pm\infty}$ are the lines ${\chi=\frac{\pi}{2}}$, ${\theta=0, \pi}$ on $\mathcal{I}$. The coordinates $p$, $q$ thus very naturally foliate the whole anti-de~Sitter universe.}\label{img:ConfadS-11pg}
\end{center}
\end{figure}

Specific character of the coordinates in metric (\ref{eq:adS-11ds}) can thus be visualized within the  anti-de~Sitter global conformal cylinder. In Figure~\ref{img:ConfadS-11pg} we plot both the surfaces  ${p=\konst}$ and the surfaces ${q=\konst}$ It can be seen that they foliate the entire anti-de~Sitter spacetime in a very natural way. Moreover, the conformal infinity~${\cal I}$, given by ${\chi=\frac{\pi}{2}}$, is now completely represented by the boundary ${p=\infty}$.

\subsection{Subcase ${\Lambda<0}$, ${\epsilon_{2}=-1}$, ${\epsilon_{0}=0}$}
\label{sbs:adS-10}

In this case, the anti-de~Sitter metric takes the form
\begin{equation}\label{eq:adS-10ds}
\dd s^{2}=p^{2}\Big(\!-q^{2}\,\dd t^{2}+\frac{\dd q^{2}}{q^{2}}\Big)
    +\frac{a^{2}\,\dd p^{2}}{p^{2}-a^{2}}+(p^{2}-a^{2})\,\dd \y^{2}\,,
\end{equation}
where
${p \in [a, \infty)}$, ${q\in\mathbb{R}\setminus\{0\}}$, $t\in\mathbb{R}$, ${\y\in[0,2\pi)}$. Again, ${p= a}$ represents the axis of symmetry, $t$~is a temporal coordinate, whereas $q$~is spatial. These coordinates are given by
\begin{equation}
\left. \begin{array}{l}
Z_0 = {\displaystyle \pul \,p\,q\big(1-t^{2}+q^{-2}\big)}\,, \\[4pt]
Z_1 = {\displaystyle \pul \,p\,q\big(1+t^{2}-q^{-2}\big)}\,, \\[4pt]
Z_2 = {\displaystyle \sqrt{p^2-a^{2}}\,\cos\y}\,, \\[4pt]
Z_3 = {\displaystyle \sqrt{p^2-a^{2}}\,\sin\y}\,, \\[4pt]
Z_4 = {\displaystyle p\,q\,t}\,,
 \end{array} \!\right\} \ \Leftrightarrow \ \left\{ \!
 \begin{array}{l}
 t  = {\displaystyle \frac{Z_4}{Z_0+Z_1}}\,, \\[8pt]
\tan \y = {\displaystyle \frac{Z_3}{Z_2}}\,, \\[6pt]
p =  {\displaystyle \sqrt{Z_0^2-Z_1^2+Z_4^2}}\,, \\[6pt]
q =  {\displaystyle \frac{Z_0+Z_1}{\sqrt{Z_0^2-Z_1^2+Z_4^2}}}\,.
 \end{array} \right.
 \label{eq:adS-10par}
\end{equation}

\begin{figure}[h!]
\begin{center}
\includegraphics[scale=1]{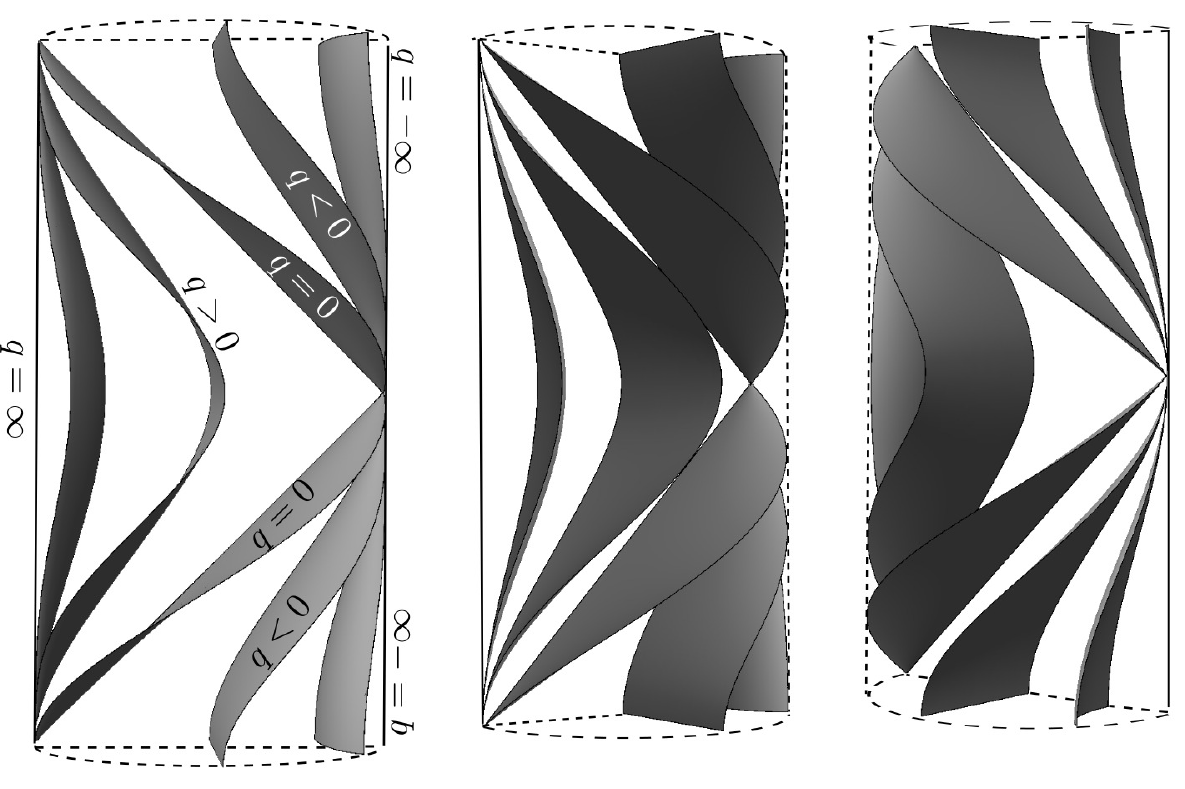}
\caption{\small
A front view (left), a view from left (middle), and a view from right (right) on surfaces ${q=\konst}$ for various values of $q$ in the global anti-de~Sitter coordinates (${\eta\in[-\pi,\pi]}$, ${\chi\in[0,\frac{\pi}{2})}$, ${\theta\in[0,\pi]}$, ${\y=\konst}$). The surfaces ${q=\pm\infty}$ degenerate to lines ${\chi=\frac{\pi}{2}}$, ${\theta=0, \pi}$ on~$\mathcal{I}$.
}\label{img:ConfadS-10q}
\end{center}
\end{figure}

\textbf{Global conformal representation} is now
  \begin{equation}
\left. \begin{array}{l}
\cotg\eta = {\displaystyle \frac{1}{2\,t}\big(1-t^{2}+q^{-2}\big)} \,, \\[7pt]
a\,\tg\chi = {\displaystyle  \sqrt{{\textstyle\frac{1}{4}}\,p^2q^2\big(1+t^{2}-q^{-2}\big)^2+p^2-a^2}}\,, \\[11pt]
\cotg\theta = {\displaystyle \frac{p\,q\,\big(1+t^{2}-q^{-2}\big)}{2\,\sqrt{p^2-a^2}}}\,,
 \end{array} \!\right\} \ \Leftrightarrow \ \left\{ \!
 \begin{array}{l}
t = {\displaystyle \frac{\sin\eta}{\cos\eta+\sin\chi\cos\theta}}\,, \\[10pt]
p = a\,\sqrt{1+\tg^2\chi\sin^2\theta}\,, \\[10pt]
q = {\displaystyle \frac{\cos\eta+\sin\chi\cos\theta}{\sqrt{1-\sin^2\chi\cos^2\theta}}}\,.
 \end{array} \right.
 \label{eq:adS-10conf}
 \end{equation}
In view of (\ref{eq:adS-11conf}), the surfaces ${p=\konst}$ are \emph{the same} as in Figure~\ref{img:ConfadS-11pg}, but the surfaces ${q=\konst}$ have a different shape shown in Figure~\ref{img:ConfadS-10q}.

\subsection{Subcase ${\Lambda<0}$, ${\epsilon_{2}=-1}$, ${\epsilon_{0}=-1}$}
\label{sbs:adS-1-1}

The anti-de~Sitter metric (\ref{eq:PDdads})--(\ref{eq:P}) in this final case reads
\begin{equation}\label{eq:adS-1-1ds}
\dd s^{2}=p^{2}\Big(\!-(q^2-1)\,\dd t^{2}+\frac{\dd q^{2}}{q^2-1}\Big)
    +\frac{a^{2}\,\dd p^{2}}{p^{2}-a^{2}}+(p^{2}-a^{2})\,\dd \y^{2}\,,
\end{equation}
where ${p\in(-\infty,-a]\cup \,[a, \infty)}$, ${q\in\mathbb{R}\setminus\{\pm1\}}$, $t\in\mathbb{R}$, ${\y\in[0,2\pi)}$, with ${p=\pm a}$ representing the axis of symmetry. For $|q|>1$ the coordinate $t$~is temporal and $q$~is spatial, whereas for $|q|<1$ the coordinate $t$~is spatial and $q$~is temporal. These two cases thus need to be discussed separately:

\noindent
$\bullet$ \textbf{For} ${|q|>1}$, the coordinates of the metric (\ref{eq:adS-1-1ds}) parametrize the anti-de~Sitter hyperboloid (\ref{D1})
 \begin{equation}
\left. \begin{array}{l}
Z_0 = {\displaystyle p\,\sqrt{q^{2}-1}\,\sinh t}\,, \\[8pt]
Z_1 = {\displaystyle p\,\sqrt{q^{2}-1}\,\cosh t}\,, \\[8pt]
Z_2 = {\displaystyle \sqrt{p^2-a^{2}}\,\cos\y}\,, \\[8pt]
Z_3 = {\displaystyle \sqrt{p^2-a^{2}}\,\sin\y}\,, \\[8pt]
Z_4 = {\displaystyle |p|\,q}\,,
 \end{array} \!\right\} \ \Leftrightarrow \ \left\{ \!
 \begin{array}{l}
\tgh t  = {\displaystyle \frac{Z_0}{Z_1}}\,, \\[8pt]
\tan \y = {\displaystyle \frac{Z_3}{Z_2}}\,, \\[6pt]
p =  {\displaystyle \sign(Z_1)\,\sqrt{Z_0^2-Z_1^2+Z_4^2}} \\[2pt]
\quad =  {\displaystyle \sign(Z_1)\,\sqrt{a^2+Z_2^2+Z_3^2}}\,, \\[6pt]
q =  {\displaystyle \frac{Z_4}{\sqrt{Z_0^2-Z_1^2+Z_4^2}}}\,.
 \end{array} \right.
 \label{eq:adS-1-1apar}
 \end{equation}
The parametrization (\ref{eq:adS-1-1apar}) represents \emph{two maps} covering the anti-de~Sitter manifold: the first one for ${p>0}$ covers the part ${Z_1>0}$, while the other for ${p<0}$ covers ${Z_1<0}$. Moreover, ${q>0}$ corresponds to ${Z_4>0}$, while ${q<0}$ corresponds to ${Z_4<0}$.

\noindent
$\bullet$ \textbf{For} ${|q|<1}$, the parametrization is the same as (\ref{eq:adS-1-1apar}), except that now
 \begin{equation}
\left. \begin{array}{l}
Z_0 = {\displaystyle p\,\sqrt{1-q^{2}}\,\cosh t}\,, \\[10pt]
Z_1 = {\displaystyle p\,\sqrt{1-q^{2}}\,\sinh t}\,,
 \end{array} \!\right\} \ \Leftrightarrow \ \left\{ \!
 \begin{array}{l}
\tgh t  = {\displaystyle \frac{Z_1}{Z_0}}\,, \\[6pt]
p =  {\displaystyle \sign(Z_0)\,\sqrt{Z_0^2-Z_1^2+Z_4^2}} \,.
 \end{array} \right.
 \label{eq:adS-1-1bpar}
 \end{equation}
Again, these are \emph{two maps}:  ${p>0}$ covers the part ${Z_0>0}$, while ${p<0}$ covers ${Z_0<0}$.

\begin{figure}[h]
\begin{center}
\includegraphics[scale=1]{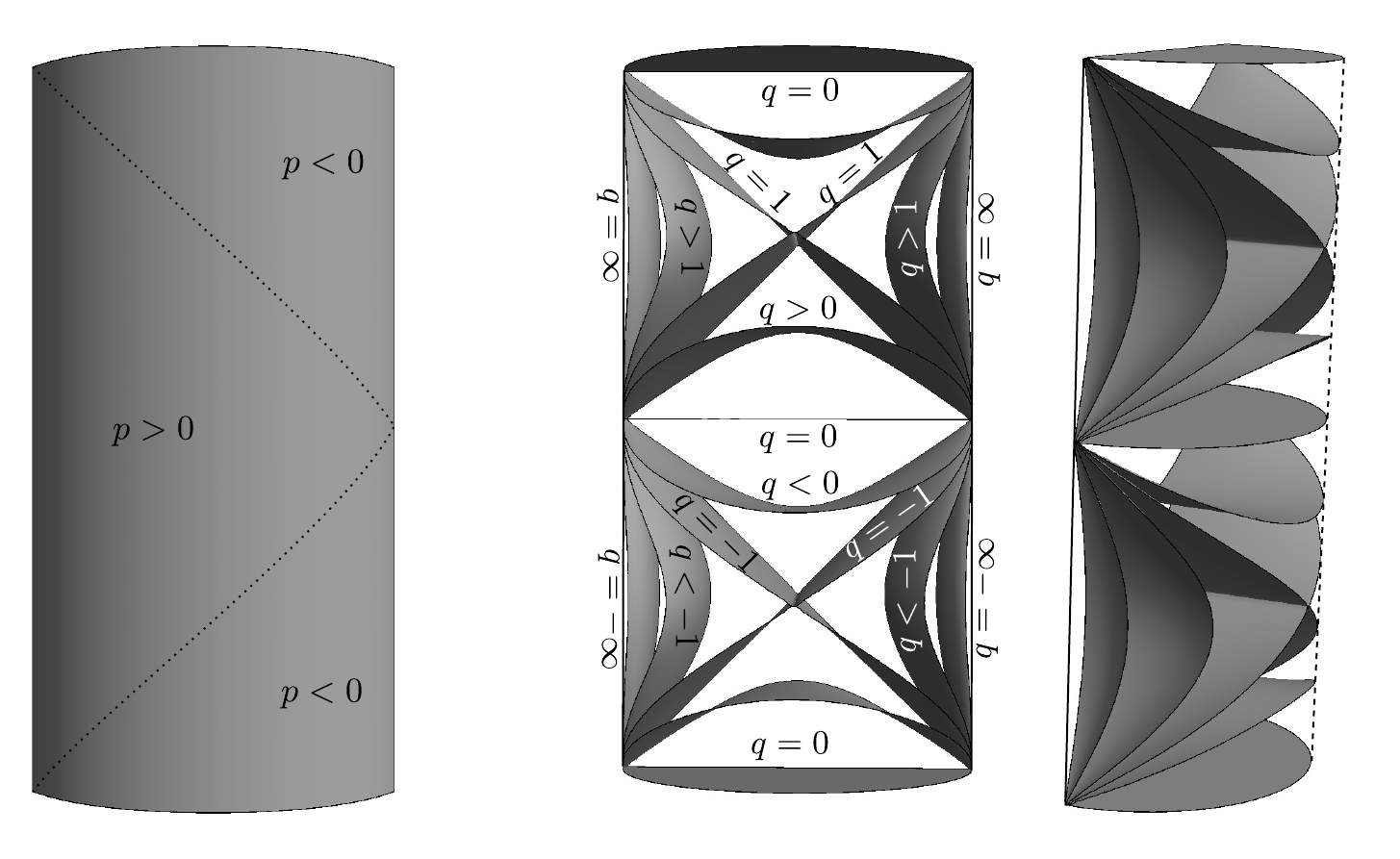}
\caption{\small
Left: A front view on the surfaces ${|p|=\infty}$ representing $\scri$ in the global  representation of the anti-de~Sitter universe for ${\y=\konst}$ The dotted lines separate the region ${p=+\infty}$ from the regions ${p=-\infty}$. Other surfaces ${p=\konst}$ are separated in a similar way into the regions ${p>0}$ and ${p<0}$.
Right: The surfaces ${q=\konst}$ for various values of $q$. The surfaces ${q=0}$ correspond to the temporal horizontal sections ${\sin\eta=0}$, while ${q=\pm\infty}$ degenerate to lines ${\chi=\frac{\pi}{2}}$, ${\theta=0, \pi}$ on $\mathcal{I}$.
The two pictures on the right are a front view and a side view, respectively.}\label{img:ConfadS-1-1q}
\end{center}
\end{figure}

\textbf{Global conformal representation} is obtained by comparing (\ref{eq:adS-1-1apar}), (\ref{eq:adS-1-1bpar}) with (\ref{confADS}):
\begin{eqnarray}\label{eq:adS-1-1cfinv}
\begin{array}{rl}
\cotg\eta&\!\!\!={\displaystyle \sqrt{1-q^{-2}}\,\sinh t} \hspace{5mm} \textrm{for} \  |q|>1\,,\\[4pt]
\cotg\eta&\!\!\!={\displaystyle \sqrt{q^{-2}-1}\,\cosh t} \hspace{5mm} \textrm{for} \ |q|<1\,,\\[10pt]
\tg\chi&\!\!\!= {\displaystyle \sqrt{\frac{p^2}{a^2}\left(q^2\,\cosh^2 t-\sinh^2 t\right)-1}} \hspace{5mm} \textrm{for} \  |q|>1\,,\\[6pt]
\tg\chi&\!\!\!= {\displaystyle \sqrt{\frac{p^2}{a^2}\left(\cosh^2 t-q^2\,\sinh^2 t\right)-1}} \hspace{5mm} \textrm{for}\    |q|<1\,,\\[12pt]
\cotg\theta&\!\!\!={\displaystyle \frac{p\,\sqrt{q^2-1}\,\cosh t}{\sqrt{p^2-a^2}}} \hspace{5mm} \textrm{for} \ |q|>1\,,\\[10pt]
\cotg\theta&\!\!\!={\displaystyle \frac{p\,\sqrt{1-q^2}\,\sinh t}{\sqrt{p^2-a^2}}} \hspace{5.5mm} \textrm{for} \ |q|<1\,,
\end{array}
\end{eqnarray}
or, inversely,
\begin{eqnarray}\label{eq:adS-1-1conf}
\tgh t \rovno  \frac{\cos\eta}{\sin\chi\cos\theta}\quad \textrm{for} \quad |q|>1\,,\,\qquad
\cotgh t =\frac{\cos\eta}{\sin\chi\cos\theta}\quad \textrm{for} \quad |q|<1\,,
\nonumber\\
p \rovno a\,\sign(\cos\theta)\,\sqrt{1+\tan^2\chi\sin^2\theta}\quad \textrm{for} \quad |q|>1,\\
p \rovno a\,\sign(\cos\eta)\,\sqrt{1+\tan^2\chi\sin^2\theta}\quad \textrm{for} \quad |q|<1,\nonumber\\
q \rovno \frac{\sin\eta}{\sqrt{1-\sin^2\chi\cos^2\theta}}\,.\nonumber
\end{eqnarray}
This can be used for plotting the coordinate surfaces of the metric (\ref{eq:adS-1-1ds}) in terms of the standard global conformal representation of the entire anti-de~Sitter spacetime. The surfaces ${p=\konst}$ are the same as in the case ${\ez=1}$ shown in Figure~\ref{img:ConfadS-11pg}, except that now there are two complementary regions ${p>0}$ and ${p<0}$, see the conformal infinity $\scri$ shown in the left part of Figure~\ref{img:ConfadS-1-1q}. Specific character of the surfaces ${q=\konst}$ is shown for two views in the right part of Figure~\ref{img:ConfadS-1-1q}.

\section{Physical application: exact gravitational field of a tachyon moving in the (anti-)de~Sitter spacetime}
\label{sec6}

To illustrate one of possible applications of the new coordinate representation (\ref{eq:PDdads})--(\ref{eq:P}) of the (anti-)de~Sitter spacetime, analyzed in this contribution, let us finally outline a physical interpretation of the family of vacuum $B$-metrics with any value of  the cosmological constant~$\Lambda$. Although this family has been known for a long time \cite{EhlersKundt1962, Stephani:2003, GriPod2009, PleDem1976, GriPod2006b}, it has not yet been systematically studied from the physical point of view.

The $B$\emph{-metrics} with $\Lambda$ can be written in the form (\ref{eq:PDdads}) with the functions $Q(q)$ and $P(p)$ given by (\ref{eq:PDsubcase}), that is
\begin{equation}\label{eq:BLambda}
 \dd s^2=p^2\Big(\!-(\ez-\epsilon_2\,q^2)\,\dd t^2+\frac{\dd q^2}{\ez-\epsilon_2\,q^2}\Big)
   +\frac{\dd p^2}{{\displaystyle\et+\frac{2n}{p}-\frac{\Lambda}{3}\, p^2}}+\Big(\et+\frac{2n}{p}-\frac{\Lambda}{3}\, p^2\Big)\,a^2\dd \y^2\,.
\end{equation}
Notice that $Q(q)$ is exactly the same as in (\ref{eq:Q}), and $P(p)$  immediately reduces to (\ref{eq:P}) when ${n=0}$. Therefore, the specific form (\ref{eq:PDdads})--(\ref{eq:P}) of the (anti-)de~Sitter metric, which we have studied in this work, can be understood as the \emph{natural background for the $B$-metrics  with $\Lambda$}: It is obtained  in the weak-field limit as ${n \to 0}$. In fact, the $B$-metrics are of algebraic type D, with a \emph{curvature singularity located at} ${p=0}$. Setting ${n=0}$, this singularity disappears and (\ref{eq:BLambda}) becomes conformally flat vacuum solution with $\Lambda$, that is the maximally symmetric \hbox{(anti-)}de~Sitter spacetime with $P(p)$ given by (\ref{eq:P}). Various subcases of the (anti-)de Sitter metrics (\ref{eq:PDdads})--(\ref{eq:P}) --- in particular, the character of the coordinates employed ---  are thus fundamental for an understanding of geometrical and physical properties of the whole $B$-metrics family (\ref{eq:BLambda}).

The $B$-metrics can be directly compared to the family of $A$\emph{-metrics}
\begin{equation}\label{eq:ALambda}
 \dd s^2=r^2\Big((\ez-\epsilon_2\,q^2)\,\dd \y^2+\frac{\dd q^2}{\ez-\epsilon_2\,q^2}\Big)
   +\frac{\dd r^2}{{\displaystyle\et+\frac{2n}{r}-\frac{\Lambda}{3}\, r^2}}-\Big(\et+\frac{2n}{r}-\frac{\Lambda}{3}\, r^2\Big)\,\dd t^2\,.
\end{equation}
It can be seen that (\ref{eq:ALambda}) is obtained from (\ref{eq:BLambda}) by \emph{formal substitutions} of the coordinates ${p \rightarrow r}$, ${t \rightarrow {\rm i}\, \y}$,  ${a\, \y \rightarrow {\rm i}\, t}$. In particular, the complex unit ``i'' formally \emph{interchanges the temporal and spatial character of the coordinates $t$ and $\y$}. For three distinct signs of the \emph{Gaussian curvature parameter} ${\et}$ of the \emph{spatial 2-surfaces} on which $r$ and $t$ are constant, there are three  subclasses of the $A$-metrics, namely $AI$ for ${\et=1}$, $AII$ for ${\et=-1}$, and $AIII$ for ${\et=0}$. These include various (topological) black holes which were studied in a number of works,  see Chapter~9 in~\cite{GriPod2009} for a review and number of references. In particular, for the $AI$ subcase ${\et=1=\ez}$ we can introduce an angular coordinate $\theta$ by ${q=\cos\theta}$ and relabel the parameter $n$ as $-m$, obtaining thus the standard form of the \emph{famous Schwarzschild--(anti-)de~Sitter metric}
\begin{equation}\label{eq:SchwAdS}
 \dd s^2=r^2\big( \dd \theta^2+\sin^2\theta\,\dd \y^2\big)
   +\frac{\dd r^2}{{\displaystyle1-\frac{2m}{r}-\frac{\Lambda}{3}\, r^2}}-\Big(1-\frac{2m}{r}-\frac{\Lambda}{3}\, r^2\Big)\,\dd t^2\,.
\end{equation}
It is also of type D, with a curvature singularity at ${r=0}$. This unique spherically symmetric solution was discovered by Kottler, Weyl, and Trefftz \cite{Kottler1918, Trefftz1922, Weyl1919}. Of course, for ${\Lambda=0}$  it is just the first exact solution to Einstien's field equations, found by Karl Schwarzschild \cite{Schwarzschild1916}.

Analogously to $A$-metrics, there exist three distinct subclasses of the $B$-metrics (\ref{eq:BLambda}), denoted as  $BI$ for ${\et=1}$, $BII$ for ${\et=-1}$, and $BIII$ for ${\et=0}$. However, ${\et}$ now determines the \emph{Gaussian curvature} of the \emph{Lorentzian 2-surfaces} on which $p$ and $\y$ are constant.

Applying the results of the present paper, we are now suggesting a physical interpretation of the $BI$-metric defined by ${\et=1}$, which is the direct \emph{counterpart of} the Schwarzschild--(anti-)de~Sitter metric (\ref{eq:SchwAdS}) representing a \emph{static} (black hole) matter source. For the convenient choice ${\ez=-1}$, a simple transformation ${q=\sinh\tau}$, $t=\chi$ puts such metric (\ref{eq:BLambda}) into the form
\begin{equation}\label{eq:BI}
 \dd s^2=p^2\big(\! -\dd \tau^2+\cosh^2\tau\,\dd \chi^2\big)
   +\frac{\dd p^2}{{\displaystyle 1+\frac{2n}{p}-\frac{\Lambda}{3}\, p^2}}+\Big(1+\frac{2n}{p}-\frac{\Lambda}{3}\, p^2\Big)\,a^2\dd\y^2\,.
\end{equation}
In fact, we are extending the idea presented in 1974 in an interesting work \cite{Gott1974} by J.~R.~Gott. In the case ${\Lambda=0}$, he argued that the $BI$-metric represents an \emph{exact gravitational field of a tachyonic source} moving with a superluminal velocity ${v>c\equiv 1}$ along the spatial axis of symmetry. We now claim that the metric (\ref{eq:BI}) generalizes this physical situation to a tachyonic-type singular source \emph{moving along the spatial axis of $\partial_\y$ at ${p=0}$ in the de~Sitter universe with ${\Lambda>0}$} (and similarly in the anti-de~Sitter universe when ${\Lambda<0}$). This is the counterpart to the static source moving along the temporal axis of $\partial_t$ at ${r=0}$ in the Schwarzschild--(anti-)de~Sitter $AI$-metric (\ref{eq:SchwAdS}).
Indeed, in the weak-field limit ${n \to 0}$ we immediately observe that the $BI$-metric (\ref{eq:BI}) for ${p=0}$ reduces to ${\dd s^2= a^2\dd\y^2}$, yielding a positive norm of $\partial_\y$  (contrary to the analogous limit ${m \to 0}$ of the $AI$-metric (\ref{eq:SchwAdS}) which for ${r=0}$ yields ${\dd s^2= -\dd t^2}$, and the norm of $\partial_t$ is thus negative).

Moreover, in the limit ${n \to 0}$ we observe that the metric (\ref{eq:BI}), which is equivalent to (\ref{eq:BLambda}) with ${\et=1}$ and ${\ez=-1}$, reduces to the metric form (\ref{eq:dS1-1ds}) of the  de~Sitter spacetime studied in Subsection~\ref{ch:1-1}. In that part of this work we have analyzed and visualized the character of the coordinates, and the corresponding foliation of the background spacetime, see Figure~\ref{img:dS1-1Z0Z1a} and Figures~\ref{img:ConfdS11p} and~\ref{img:ConfdS1-1q}.

\begin{figure}[h!]
\begin{center}
\includegraphics[scale=1]{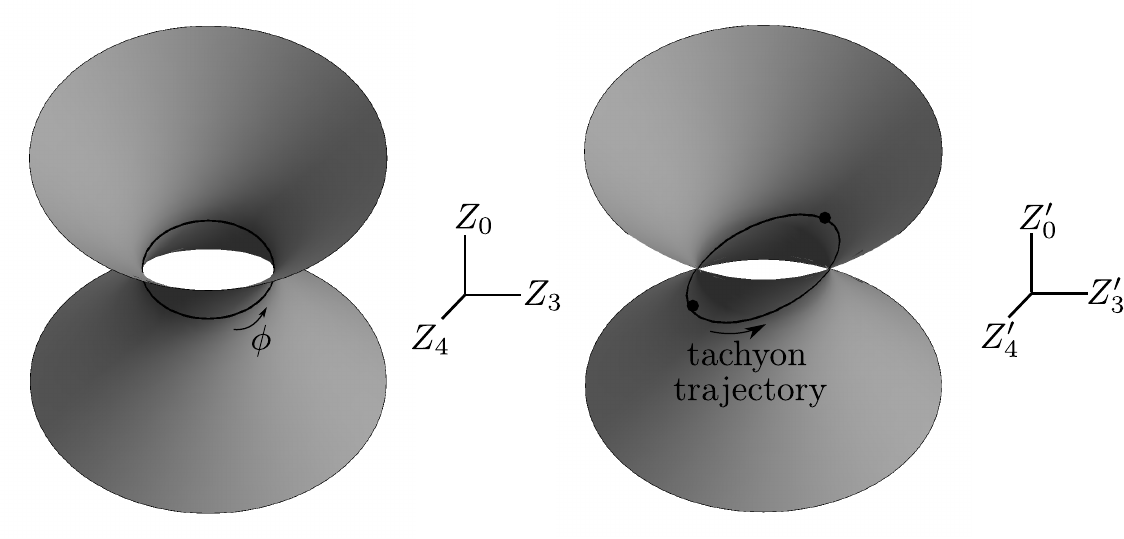}
\caption{\small
Left: Special tachyon trajectory in the de~Sitter spacetime (trivial coordinates $Z_1$, $Z_2$ are suppressed). The tachyon moves with an \emph{infinite velocity} around the ``neck'' of the de~Sitter hyperboloid located at ${Z_0}$.
Right: Trajectory of the tachyon moving with a \emph{finite} superluminal velocity, shown in the boosted frame ${Z_0', Z_3', Z_4'}$. This picture can be understood in two ways: For ${Z_4'>0}$ the tachyon moves up the de~Sitter hyperboloid, while on its opposite side ${Z_4'<0}$ it returns back in time to the initial point, making thus a \emph{closed loop in time} ${Z_0'}$. An alternative interpretation of the same picture is that the tachyon moves from the initial (lowest) point to the final (upmost) point on \emph{both sides} of the hyperboloid. This is possible since in the Lorentzian geometry (as opposed to the Riemannian one), the geodesics between two events are not unique.}\label{img:dS_Z0Z3Z4}
\end{center}
\end{figure}

In particular, we immediately obtain from the specific parametrization (\ref{eq:dS1-1par}) of the de~Sitter hyperboloid (\ref{C1}) in (\ref{C2}) by these coordinates that the source at ${p=0}$ is, in fact, located at
\begin{equation}
\begin{array}{l}
 Z_0 = 0 \,, \qquad
 Z_1 = 0 = Z_2\,, \\
 Z_3 = a\,\cos\,\y\,, \\
 Z_4 = a\,\sin\,\y\,.
 \end{array}
\label{eq:dS1-1parp=0}
\end{equation}
In other words, the \emph{trajectory of the source is given by}
\begin{equation}
 Z_3^2+Z_4^2 = a^2\,,
\label{eq:tachyontrajecotry}
\end{equation}
with ${Z_0=Z_1=Z_2= 0}$. This fully confirms that the trajectory of the source of the $BI$-metric is indeed spacelike, and thus ``tachyonic''. In fact, it is exactly the  \emph{closed spatial loop around the ``neck'' of the de~Sitter hyperboloid} at ${Z_0}$, see the left part of Figure~\ref{img:dS_Z0Z3Z4}. Geometrically, it is one of the main circles of the spatial 3-sphere. Notice also that in the global conformal representation (\ref{eq:dS1-1conf}) this geodesic corresponds to a special curve given by ${\eta = \frac{\pi}{2}}$, ${\chi = \frac{\pi}{2}}$, ${\theta = \frac{\pi}{2}}$ with the angular coordinate taking the range ${\y\in[0,2\pi)}$, i.e., it is a closed circle around ``the middle'' of the conformal 4D representation of the whole universe.

Since this source of the field (\ref{eq:BI}) moves with a \emph{superluminal velocity}, it generates gravitational \emph{Cherenkov shock cone}. Its location is obtained by substituting (\ref{eq:tachyontrajecotry}) into (\ref{C1}), i.e.,
\begin{equation}
Z_1^2+Z_2^2=Z_0^2 \,.
\label{CerenkovCone}
\end{equation}
In fact, there are \emph{two} Cherenkov cones: the first given by ${\sqrt{Z_1^2+Z_2^2}=Z_0}$ (rear one) is \emph{expanding}, while the second given by ${\sqrt{Z_1^2+Z_2^2}=-Z_0}$ (front one) is \emph{contracting}. The tachyonic source is located at their common vertex. The background de~Sitter spacetime is divided by these Cherenkov cones into three distinct regions, namely \emph{region~1} given by ${Z_0>\sqrt{Z_1^2+Z_2^2}}$, \emph{region~2} given by ${Z_0<-\sqrt{Z_1^2+Z_2^2}}$, and \emph{region~3} defined by ${Z_0^2<Z_1^2+Z_2^2}$. As in the case ${\Lambda=0}$ studied by Gott in \cite{Gott1974}, detailed analysis shows that for ${n\not=0}$ the curved region~3 outside the Cherenkov cones is covered by the $BI$-metric (\ref{eq:BI}), while each of the regions~1 and~2 is covered by the $AII$-metric (\ref{eq:ALambda}).

\begin{figure}[h]
\begin{center}
\includegraphics[scale=1]{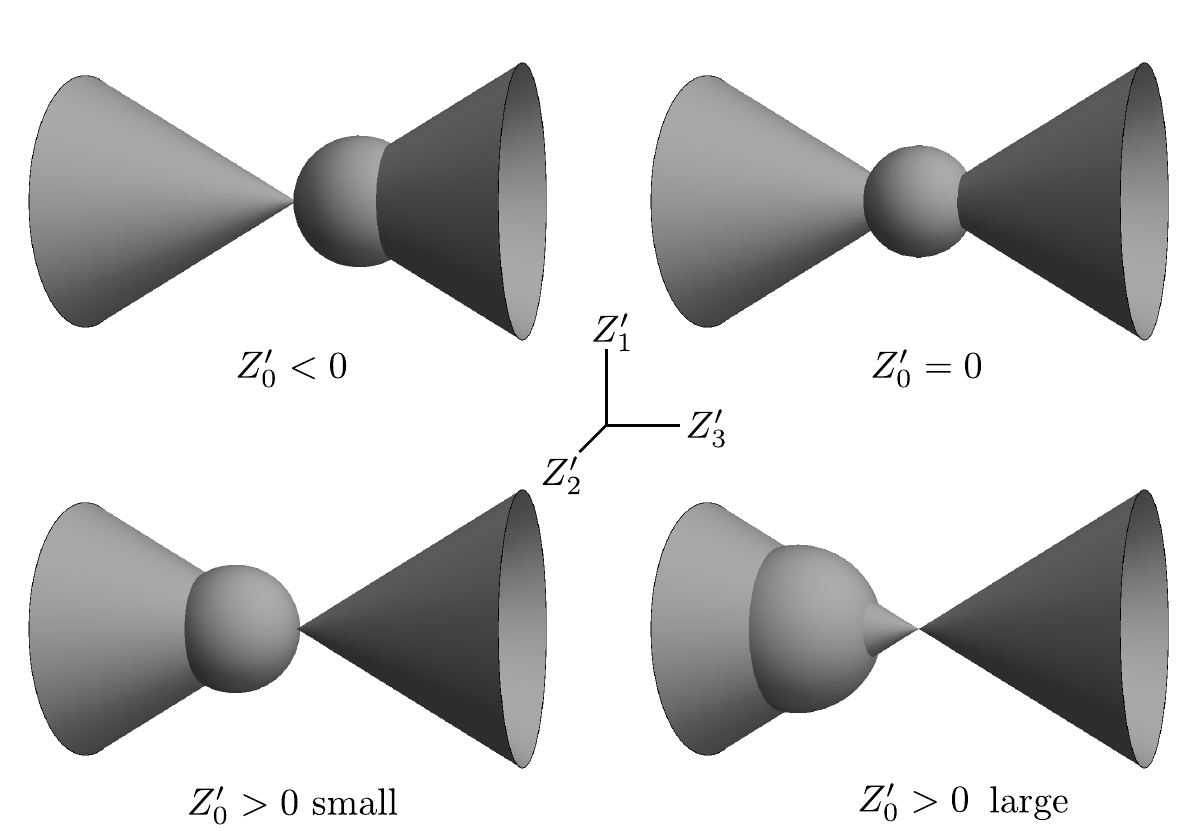}
\caption{\small
Visualization of the Cherenkov shock waves caused by the tachyon moving in the de~Sitter spacetime (represented as a sphere in this section ${Z_1', Z_2', Z_3'}$ ), for four different times ${Z_0'=\,}$const. (with a fixed ${Z_4'<a}$). The tachyon is at the joint vertex of the two parts of the Cherenkov cone. The intersections of these cones with the sphere gives the actual position of the two Cherenkov shocks in the de~Sitter universe  (the rear one on the left is expanding, while the front one on the right is contracting). Top left: At this special time ${Z_0'}$ the tachyon is in the North Pole of the space, creating the expanding shock wave, while there is already  the contracting shock on the southern hemisphere. Top right: At ${Z_0'=0}$ the closed de~Sitter universe has a minimal radius $a$, both shock waves are of the same size, and are located symmetrically with respect to the equator. Bottom left: A complementary situation to the top left part --- the tachyon is in the South Pole, and the contracting shock has just shrinked to zero. Bottom right: Later, after the contracting shock has crossed the South Pole, its starts to expand. There are thus two expanding shock waves in the de~Sitter universe, both approaching the equator.}\label{img:Tach_motion_dS}
\end{center}
\end{figure}

The source of the gravitational field (\ref{eq:BI}) moves along (\ref{eq:dS1-1parp=0}) with an \emph{infinite velocity} ${v=\infty}$.  Its Cherenkov cone (\ref{CerenkovCone}) at ${Z_0=0}$ thus \emph{degenerates} to the single circular axis along which the tachyon moves. To obtain a nondegenerate Cherenkov cone, it is necessary to ``slow down'' the tachyon to a \emph{finite velocity} $v$ such that ${v>1}$. This is achieved by performing an appropriate \emph{Lorentz boost} in the representation of the de~Sitter spacetime as the hyperboloid (\ref{C1}) embedded in the five-dimensional Minkowski space (\ref{C2}), for example in the $Z_3$-direction,
\begin{equation}
\begin{array}{l}
 Z_0' = v(v^2-1)^{-\frac{1}{2}}\,Z_0+(v^2-1)^{-\frac{1}{2}}\,Z_3 \,, \\[1pt]
 Z_3' = (v^2-1)^{-\frac{1}{2}}\,Z_0+v(v^2-1)^{-\frac{1}{2}}\,Z_3 \,, \\[1pt]
 Z_1' = Z_1\,,\quad Z_2' = Z_2\,,\quad Z_4' = Z_4 \,.
 \end{array}
\label{eq:CerenkovCone2}
\end{equation}
With respect to the new frame, the tachyonic source (\ref{eq:dS1-1parp=0}) localized along ${Z_0 = 0}$ now moves along ${Z_3'/Z_0' = v}$. This is illustrated in the right part of Figure~\ref{img:dS_Z0Z3Z4}. The corresponding Cherenkov cone (\ref{CerenkovCone}) then takes the form
\begin{equation}
Z_1'^2+Z_2'^2=\frac{1}{v^2-1}\left(Z_3'-v\,Z_0' \right)^2\,,
\label{CerenkovConeboost}
\end{equation}
with its source (vertex) given by ${Z_3'=vZ_0' }$, i.e., moving with the velocity $v$ along the axis~$Z_3'$. Both parts of this Cherenkov shock (the expanding part of the cone and the contracting one) are visualized in Figure~\ref{img:Tach_motion_dS} for four typical values of the fixed global time ${Z_0'=}$const., together with the spatial representation of the de~Sitter spacetime as a sphere. As explained in Subsection~\ref{sec2a}, the spatial geometry of the de~Sitter spacetime is a 3-sphere $S^3$ which contracts to a minimal size $a$ at ${Z_0'=0}$ and then re-expands. At a given time $Z_0'$, the position of  both the Cherenkov shocks are two intersections of the cone with this sphere. In the full de~Sitter space $S^3$, the Cherenkov shocks are two 2-spheres (one of which is expanding while the other is contracting) but since in Figure~\ref{img:Tach_motion_dS} we have suppressed one angular coordinate, the de~Sitter space is represented just by a 2-sphere while the Cherenkov shocks are circles.
\clearpage

\clearpage

\section{Summary and conclusions}

In our contribution we have presented 3~interesting forms (\ref{eq:PDdads})--(\ref{eq:P}) of the de~Sitter spacetime and 8~forms of the anti-de~Sitter spacetime in four dimensions, according to the values of two discrete parameters ${\ez, \et = 1,0,-1}$, see Tables~\ref{tbl:dS} and \ref{tbl:adS}. As far as we know, these have not yet been explicitly identified, described and analyzed in (otherwise vast) literature on these fundamental solutions to Einstein's field equations with a cosmological constant $\Lambda$.

Such new forms of maximally symmetric vacuum spacetimes with $\Lambda$ naturally arise in the context of $B$-metrics with a cosmological constant, when they are expressed in a unified Pleba\'nski--Demia\'nski form of the Kundt type D  solutions. Quite surprisingly, all the metric forms (\ref{eq:PDdads})--(\ref{eq:P}) of the (anti-)de~Sitter spacetime are very simple: they are diagonal, represent a warped product of two 2-dimensional geometries of constant curvature, and many of them explicitly exhibit stationarity and axial symmetry. The coordinates of (\ref{eq:PDdads}) are naturally adapted to a ${2+2}$ foliation of the spacetime, in contrast to the coordinates of standard well-known forms, which correspond to a ${1+3}$ foliation with the 3-space of a constant spatial curvature.

In each subcase, determined by ${\ez, \et = 1,0,-1}$, we found the corresponding parametrizations of the de Sitter  or anti-de Sitter hyperboloid, and then we plotted the coordinate lines and surfaces in the global conformal cylinder. This clearly demonstrated specific character of the new coordinates. In the appendices we also present mutual relations between various subcases, some other related forms of the metrics, and also transformations to standard coordinates of the de~Sitter and anti-de~Sitter spacetimes.

Our initial motivation was to investigate geometrical and physical properties of the $B$-metrics with a cosmological constant, which can be understood as type D extensions of the conformally flat (anti-)de~Sitter metric (\ref{eq:PDdads}). In the final Section~\ref{sec6} we outlined such a study. Using the results given in the main part of this paper, we demonstrated that the family of $B$-metrics can be physically interpreted as representing exact gravitational field of a tachyon moving in the de~Sitter or anti-de~Sitter spacetime. A detailed analysis will be performed in our future work, as we think that the new coordinate representations of the de~Sitter and anti-de~Sitter spacetime presented in the current contribution can be interesting by itself. We hope that they could also be employed for various other investigations of these fundamental spacetimes, discovered 100 years ago.

\section*{Acknowledgements}
We are grateful to Robert~\v{S}varc for reading the manuscript and some useful suggestions.
This work has been supported by the grant GA\v{C}R 17-01625S. O.H. also acknowledges the support by the Charles University Grant GAUK~196516.

\clearpage

\section{Appendix A: Unified form of all parametrizations}\label{sc:dSadS}
All new parametrizations of the de~Sitter or anti-de~Sitter hyperboloid, presented and studied in  previous Sections~\ref{sc:dS} and~\ref{sc:adS}, can be written in a  unified way as:
\begin{eqnarray}\label{eq:allpar}
Z_{a} \rovno p\,\sqrt{|Q|}\left(1+S\right)T_{1}\,,\nonumber\\
Z_{b} \rovno p\,\sqrt{|Q|}\left(1-S\right)T_{2}\,,\nonumber\\
Z_{c} \rovno |p|\,q\,t^{1-|\epsilon_{0}|}\,,\\
Z_{d} \rovno a\,\sqrt{P}\,Y_{1}\,,\nonumber\\
Z_{e} \rovno a\,\sqrt{P}\,Y_{2}\,,\nonumber
\end{eqnarray}
where ${Q=\ez-\epsilon_{2}\,q^{2}}$, ${P =\et-\frac{\Lambda}{3}\, p^{2}}$, as in (\ref{eq:Q}), (\ref{eq:P}). The indices $a,b,c,d,e$ and also the functions $S,T_{1},T_{2},Y_{1},Y_{2}$ for various $\epsilon_{2}$ and $\epsilon_{0}$ are listed in Tables~\ref{tbl:alldS} and~\ref{tbl:alladS} for ${\Lambda>0}$ or ${\Lambda<0}$, respectively. The last column refers to the corresponding equation in the main text.

\begin{table}[h]
\begin{center}
\begin{tabular}{| c | c | c | c | c | c | c | c | c | c | c | c | c|}
\hline
$\epsilon_{2}$ & $\epsilon_{0}$ & $T_{1}$ & $T_{2}$ & $S$ & $Y_{1}$ & $Y_{2}$ & $a$ & $b$ & $c$ & $d$ & $e$ & Eq. \\
\hline
1 & 1 & $\sinh t$ & $\cosh t$ & 0 & $\cos\y$ & $\sin\y$ & 0 & 1 & 2 & 3 & 4 & (\ref{eq:dS11para})\\
\hline
1 & 1 & $\cosh t$ & $\sinh t$ & 0 & $\cos\y$ & $\sin\y$ & 0 & 1 & 2 & 3 & 4 & (\ref{eq:dS11bpar})\\
\hline
1 & 0 & $\frac{1}{2}\sign q$ & $\frac{1}{2}\sign q$ & $t^{2}-q^{-2}$ & $\cos\y$ & $\sin\y$ & 0 & 1 & 2 & 3 & 4 & (\ref{eq:dS10par})\\
\hline
1 & $-1$ & $\sin t$ & $\cos t$ & 0 & $\cos\y$ & $\sin\y$ & 2 & 1 & 0 & 3 & 4 & (\ref{eq:dS1-1par})\\
\hline
\end{tabular}
\caption{\small
Specific functions used in the unified parametrization (\ref{eq:allpar}) for ${\Lambda>0}$.}
\label{tbl:alldS}
\end{center}
\end{table}

\begin{table}[h]
\begin{center}
\begin{tabular}{| c | c | c | c | c | c | c | c | c | c | c | c | c|}
\hline
$\epsilon_{2}$ & $\epsilon_{0}$ & $T_{1}$ & $T_{2}$ & $S$ & $Y_{1}$ & $Y_{2}$ & $a$ & $b$ & $c$ & $d$ & $e$ & Eq. \\
\hline
1 & 1 & $\sinh t$ & $\cosh t$ & 0 & $\sinh\yy$ & $\cosh\yy$ & 0 & 1 & 2 & 3 & 4 & (\ref{eq:adS11apar})\\
\hline
1 & 1 & $\cosh t$ & $\sinh t$ & 0 & $\sinh\yy$ & $\cosh\yy$ & 0 & 1 & 2 & 3 & 4 & (\ref{eq:adS11bpar})\\
\hline
1 & 0 & $\frac{1}{2}\sign q$ & $\frac{1}{2}\sign q$ & $t^{2}-q^{-2}$ & $\sinh\yy$ & $\cosh\yy$ & 0 & 1 & 2 & 3 & 4 & (\ref{eq:adS10par})\\
\hline
1 & $-1$ & $\sin t$ & $\cos t$ & 0 & $\sinh\yy$ & $\cosh\yy$ & 2 & 1 & 0 & 3 & 4 & (\ref{eq:adS1-1par})\\
\hline
0 & 1 & $\frac{1}{2}$ & $\frac{1}{2}$ & $s/a^2$ & $\yy$ & $t$ & 0 & 1 & 2 & 3 & 4 & (\ref{eq:adS01par})\\
\hline
$-1$ & 1 & $\cos t$ & $\sin t$ & 0 & $\sin\y$ & $\cos\y$ & 0 & 4 & 1 & 3 & 2 & (\ref{eq:adS-11par})\\
\hline
$-1$ & 0 & $\frac{1}{2}\sign q $ & $\frac{1}{2}\sign q $ & $q^{-2}-t^{2}$ & $\sin\y$ & $\cos\y$ & 0 & 1 & 4 & 3 & 2 & (\ref{eq:adS-10par})\\
\hline
$-1$ & $-1$ & $\sinh t$ & $\cosh t$ & 0 & $\sin\y$ & $\cos\y$ & 0 & 1 & 4 & 3 & 2 & (\ref{eq:adS-1-1apar})\\
\hline
$-1$ & $-1$ & $\cosh t$ & $\sinh t$ & 0 & $\sin\y$ & $\cos\y$ & 0 & 1 & 4 & 3 & 2 & (\ref{eq:adS-1-1bpar})\\
\hline
\end{tabular}
\caption{\small
Specific functions used in the unified parametrizations (\ref{eq:allpar}) for ${\Lambda<0}$. Here ${s=a^2\left(-t^2+q^2+\yy^2+a^2/p^2\right)}$. For ${\et=1}$ we also consider the second chart with $-Y_1$ and $-Y_2$ in order to obtain ${Z_e<0}$. For ${\et=0}$ we also consider $-Z_2$, $-Z_3$ and $-Z_4$.}
\label{tbl:alladS}
\end{center}
\end{table}

\noindent
The coordinates $(p,q,t,\y)$ only cover \emph{the part} ${Z_3,Z_4\in(-a,a)}$ of de~Sitter spacetime, while they cover \emph{the whole} anti-de~Sitter spacetime (except in the cases ${\et=1}$ for which  ${Z_4\in\mathbb{R}\setminus(-a,a)}$).

\section{Appendix B: Mutual relations}
\label{apC}

\noindent
\textbf{Case ${\Lambda>0}$:}
The de~Sitter metric (\ref{eq:dS11ds}) for ${\et=1}$, ${\ez=1}$  (in which the coordinates ${p, q, t}$ are relabeled to ${\tilde{p}, \tilde{q}, \tilde{t}}$) can be explicitly transformed to the metric (\ref{eq:dS10ds}) for ${\et=1}$, ${\ez=0}$ by applying
\begin{equation}\label{relDS1}
|\tilde{p}| = p \,,\qquad
\tilde{q} = q\,t\,,\qquad
\exp \tilde{t} = \frac{1}{\sqrt{|t^2-q^{-2}|}} \,.
\end{equation}
Similarly, the metric (\ref{eq:dS11ds}) for ${|q|<1}$ can be explicitly transformed to the metric (\ref{eq:dS1-1ds}) for ${\et=1}$, ${\ez=-1}$ by applying
\begin{equation}\label{relDS2}
|\tilde{p}| = p \,,\qquad
\tilde{q} = \sqrt{1+q^{2}}\,\sin t\,,\qquad
\cotgh\tilde{t} = \frac{\sqrt{1+q^{2}}}{q}\,\cos t \,.
\end{equation}
These relations are easily obtained by comparing (\ref{eq:dS11para}) to (\ref{eq:dS10par}), and (\ref{eq:dS11para}) to (\ref{eq:dS1-1par}), respectively.
\vspace{2mm}

\noindent
\textbf{Case ${\Lambda<0}$:}
Interestingly, the same transformations (\ref{relDS1}) and (\ref{relDS2}) relate the anti-de~Sitter metric (\ref{eq:adS11ds}) for ${\et=1}$, ${\ez=1}$  (with the coordinates relabeled as ${\tilde{p}, \tilde{q}, \tilde{t}}$) to (\ref{eq:adS10ds}) for ${\et=1}$, ${\ez=0}$ and (\ref{eq:adS1-1ds}) for ${\et=1}$, ${\ez=-1}$, respectively.

Comparing the parameterizations (\ref{eq:adS10par}) and (\ref{eq:adS01par}) we find a transformation between the anti-de~Sitter metric (\ref{eq:adS10ds}) for ${\et=1}$, ${\ez=0}$  and the form (\ref{eq:adS01ds}) for ${\et=0}$, ${\ez=1}$ (relabeled to ${\tilde{p}, \tilde{q}, \tilde{t}, \tilde{\yy}}$):
\begin{equation}\label{relADS1}
\tilde{p} = p\,q \,,\qquad
\tilde{q} = t\,,\qquad
\tilde{t}  = \frac{\sqrt{a^2+p^2}}{p\,q}\,\cosh\yy \,,\qquad
\tilde{\yy} = \frac{\sqrt{a^2+p^2}}{p\,q}\,\sinh\yy \,.
\end{equation}

Also, the anti-de~Sitter metric (\ref{eq:adS-11ds}) for ${\et=-1}$, ${\ez=1}$  (relabeled to ${\tilde{p}, \tilde{q}, \tilde{t}}$) is  transformed to the metric (\ref{eq:adS-10ds}) for ${\et=-1}$, ${\ez=0}$ by
\begin{equation}\label{relADS2}
\tilde{p} = p \,,\qquad
\tilde{q} = {\textstyle\frac{1}{2}}\, q\,(1+t^2-q^{-2})\,,\qquad
\tan \tilde{t} = \frac{2t}{1-t^2+q^{-2}} \,,
\end{equation}
while (\ref{eq:adS-11ds}) for ${\et=-1}$, ${\ez=1}$ is  transformed to the metric (\ref{eq:adS-1-1ds}) for ${\et=-1}$, ${\ez=-1}$ by
\begin{eqnarray}\label{relADS3}
&&\tilde{p} = |p| \,,\qquad
\tilde{q} = \sqrt{1-q^2}\,\sinh t\,,\qquad
\tan \tilde{t} = \frac{q\,\sign p}{\sqrt{1-q^2}\,\cosh t} \hspace{5mm} \textrm{for}\ |q|<1 \,, \nonumber\\
&&\tilde{p} = |p| \,,\qquad
\tilde{q} = \sqrt{q^2-1}\,\cosh t\,,\qquad
\tan \tilde{t} = \frac{q\,\sign p}{\sqrt{q^2-1}\,\sinh t} \hspace{5mm} \textrm{for}\ |q|>1 \,.
\end{eqnarray}

\section{Appendix C: Further related metric forms of the de~Sitter and anti-de~Sitter spacetimes}

The new metrics analyzed in this work can also be expressed in closely related forms. In these, both (warped) parts representing the constant-curvature 2-spaces are more clearly seen because they are all put into the ``canonical'' forms using triginometric and/or hyperbolic functions. In this appendix we describe all such related metrics. For each subcase, we present the transformation, the resulting metric, and the corresponding paramaterization of the 5D hyperboloid:

\subsection{The de~Sitter spacetime (${\Lambda>0}$)}

\noindent
\textbf{Subcase} ${\epsilon_{2}=1}$, ${\epsilon_{0}=1}$:
\begin{eqnarray}\label{eq:dS11atransf}
&&
p = a\,\cos \p\,,\qquad
q = \sin\q\,,\qquad
t = \frac{T}{a}\,,\hspace{5mm} \textrm{for}\ |q|<1\,,  \nonumber\\
&&
p = a\,\cos \p\,,\qquad
|q| = \cosh\frac{T}{a}\,,\qquad
t = \q\,,\hspace{5mm} \textrm{for}\ |q|>1\,,
\end{eqnarray}
transforms (\ref{eq:dS11ds}) to
\begin{eqnarray}\label{eq:dS11ads}
&&
\dd s^{2}=\cos^{2}\p\,\Big(-\cos^{2}\q\, \dd T^{2}+a^{2}\,\dd \q^{2}\Big) +a^{2}\left(\dd \p^{2} + \sin^{2}\p\,\dd \y^{2}\right), \nonumber\\
&&
\dd s^{2}=\cos^{2}\p\,\Big(-\dd T^{2}+a^{2}\sinh^{2}\frac{T}{a}\,\dd \q^{2}\Big) +a^{2}\left(\dd \p^{2} + \sin^{2}\p\,\dd \y^{2}\right),
\end{eqnarray}
respectively, corresponding to the parametrizations
 \begin{equation}
\left. \begin{array}{l}
Z_0 = {\displaystyle a\cos \p \cos\q\sinh\frac{T}{a}}\,, \\[8pt]
Z_1 = {\displaystyle a\cos \p \cos\q \cosh\frac{T}{a}}\,, \\[8pt]
Z_2 = {\displaystyle a|\cos \p| \sin\q}\,, \\[8pt]
Z_3 = {\displaystyle a\sin \p \cos\y}\,, \\[8pt]
Z_4 = {\displaystyle a\sin \p \sin\y}\,,
 \end{array} \!\right\} \hspace{0mm} \textrm{for}\ |q|<1\,, \ \left. \!
 \begin{array}{l}
Z_0 = {\displaystyle a\cos \p \sinh\frac{T}{a}\cosh \q}\,, \\[8pt]
Z_1 = {\displaystyle  a\cos \p \sinh\frac{T}{a}\sinh \q}\,, \\[6pt]
Z_2 = {\displaystyle \pm a|\cos \p| \cosh\frac{T}{a}} \\[2pt]
Z_3 = {\displaystyle a\sin \p \cos\y}\,, \\[6pt]
Z_4 = {\displaystyle a\sin \p \sin\y}\,,
 \end{array} \!\right\} \hspace{0mm} \textrm{for}\ |q|>1\,.
 \label{eq:dS11apar}
 \end{equation}

\noindent
\textbf{Subcase} ${\epsilon_{2}=1}$, ${\epsilon_{0}=0}$:
\begin{eqnarray}\label{eq:dS10atransf}
&&
p = a\,\cos \p\,,\qquad
|q| = \exp\frac{T}{a}\,,\qquad
t = \q \,,
\end{eqnarray}
transforms (\ref{eq:dS10ds}) to
\begin{equation}\label{eq:dS10ds1}
\dd s^{2}=\cos^{2}\p\,\Big(-\dd T^{2}+a^2\exp^2\frac{T}{a}\,\dd\q^{2}\Big) +a^{2}\left(\dd \p^{2} + \sin^{2}\p\,\dd \y^{2}\right),
\end{equation}
corresponding to
\begin{eqnarray}
Z_{0} \rovno \pm\frac{1}{2a}\,\cos \p\,\exp\frac{T}{a}\,(a^{2}+s)\,,\nonumber\\
Z_{1} \rovno \pm\frac{1}{2a}\,\cos \p\,\exp\frac{T}{a}\,(a^{2}-s)\,,\\
Z_{2} \rovno \pm\,a\,\q\,\cos \p\,\exp\frac{T}{a}\,,\nonumber\\
Z_{3} \rovno a\,\sin \p\cos\y\,,\nonumber\\
Z_{4} \rovno a\,\sin \p\sin\y\,, \hspace{10mm} s=a^2\big(\q^2-\exp^{-2}(T/a)\big)\,.\nonumber
\end{eqnarray}

\noindent
\textbf{Subcase} ${\epsilon_{2}=1}$, ${\epsilon_{0}=-1}$:
\begin{eqnarray}\label{eq:dS1-1atransf}
&&
p = a\,\cos \p\,,\qquad
q = \sinh\frac{T}{a}\,,\qquad
t = \q\,,
\end{eqnarray}
transforms (\ref{eq:dS1-1ds}) to
\begin{equation}\label{eq:dS1-1ds1}
\dd s^{2}=\cos^{2}\p\,\Big(-\dd T^{2}+a^{2}\cosh^{2}\frac{T}{a}\,\dd \q^{2}\Big) +a^{2}\left(\dd \p^{2} + \sin^{2}\p\,\dd \y^{2}\right),
\end{equation}
corresponding to
\begin{eqnarray}\label{eq:dS1-1par1}
Z_{0} \rovno a\cos \p \sinh\frac{T}{a}\,,\nonumber\\
Z_{1} \rovno a\cos \p \cosh\frac{T}{a}\cos \q\,,\nonumber\\
Z_{2} \rovno a\cos \p \cosh\frac{T}{a}\sin \q\,,\\
Z_{3} \rovno a\sin \p \cos\y,\nonumber\\
Z_{4} \rovno a\sin \p \sin\y\,.\nonumber
\end{eqnarray}

\subsection{The anti-de~Sitter spacetime  (${\Lambda<0}$)}

\noindent
\textbf{Subcase} ${\epsilon_{2}=1}$, ${\epsilon_{0}=1}$:
\begin{eqnarray}\label{eq:adS11atransf}
&&
p = a\,\sinh \p\,,\qquad
q = \sin\q\,,\qquad
t = \frac{T}{a}\,,\hspace{5mm} \textrm{for}\ |q|<1\,,  \nonumber\\
&&
p = a\,\sinh \p\,,\qquad
|q| = \cosh\frac{T}{a}\,,\qquad
t = \q\,,\hspace{5mm} \textrm{for}\ |q|>1\,,
\end{eqnarray}
transforms (\ref{eq:adS11ds}) to
\begin{eqnarray}\label{eq:adS11ads}
&&
\dd s^{2}=\sinh^{2}\p\,\Big(-\cos^{2}\q\, \dd T^{2}+a^{2}\,\dd \q^{2}\Big) +a^{2}\left(\dd \p^{2} + \cosh^{2}\p\,\dd \yy^{2}\right), \nonumber\\
&&
\dd s^{2}=\sinh^{2}\p\,\Big(-\dd T^{2}+a^{2}\sinh^{2}\frac{T}{a}\,\dd \q^{2}\Big) +a^{2}\left(\dd \p^{2} + \cosh^{2}\p\,\dd \yy^{2}\right),
\end{eqnarray}
respectively, corresponding to the parametrizations
 \begin{equation}
\left. \begin{array}{l}
Z_0 = {\displaystyle a\sinh \p \cos\q \sinh\frac{T}{a}}\,, \\[8pt]
Z_1 = {\displaystyle a\sinh \p \cos\q \cosh\frac{T}{a}}\,, \\[8pt]
Z_2 = {\displaystyle a\sinh |\p| \sin\q}\,, \\[8pt]
Z_3 = {\displaystyle \pm a\cosh \p \sinh\yy}\,, \\[8pt]
Z_4 = {\displaystyle \pm a\cosh \p \cosh\yy}\,,
 \end{array} \!\right\} \hspace{0mm} \textrm{for}\ |q|<1\,, \ \left. \!
 \begin{array}{l}
Z_0 = {\displaystyle a\sinh \p \sinh\frac{T}{a} \cosh\q}\,, \\[8pt]
Z_1 = {\displaystyle a\sinh \p \sinh\frac{T}{a} \sinh\q}\,, \\[6pt]
Z_2 = {\displaystyle \pm a\sinh |\p| \cosh\frac{T}{a}} \\[2pt]
Z_3 = {\displaystyle \pm a\cosh \p \sinh\yy}\,, \\[6pt]
Z_4 = {\displaystyle \pm a\cosh \p \cosh\yy}\,,
 \end{array} \!\right\} \hspace{0mm} \textrm{for}\ |q|>1\,.
 \label{eq:adS11aparpl}
 \end{equation}

\noindent
\textbf{Subcase} ${\epsilon_{2}=1}$, ${\epsilon_{0}=0}$:
\begin{eqnarray}\label{eq:adS10atransf}
&&
p = a\,\sinh \p\,,\qquad
|q| = \exp\frac{T}{a}\,,\qquad
t = \q \,,
\end{eqnarray}
transforms (\ref{eq:adS10ds}) to
\begin{equation}\label{eq:adS10ds1}
\dd s^{2}=\sinh^{2}\p\,\Big(-\dd T^{2}+a^2\exp^2\frac{T}{a}\,\dd \q^{2}\Big) +a^{2}\left(\dd \p^{2} + \cosh^{2}\p\,\dd \yy^{2}\right),
\end{equation}
corresponding to
\begin{eqnarray}\label{eq:adS10par1}
Z_{0} \rovno \pm\frac{1}{2a}\sinh \p\,\exp\frac{T}{a}\left(a^{2}+s\right)\,,\nonumber\\
Z_{1} \rovno \pm\frac{1}{2a}\sinh \p\,\exp\frac{T}{a}\left(a^{2}-s\right)\,,\nonumber\\
Z_{2} \rovno \pm a\,\q\,\sinh \p\,\exp\left(\frac{T}{a}\right)\,,\\
Z_{3} \rovno \pm a\,\cosh \p\sinh\yy\,,\nonumber\\
Z_{4} \rovno \pm a\,\cosh \p\cosh\yy\,,  \hspace{10mm} s=a^2\big(\q^2-\exp^{-2}(T/a)\big)\,.\nonumber
\end{eqnarray}

\noindent
\textbf{Subcase} ${\epsilon_{2}=1}$, ${\epsilon_{0}=-1}$:
\begin{eqnarray}\label{eq:adS1-1atransf}
&&
p = a\,\sinh \p\,,\qquad
q = \sinh\frac{T}{a}\,,\qquad
t = \q\,,
\end{eqnarray}
transforms (\ref{eq:adS1-1ds}) to
\begin{equation}\label{eq:adS1-1ds1}
\dd s^{2}=\sinh^{2}\p\,\Big(-\dd T^{2}+a^{2}\cosh^{2}\frac{T}{a}\,\dd \q^{2}\Big) +a^{2}\left(\dd \p^{2} + \cosh^{2}\p\,\dd \yy^{2}\right),
\end{equation}
corresponding to
\begin{eqnarray}\label{eq:adS1-1par1}
Z_{0} \rovno a\sinh \p \sinh\frac{T}{a}\,,\nonumber\\
Z_{1} \rovno a\sinh \p \cosh\frac{T}{a} \cos\q\,,\nonumber\\
Z_{2} \rovno a\sinh \p \cosh\frac{T}{a} \sin\q\,,\\
Z_{3} \rovno \pm a\cosh \p \sinh\yy\,,\nonumber\\
Z_{4} \rovno \pm a\cosh \p \cosh\yy\,.\nonumber
\end{eqnarray}

\noindent
\textbf{Subcase} ${\epsilon_{2}=0}$, ${\epsilon_{0}=\pm 1}$:

\noindent
The transformation (\ref{relads01}) puts (\ref{eq:adS01ds}) into (\ref{eq:adScoflat}), and similarly for (\ref{eq:adS0-1ds}).

\noindent
\textbf{Subcase} ${\epsilon_{2}=-1}$, ${\epsilon_{0}=1}$:
\begin{eqnarray}\label{eq:adS-11atransf}
&&
p = a\,\cosh \p\,,\qquad
q = \sinh\q\,,\qquad
t = \frac{T}{a}\,,
\end{eqnarray}
transforms (\ref{eq:adS-11ds}) to
\begin{equation}\label{eq:adS-11ds1}
\dd s^{2}=\cosh^{2}\p\,\Big(-\cosh^{2}\q\, \dd T^{2}+a^{2}\,\dd \q^{2}\Big) +a^{2}\left(\dd \p^{2} + \sinh^{2}\p\,\dd \y^{2}\right),
\end{equation}
corresponding to
\begin{eqnarray}
Z_{0} \rovno a\cosh \p \cosh\q \cos\frac{T}{a}\,,\nonumber\\
Z_{1} \rovno a\cosh \p \sinh\q\,,\nonumber\\
Z_{2} \rovno a\sinh \p \cos\y\,,\\
Z_{3} \rovno a\sinh \p \sin\y\,,\nonumber\\
Z_{4} \rovno a\cosh \p \cosh\q\sin\frac{T}{a}\,.\nonumber
\end{eqnarray}

\noindent
\textbf{Subcase} ${\epsilon_{2}=-1}$, ${\epsilon_{0}=0}$:
\begin{eqnarray}\label{eq:adS-10atransf}
&&
p = a\,\cosh \p\,,\qquad
|q| = \exp\q\,,\qquad
t = \frac{T}{a}\,,
\end{eqnarray}
transforms (\ref{eq:adS-10ds}) to
\begin{equation}\label{eq:adS-10ds1}
\dd s^{2}=\cosh^{2}\p\,\Big(-\exp^{2}\q\, \dd T^{2}+a^2\dd \q^{2}\Big) +a^{2}\left(\dd \p^{2} + \sinh^{2}\p\,\dd \y^{2}\right),
\end{equation}
corresponding to
\begin{eqnarray}
Z_{0} \rovno \pm\frac{1}{2a}\cosh \p\,\exp\q\left(a^{2}+s\right)\,,\nonumber\\
Z_{1} \rovno \pm\frac{1}{2a}\cosh \p\,\exp\q\left(a^{2}-s\right)\,,\nonumber\\
Z_{2} \rovno a\sinh \p\cos\y\,,\\
Z_{3} \rovno a\sinh \p\sin\y\,,\nonumber\\
Z_{4} \rovno \pm \,T\cosh \p\,\exp \q\,,  \hspace{10mm} s=a^2\exp^{-2}\q-T^2\,.\nonumber
\end{eqnarray}

\noindent
\textbf{Subcase} ${\epsilon_{2}=-1}$, ${\epsilon_{0}=-1}$:
\begin{eqnarray}\label{eq:adS-1-1atransf}
&&
|p| = a\,\cosh \p\,,\qquad
q = \sin\frac{T}{a}\,,\qquad
t = \q\,,\hspace{5mm} \textrm{for}\ |q|<1\,,  \nonumber\\
&&
|p| = a\,\cosh \p\,,\qquad
|q| = \cosh\q\,,\qquad
t = \frac{T}{a}\,,\hspace{5mm} \textrm{for}\ |q|>1\,,
\end{eqnarray}
transforms (\ref{eq:adS-1-1ds}) to
\begin{eqnarray}\label{eq:adS-1-1ads}
&&
\dd s^{2}=\cosh^{2}\p\,\Big(-\dd T^{2}+a^{2}\cos^{2}\frac{T}{a}\,\dd \q^{2}\Big) +a^{2}\left(\dd \p^{2} + \sinh^{2}\p\,\dd \y^{2}\right), \nonumber\\
&&
\dd s^{2}=\cosh^{2}\p\,\Big(-\sinh^{2}\q\, \dd T^{2}+a^{2}\,\dd \q^{2}\Big) +a^{2}\left(\dd \p^{2} + \sinh^{2}\p\,\dd \y^{2}\right),
\end{eqnarray}
respectively, corresponding to the parametrizations
 \begin{equation}
\left. \begin{array}{l}
Z_0 = {\displaystyle \pm a\cosh \p \cos\frac{T}{a}\cosh \q}\,, \\[8pt]
Z_1 = {\displaystyle \pm a\cosh \p \cos\frac{T}{a}\sinh \q}\,, \\[8pt]
Z_2 = {\displaystyle a\sinh \p \cos\y}\,, \\[8pt]
Z_3 = {\displaystyle a\sinh \p \sin\y}\,, \\[8pt]
Z_4 = {\displaystyle a\cosh \p \sin\frac{T}{a}}\,,
 \end{array} \!\right\} \hspace{0mm} \textrm{for}\ |q|<1\,, \ \left. \!
 \begin{array}{l}
Z_0 = {\displaystyle \pm a\cosh \p \sinh\q \sinh\frac{T}{a}}\,, \\[8pt]
Z_1 = {\displaystyle \pm a\cosh \p \sinh\q \cosh\frac{T}{a}}\,, \\[6pt]
Z_2 = {\displaystyle a\sinh \p \cos\y} \\[2pt]
Z_3 = {\displaystyle a\sinh \p \sin\y}\,, \\[6pt]
Z_4 = {\displaystyle \pm a\cosh \p \cosh\q}\,,
 \end{array} \!\right\} \hspace{0mm} \textrm{for}\ |q|>1\,.
 \label{eq:adS-1-1aparpl}
 \end{equation}

\section{Appendix D: Transformations to standard metric forms of (anti-)de~Sitter spacetime}\label{apD}

\subsection{The de~Sitter spacetime (${\Lambda>0}$)}

\noindent
\textbf{Subcase} ${\epsilon_{2}=1}$, ${\epsilon_{0}=1}$:
By applying transformation
\begin{equation}\label{DStranstand1}
|p| = \sqrt{a^2-R^2\sin^2\theta}\,,\qquad
q = \frac{R\cos\theta}{\sqrt{a^2-R^2\sin^2\theta}}\,,\qquad
t=\frac{T}{a}\,,
\end{equation}
the metric (\ref{eq:dS11ds}) is put into standard spherical form of the de~Sitter spacetime (see (4.9) in~\cite{GriPod2009})
\begin{equation}\label{eq:dSsf}
\dd s^2=-\left(1-\frac{R^2}{a^2}\right)\dd T^{2}
+\left(1-\frac{R^2}{a^2}\right)^{-1} \dd R^2
+R^2 \left(\dd\theta^{2}+\sin^{2}\theta\,\dd\phi^{2}\right).
\end{equation}

\noindent
\textbf{Subcase} ${\epsilon_{2}=1}$, ${\epsilon_{0}=0}$:
With
\begin{equation}\label{DStranstand2}
p = \sqrt{a^2-\rho^2\exp^2(\tau/a)}\,,\qquad
|q| = \frac{a}{\sqrt{a^2\exp^2(-\tau/a)-\rho^2}}\,,\qquad
t=\frac{x}{a}\,,
\end{equation}
the metric (\ref{eq:dS10ds}) is put into the well-known exponentially expanding flat FLRW form of the de~Sitter spacetime (see (4.14) in~\cite{GriPod2009}) with ${y=\rho\cos\phi}$, ${z=\rho\sin\phi}$:
\begin{equation}\label{eq:dSexpds}
\dd s^2=-\dd \tau^2+\exp^2\frac{\tau}{a}\,
\left(\dd x^2+\dd\rho^2+\rho^2\dd \y^2 \right).
\end{equation}

\noindent
\textbf{Subcase} ${\epsilon_{2}=1}$, ${\epsilon_{0}=-1}$:
With
\begin{equation}\label{DStranstand3}
p = a\,\sqrt{1-\sinh^2(\tau/a)\,\sinh^2\theta}\,,\qquad
q = \frac{\sinh(\tau/a)\,\cosh\theta}{\sqrt{1-\sinh^2(\tau/a)\,\sinh^2\theta}}\,,\qquad
t=\q\,,
\end{equation}
the de~Sitter metric (\ref{eq:dS1-1ds}) transforms into the Bianchi III form (see (4.21) in~\cite{GriPod2009}):
\begin{equation}\label{eq:dSBIIIds}
\dd s^2=-\dd \tau^2+a^2\cosh^2\frac{\tau}{a}\,\dd \q^2+a^2\sinh^2\frac{\tau}{a}\left(\dd\theta^{2}+\sinh^{2}\theta\,\dd\y^{2}\right).
\end{equation}

Interestingly, inverse transformations to (\ref{DStranstand1}), (\ref{DStranstand2}), (\ref{DStranstand3}) have very similar structure, namely
\begin{eqnarray}
\sqrt{1-\frac{R^2}{a^2}} \rovno \frac{|p|}{a}\,\sqrt{1-q^2}\,,\qquad \tan\theta=\frac{\sqrt{a^2-p^2}}{|p|\,q}\,,\nonumber\\
\exp\frac{\tau}{a} \rovno \frac{p}{a}\,|q|\,,\hspace{26mm} \frac{\rho}{a}=\,\frac{\sqrt{a^2-p^2}}{p\,|q|}\,,\\
\cosh\frac{\tau}{a} \rovno \frac{p}{a}\,\sqrt{1+q^2}\,,\hspace{6.5mm} \tanh\theta=\frac{\sqrt{a^2-p^2}}{p\,q}\,.\nonumber
\end{eqnarray}

\subsection{The anti-de~Sitter spacetime (${\Lambda<0}$)}

\noindent
\textbf{Subcase} ${\epsilon_{2}=1}$, ${\epsilon_{0}=1}$:
Using
\begin{eqnarray}\label{DStranstand4}
&& |p| = \sqrt{\tau^2\exp^2(x/a)-a^2}\,,\qquad
q = \frac{y}{\sqrt{\tau^2-a^2\exp^2(-x/a)}}\,,\nonumber\\
&&\tanh t=-\frac{(\tau^2-y^2-a^2)\exp^2(x/a)-a^2}{(\tau^2-y^2+a^2)\exp^2(x/a)-a^2}\,,
\end{eqnarray}
the metric (\ref{eq:adS11ds}) for ${|q|<1}$ is transformed into
\begin{equation}\label{eq:adSexp2}
\dd s^{2}=\dd x^{2}+\exp^2\frac{x}{a}\,\Big(-\dd \tau^{2}+\tau^{2}\dd\yy^{2}+\dd y^{2}\Big)\,,
\end{equation}
which is simply related via
${\eta_{_0} = \tau\cosh\yy}$, ${z= \tau\sinh\yy}$
to standard form of the anti-de~Sitter spacetime (see (5.16) in~\cite{GriPod2009})
\begin{equation}\label{eq:adSexp1}
\dd s^{2}=\dd x^{2}+\exp^2\frac{x}{a}\,\Big(-\dd \eta_{_0}^{2}+\dd y^{2}+\dd z^{2}\Big)\,.
\end{equation}

\noindent
\textbf{Subcase} ${\epsilon_{2}=1}$, ${\epsilon_{0}=0}$:
Similarly, with
\begin{equation}\label{DStranstand5}
p = \sqrt{\tau^2\exp^2(x/a)-a^2}\,,\qquad
|q| = \frac{a}{\sqrt{\tau^2-a^2\exp^2(-x/a)}}\,,\qquad
t = \frac{y}{a}\,,
\end{equation}
the metric (\ref{eq:adS10ds}) is transformed into the metric (\ref{eq:adSexp2}).

\noindent
\textbf{Subcase} ${\epsilon_{2}=1}$, ${\epsilon_{0}=-1}$:
By
\begin{eqnarray}\label{DStranstand6}
&& p = \sqrt{\tau^2\exp^2(x/a)-a^2}\,,\qquad
q = \pm\frac{(\tau^2-y^2-a^2)-a^2\exp^2(-x/a)}{2a\,\sqrt{\tau^2-a^2\exp^2(-x/a)}}\,,\nonumber\\
&&\tan t=\frac{2a\,y\,\exp^2(x/a)}{(\tau^2-y^2+a^2)\exp^2(x/a)-a^2}\,,
\end{eqnarray}
the metric (\ref{eq:adS1-1ds}) is also put into (\ref{eq:adSexp2}).

\noindent
\textbf{Subcase} ${\epsilon_{2}=-1}$, ${\epsilon_{0}=1}$:
With
\begin{equation}\label{DStranstand7}
p = a\,\sqrt{1+\sinh^2r\,\sin^2\theta}\,,\qquad
q = \frac{\tanh r\,\cos\theta}{\sqrt{1-\tanh^2r\,\cos^2\theta}}\,,\qquad
t=\frac{T}{a}\,,
\end{equation}
the metric (\ref{eq:adS-11ds}) transforms into the anti-de~Sitter metric (\ref{D4}) in the global static coordinates (see (5.4) in~\cite{GriPod2009})
\begin{equation}\label{eq:adSglob2}
\dd s^2=-\cosh^2 r\> \dd T^2+a^2 \Big(\dd r^2+\sinh^2 r\>(\dd\theta^2+\sin^2\theta\> \dd\phi^2)\Big)\,.
\end{equation}

\noindent
\textbf{Subcase} ${\epsilon_{2}=-1}$, ${\epsilon_{0}=0}$:
By applying
\begin{equation}\label{DStranstand8}
p = \sqrt{a^2+\rho^2\exp^2(x/a)}\,,\qquad
|q| = \frac{a}{\sqrt{a^2\exp^2(-x/a)+\rho^2}}\,,\qquad
t = \frac{\eta_{_0}}{a}\,,
\end{equation}
the metric (\ref{eq:adS-10ds}) is put into the axially symmetric form
\begin{equation}\label{eq:adSexp3}
\dd s^{2}=\dd x^{2}+\exp^2\frac{x}{a}\,(-\eta_{_0}^{2}+\dd\rho^{2}+\rho^{2}\dd\y^{2})\,.
\end{equation}
which is related via
${y = \rho\cos\y}$, ${z= \rho\sin\y}$
to the standard form (\ref{eq:adSexp1}).

\noindent
\textbf{Subcase} ${\epsilon_{2}=-1}$, ${\epsilon_{0}=-1}$:
With
\begin{equation}\label{DStranstand9}
|p| = a\,\sqrt{1+\sinh^2r\,\sin^2\theta}\,,\qquad
q = \frac{\cosh r\,\sin(T/a)}{\sqrt{1+\sinh^2r\,\sin^2\theta}}\,,\qquad
\tanh t=\frac{\coth r\,\cos(T/a)}{\cos\theta}\,,
\end{equation}
for ${|q|>1}$, (\ref{eq:adS-1-1ds}) is transformed into the anti-de~Sitter metric (\ref{D4}) in the global static coordinates (for ${|q|<1}$, the function $\tanh t$ is replaced by $\coth t$).

\end{document}